\documentclass[a4paper,11pt]{article}
\pdfoutput=1 
\usepackage{jcappub}

\usepackage[T1]{fontenc} 
\usepackage{ulem}

\title{Anisotropic separate universe and Weinberg's adiabatic mode}

\author[a,b]{Takahiro Tanaka,}
\emailAdd{t.tanaka@tap.scphys.kyoto-u.ac.jp}
\affiliation[a]{Department of Physics, Kyoto University, Kyoto 606-8502, Japan}
\affiliation[b]{Center for Gravitational Physics, Yukawa Institute for Theoretical Physics, Kyoto University, Kyoto 606-8502, Japan}
\author[c, d]{Yuko Urakawa}
\emailAdd{yuko@physik.uni-bielefeld.de}
\affiliation[d]{Fakult\"at f\"ur Physik, Universit\"at Bielefeld, 33501 Bielefeld, Germany}
\affiliation[c]{Department of Physics and Astrophysics, Nagoya University,
Chikusa, Nagoya 464-8602, Japan}

\abstract{In the separate universe approach, an inhomogeneous universe is rephrased as a set of glued numerous homogeneous local patches. This is the essence of the gradient expansion and the $\delta N$ formalism, which have been widely used in solving a long wavelength evolution of the universe. In this paper, we show that the separate universe approach can be generically used, as long as a theory under consideration is local and preserves the spatial diffeomorphism invariance. Focusing on these two conditions, we also clarify the condition for the existence of the so-called Weinberg's adiabatic mode. Remarkably, the separate universe approach and the $\delta N$ formalism turn out to be applicable also to models with shear on large scales and also to modified theories of gravity, accepting violation of four-dimensional diffeomorphism invariance. The generalized $\delta N$ formalism enables us to calculate all the large scale fluctuations, including gravitational waves. We also argue several implications on anisotropic inflation and ultra slow-roll inflation.}

\newcommand{\sDif}{$[$sDiff$]~$}
\newcommand{\locality}{$[$locality$]~$}
\newcommand{\approxGe}{$[$approx${\cal G}_{\epsilon}$$]~$}
\newcommand{\Lma}{{\cal L}_{\rm matter}}
\newcommand{\lHL}{\lambda_{\rm HL}}
\newcommand{\bm}[1]{\hbox{\boldmath{$#1$}}}
\newcommand{\sbm}[1]{\hbox{\boldmath{\scriptsize$#1$}}}
\newcommand{\oep}[1]{{\cal O}(\epsilon^{#1})}
\newcommand{\dNex}{g$\delta N$ formalism}
\newcommand{\Lee}{{{\cal L}\hspace{-6pt}\raise 1.5pt \hbox{-}\,}}

\begin{document} 
\maketitle
\flushbottom

\section{Introduction and summary}  \label{sec:intro}
The crucial window to probe the physics of cosmic inflation is observing the primordial fluctuations which were generated during inflation and subsequently stretched far beyond the Hubble horizon scales. In order to connect the theoretical prediction of inflation to various observations, we need to solve the evolution, when the scale of interest is much larger than the accessible scale by any causal propagation. The hierarchical difference between these two scales provides a new expansion parameter $\epsilon$, enabling a completely different expansion scheme, called the gradient expansion~\cite{Salopek:1990jq, Shibata:1999zs, Deruelle:1994iz}, from the standard cosmological perturbation theory.

At the leading order of the gradient expansion, two different local regions in the universe which are separated more than the scale of the causal communication evolve mutually independently. This is the basic assumption in {\it the separate universe approach}~\cite{Salopek:1990jq, Wands:2000dp, Lyth:2004gb}, which identifies the inhomogeneous universe with glued numerous local regions which evolve independently. With this identification, the time evolution of the inhomogeneous universe can be computed merely by solving the time evolution of the background homogeneous and isotropic universe with different initial conditions without solving the partial differential equations. This largely facilitates the computation of the non-linear evolution at large scales, which is necessary to evaluate the primordial non-Gaussianity. Based on the gradient expansion, the $\delta N$ formalism \cite{Starobinsky:1982ee, Starobinsky:1986fxa, Sasaki:1995aw, Lyth:2004gb, Sasaki:1998ug} was developed to calculate the primordial spectrum of the adiabatic curvature perturbation $\zeta$, including its non-Gaussian spectrums. In Refs.~\cite{Tanaka:2006zp, Weinberg:2008nf, Weinberg:2008si, Takamizu:2010xy, Naruko:2012fe}, it was later extended to the next to leading order of the gradient expansion.

Observing the primordial non-Gaussianity through the cosmic microwave background (CMB) and the large scale structure (LSS) can uncover a detailed property of the inflaton. Furthermore, these observations can also explore a possible imprint of spectator fields, which are not the driving force of inflation but still contribute to the primordial perturbations. Apart from spectator scalar fields, which has attracted vast attention also as a source of the primordial non-Gaussianity or the isocurvature perturbation, non-zero spin fields might be excited during inflation. The generation of a vector field perturbation through the coupling with the inflaton $\phi$, e.g., $I(\phi) F_{\mu \nu} F^{\mu \nu}$~\cite{Ratra:1991bn, Caldwell_2011, Martin:2007ue} or non-minimal coupling with gravity~\cite{Turner:1987bw}, has been discussed as a mechanism of the primordial magnetogenesis. Here, $F_{\mu \nu}$ denotes the field strength of the $U(1)_Y$ gauge field. A significant growth of the vector field results in a large backreaction on the dynamics of the inflaton and the geometry of the universe. While a too large backreaction can spoil inflation~\cite{Demozzi:2009fu}, a mild backreaction can lead to a sustainable anisotropic inflation with shear on large scales~\cite{Watanabe:2009ct, Watanabe:2010fh, Kanno:2010nr, Soda:2012zm}, providing a counterexample of cosmic no-hair conjecture~\cite{Wald:1983ky}. Recently, Gong el al. investigated the effective field theory prescription~\cite{Cheung:2007st} of the anisotropic inflation~\cite{Gong:2019hwj}. In Refs.~\cite{Graham:2015rva, Nakayama:2019rhg}, the possibility that the generated vector field on large scales serves as a vector oscillating dark matter was explored, while keeping the backreaction negligible. In Ref.~\cite{Kehagias:2017cym}, the model in Refs.~\cite{Ratra:1991bn, Caldwell_2011, Martin:2007ue} with the kinetic coupling $I(\phi) F_{\mu \nu} F^{\mu \nu}$ was generalized to discuss the generation of general integer spin fields through the kinetic coupling with the inflaton as a framework of cosmological collider physics~\cite{Arkani-Hamed:2015bza, Lee:2016vti, Ghosh:2014kba}. The imprints of these non-zero spin fields encoded in the primordial curvature perturbation might be detectable from measurements of CMB~\cite{Akrami:2019izv, Bartolo:2017sbu, Bordin:2019tyb,Franciolini:2018eno} and LSS~\cite{MoradinezhadDizgah:2018pfo, MoradinezhadDizgah:2018ssw, Schmidt:2015xka, Chisari:2016xki, Kogai:2018nse, Kogai:2020vzz}.

The original $\delta N$ formalism~\cite{Starobinsky:1982ee, Starobinsky:1986fxa, Sasaki:1995aw, Lyth:2004gb, Sasaki:1998ug} only applies to a model composed of scalar fields as matter contents. Although the $\delta N$ formalism has been used more broadly~\cite{Karciauskas:2008bc, Abolhasani:2013zya}, the range of the validity remains unclear. Having considered the recent increasing interest in models with non-zero (integer) spin fields, in this paper, we upgrade the gradient expansion and the $\delta N$ formalism, enabling an application to such models. When the contribution of the non-zero spin fields to the geometry is not negligible, the conventional argument which verifies the separate universe picture (by ensuring the approximate validity of the momentum constraints)~\cite{Sasaki:1998ug, Sugiyama:2012tj, Garriga:2015tea} cannot apply.

In this paper, we will show that the separate universe approach can be justified, as long as the two fundamental conditions are satisfied, including a model with non-zero spin fields. This enables a generalization of the $\delta N$ formalism, which we dub the generalized $\delta N$ formalism (the \dNex). Another assumption which is usually employed in the gradient expansion and the $\delta N$ formalism, leaving aside a few exceptions~\cite{Izumi:2011eh, Gumrukcuoglu:2011ef}, is the spacetime diffeomorphism (Diff) invariance. It will turn out that our new formalism can apply even when the spacetime Diff is broken down to the spatial Diff.

The final time of the $\delta N$ formalism, at which we evaluate the curvature perturbation $\zeta$, is usually set after the background trajectory in the field space converges. This is because $\zeta$ is conserved in time after the convergence of the trajectories~\cite{Wands:2000dp, Salopek:1990jq, Sasaki:1995aw, Lyth:2004gb}, when $\zeta$ has the constant solution in the large scale limit, called the Weinberg's adiabatic mode (WAM)~\cite{Weinberg:2003sw}. In this paper, we seek for a general condition that ensures the existence of the WAM. To make a systematic argument possible, going beyond case studies, we distinguish the existence of the WAM from the conservation of the adiabatic curvature perturbation $\zeta$. When $\zeta$ is conserved the WAM should exist, while even if the WAM exists, $\zeta$ is not always conserved, since it is not necessarily the dominant solution.

\begin{figure}
    \centering
    \includegraphics[
    width=\textwidth]{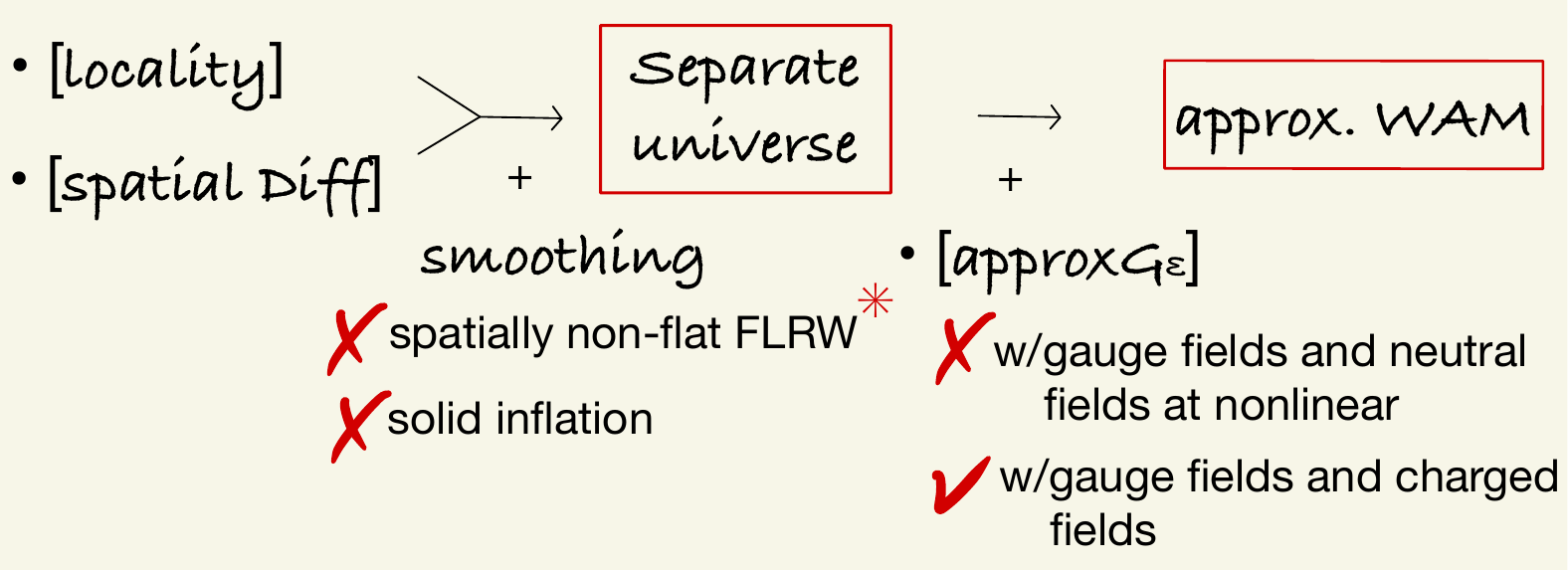}
    \hfill
    \caption{\label{fig:chart}This is the flowchart of the main claim. The first arrow leads to the validity of the separate universe evolution, $(\star)$ (see Sec.~\ref{SSec:smoothing}) and the second one leads to the approximate existence of the Weinberg's adiabatic mode (WAM). The models with a cross are excluded and the one with a check is allowed at each step. We put $\ast$ on the spatially non-flat FLRW, since $(\star)$ still holds, when the curvature scale is much larger than the size of the causal patch
     (see Sec.~\ref{SSec:homogeneitysmoothing}).}
\end{figure}

The main claim of this paper is summarized in Fig.~\ref{fig:chart}. In Sec.~\ref{SSec:conditions}, we show that the locality of the theory and the spatial Diff invariance ensure the separate universe picture, on which the gradient expansion is based~\cite{Salopek:1990jq}. When these two fundamental conditions are valid, we can capture the large scale evolution of the universe just by solving the field equations in the homogeneous limit, which is nothing more than solving the background evolution. This enables us to apply the $\delta N$ formalism also to models with anisotropies on large scales \cite{Watanabe:2009ct, Watanabe:2010fh, Kanno:2010nr, Soda:2012zm} and also to models in which the spacetime Diff is broken down to the spatial Diff. \cite{Horava:2009uw}. Furthermore, in the generalized formalism, 
not only the curvature perturbation, we can compute all the quantities which do not vanish in the large scale limit, including gravitational waves. In Sec.~\ref{Sec:WAM}, we show that when one more condition, which we call the \approxGe condition, is also satisfied, the WAM exists as an approximate large scale solution. In Sec.~\ref{SSec:deltaN}, we apply the generalized $\delta N$ formalism to a model of anisotropic inflation. In Ref.~\cite{Watanabe:2009ct}, it was shown that shear on large scales can survive without being diluted by the cosmic expansion, but the amplitude is bounded by the slow-roll parameter $\varepsilon$. We generalize their result, removing the slow-roll approximation. We also make a brief comment on an inflection type inflation model, which has attracted attention also as a model which can predict primordial black holes. In this paper, we only consider a classical theory. In Sec.~\ref{Sec:futureissues}, we argue possible implications of this paper on quantized system. Table \ref{tab:my_label} summarizes the symbols introduced in this paper.

\section{Gradient expansion and separate universe} \label{SSec:conditions}
In this paper, we express the $(d+1)$-dimensional line element as
\begin{align}
    ds^2 = - N^2 dt^2 + g_{ij} (dx^i + N^i dt) (dx^j + N^j dt)\,, 
\end{align}
with $i, j = 1, \cdots, d$. We express the spatial metric as
\begin{align}
    g_{ij} = e^{2 \psi}\, \gamma_{ij}\,,
\end{align}
where $\gamma_{ij}$ satisfies $\det[\gamma]=1$. Using $\psi$, the determinant of $g_{ij}$ is given by $g= e^{2 d\psi}$.

\subsection{Smoothing and separate universe evolution}  \label{SSec:smoothing}
The gradient expansion \cite{Salopek:1990jq, Shibata:1999zs} is an expansion scheme with respect to the spatial gradient at each given time, which provides a useful tool to address the long wavelength evolution of the fluctuations in a cosmological setup (see also Ref.~\cite{Deruelle:1994iz}). The gradient expansion starts with smoothing out small scale fluctuations~\cite{Salopek:1990jq}. We express a set of the coarse grained fields which are smoothed out at a physical scale $\lambda_{\rm s}(t)$ as $\{ \varphi^a \}$. We require that the smoothing scale $\lambda_s$ should be equal to or larger than the scale of the causal propagation during the time scale of an expanding universe\footnote{In this paper, as a typical time scale of the causal propagation, we consider the time scale of the cosmic expansion, $1/K$. When we only consider a much shorter time scale than $1/K$ with $\delta t \ll 1/K$, obviously the size of the causal patch becomes smaller. Then, $\lambda_s$ can be chosen as $c_{\rm max}/K > \lambda_{\rm s} > c_{\rm max} \delta t$.} and much smaller than the scale under consideration, $\lambda$, satisfying~\cite{Wands:2000dp}
\begin{align}
    \lambda \gg \lambda_{\rm s} \geq \frac{c_{\rm max}}{K} \,. \label{SH}
\end{align}
Here, $K$ denotes the trace of the extrinsic curvature of the time constant hypersurface, which corresponds to ($d$ times) the Hubble parameter for the FLRW spacetime (the definition will be given in Sec.~\ref{SSec:shear}) and $c_{\rm max}$ denotes the maximum speed of propagation in the system under consideration, i.e., $c_{\rm max} \equiv {\rm max}\, \{c_a\}$, where $c_a$ denotes the propagation speed of $\varphi^a$. The fields $\{ \varphi^a \}$ include both the metric and matter fields. 

\begin{table}[]
    \centering
    \begin{tabular}{|c|l|}
    \hline
       Symbol  & ~~~~~~~~~~~~~~~~~~~~~~~~~~~~~~~~~~~~~~Definition  \\
\hline
      $(\star)$   & Separate universe evolution, defined in Sec.~\ref{SSec:smoothing}\\
      ${\cal C}$   & Set of constraints\\
      ${\cal G}$   & Set of gauge constraints, which serve generators of gauge symmetries \\
      ${\cal G}_{\epsilon}$   & Set of gauge constraints in which all terms vanish in the limit $\epsilon \to 0$ \\
      ${\cal X}$   & Set of gauge conditions\\
      ${\cal X}_{\epsilon}$ & Set of gauge conditions that vanish in the limit $\epsilon \to 0$\\
      ${\cal H}$   & Hamiltonian constraint\\
      ${\cal H}_i$   & Momentum constraints\\
      ${\cal H}_{\rm U(1)}$   & Gauge constraint of U(1) gauge symmetry, Eq.~(\ref{Exp:GC_U1})\\
      ${\cal E}_{\rm all}$   & All field equations after solving non-gauge constraints ${\cal C} \cap {\cal G}^c$\\
      ${\cal E}_{\rm ev}$   & Set of all evolution equations\\
      ${\cal E}$   & Sum of ${\cal E}_{\rm ev}$ and ${\cal G} \cap {\cal G}_\epsilon^c$, i.e., ${\cal E}_{\rm ev} \cup ({\cal G} \cap {\cal G}_\epsilon^c)$\\
      $\{\varphi^a\}$ & All fields after solving ${\cal C} \cap {\cal G}^c$ and imposing ${\cal X}$ which are not ${\cal X}_\epsilon$\\
      $\{\varphi^{a'}\}'$ & Set of remaining fields after  solving ${\cal C} \cap {\cal G}^c_{\epsilon}$ for  $\{\varphi^a\}$\\
      $\{\varphi^{a_{\rm phys}}\}_{\rm phys}$ & Set of all physical degrees of freedom\\
\hline
    \end{tabular}
    \caption{This table summarizes the symbols introduced in this paper. Here $c$ denotes a complementary set. The detailed definition for ${\cal G}_\epsilon$ is given in Sec.~\ref{SSSec:GC} and the ones for $\{\varphi^{a}\}$, $\{\varphi^{a'} \}'$, and $\{\varphi^{a_{\rm phys}}\}_{\rm phys}$ are given in Sec.~\ref{SSec:proofGE}. By definition, ${\cal C} \supseteq {\cal G} \supseteq {\cal G}_{\epsilon}$, ${\cal E}_{\rm all} \supseteq {\cal E} \supseteq {\cal E}_{\rm ev}$, and $\{\varphi^a\} \supseteq \{\varphi^{a'}\}' \supseteq \{\varphi^{a_{\rm phys}}\}_{\rm phys}$. We can reduce $\{\varphi^{a'}\}'$ to $\{\varphi^{a_{\rm phys}}\}_{\rm phys}$ by solving ${\cal G}_\epsilon$ and removing residual gauge degrees of freedom (if any).}
    \label{tab:my_label}
\end{table}

\begin{table}[]
    \centering
    \begin{tabular}{|c|l|}
    \hline
       Symbol  & ~~~~~~~~~~~~~~~~~~~~~~~~~~~~~~~~~~~~~~Examples  \\
\hline
      ${\cal G}$   & ${\cal H}$ for a $(d+1)$-dim Diff theory,\, ${\cal H}_i$,\,${\cal H}_{\rm U(1)}$ \\
      ${\cal G}_{\epsilon}$   & ${\cal H}_i$,\, ${\cal H}_{\rm U(1)}$ in the absence of charged fields\\
      ${\cal G} \cap {\cal G}^c_{\epsilon}$   & ${\cal H}$ for a $(d+1)$-dim Diff theory,\, ${\cal H}_{\rm U(1)}$ in the presence of charged fields\\
      ${\cal X}_{\epsilon}$ & Transverse condition $\partial_i {\gamma^i}_j=0$, Coulomb gauge condition $\partial_i A^i=0$\\
      ${\cal X} \cap {\cal X}^c_{\epsilon}$ & $N_i=0$,\, $A_0 =0$\\
      ${\cal E}_{\rm ev}$   & $(i, \, j)$ component of Einstein equations, Klein-Gordon equation\\
\hline
    \end{tabular}
    \caption{This table summarizes examples for sets of the equations listed in Table \ref{tab:my_label}.}
    \label{tab:my_label2}
\end{table}


As a consequence of the smoothing, operating the spatial gradient gives rise to the suppression by a small quantity, which is usually characterized by the spatial variation of $\varphi^a$ within each causally connected patch with the size $c_{\rm max}/(e^{\psi} K)$, 
\begin{align}
    \epsilon  \equiv {\rm max} \left\{  \frac{1}{|\varphi^a|} \left| \frac{c_a}{e^{\psi} K} \partial_i \varphi^a \right| \right\} \,. \label{Cond:graex1}
\end{align}
The value of $\epsilon$ depends on the smoothing scheme such as the sharpness of the window function which smooths out the small scale fluctuation. Without going into the detail of the smoothing scheme, we simply assume that the window function is an analytic function and $\epsilon$ takes a value of $(0 \leq )\, \epsilon \ll 1$. Then, $|c_a \partial_i \varphi^a/(e^{\psi} K)|$ is approximately bounded by $\epsilon |\varphi^a|$, allowing us to expand the field equations in terms of the small parameter $\epsilon$, which is the gradient expansion.
In the standard cosmological perturbation theory, the fields $\varphi^a$ are divided into the background configuration and perturbations. The field equations are solved order by order after expanding them with respect to the perturbations. 
The $\epsilon$ expansion, which is the spatial gradient expansion 
on each time slice, provides a completely different expansion scheme from the standard cosmological perturbation theory. The validity of the gradient expansion does not require that the deviation from the background should be perturbatively small.

Since the field values in each causal patch are almost homogeneous, we can identify an inhomogeneous universe with glued numerous nearly homogeneous patches. The basic assumption of the separate universe approach is the validity of the following identity~\cite{Salopek:1990jq, Lyth:2004gb},
\begin{align*}
&[{\rm Separate~universe~evolution}~(\star)]:\cr
    &\quad \,\,(\mbox{Solving the dynamics of an inhomogeneous universe}) \cr
& \quad \,\,\,= (\mbox{Solving the dynamics of glued nearly homogeneous patches independently})\,
\end{align*}
which enables us to avoid solving a set of the partial differential equations for the coarse-grained fields, $\{\varphi^a (t,\, \bm{x})\}$. When $(\star)$ holds, the time evolution of the inhomogeneous universe is determined solely by solving a set of the corresponding ordinary differential equations, where the spatial gradient terms are simply dropped, for a corresponding initial condition. We express a set of solutions which satisfy the field equations in this limit as $\{\bar{\varphi}^a (t)\}$. In order to emphasize that $(\star)$ is the identification of the evolution in two pictures for a common initial condition, we call it separate universe evolution instead of separate universe approach, which has been widely used.

Since $\{\varphi^a (t,\, \bm{x})\}$ is inhomogeneous, these fields in different patches, in general, take different values. This is incorporated by assigning different initial conditions, correspondingly, to these fields in different causal patches as
\begin{align}
   \lim_{\epsilon \to 0}  \varphi^a (t,\, \bm{x}) = \bar{\varphi}^a (t; \{\bar{\varphi}^{a'}(t_*)= \varphi^{a'}(t_*,\, \bm{x} ) \}')\,, \label{Cond:asymp}
\end{align}
where $t_*$ denotes the initial time. Here, $\{ \varphi^{a'}(t_*,\, \bm{x} )\}'$ denotes a set of metric and matter fields by imposing gauge conditions and also by solving constraint equations, which will be specified in Sec.~\ref{SSec:proofGE}. For clarity, we use the index with a dash to denote the reduced set of the fields. The first line of $(\star)$ corresponds to the left hand side of Eq.~(\ref{Cond:asymp}) and the second line corresponds to the right hand side. Equation (\ref{Cond:asymp}) equates the leading terms of the solution expanded with respect to the small parameter of the gradient expansion $\epsilon$ (the left hand side) with the solution of the equations in the limit $\epsilon \to 0$ (the right hand side).

As is clear from the definition, (\ref{Cond:graex1}), the expansion in terms of $\epsilon$ is the expansion with respect to the spatial gradient. Lyth, Malik and Sasaki introduced a small parameter for their gradient exapsion, $\epsilon_{\rm LMS}$, as the ratio between the Fourier mode $k$ which corresponds to the scale of interest and the scale of the cosmic expansion, i.e., $\epsilon_{\rm LMS} \equiv c_{\rm max} k/(e^{\psi} K)$~\cite{Lyth:2004gb}. Their expansion scheme with respect to $\epsilon_{\rm LMS}$ is slightly different from our expansion in terms of $\epsilon$. When the coarse-grained field $\varphi^a$ is expressed by the Fourier modes, our scheme includes all the modes with $k/e^{\psi} < 1/\lambda_{\rm s}\, (\leq K/c_{\rm max})$. By contrast, $\epsilon_{\rm LMS}$ focuses on a single Fourier mode under consideration. When we consider non-linear perturbation, different Fourier modes need to be considered altogether. 
Therefore, in this paper, we use $\epsilon$ as an expansion parameter, which takes into account all the Fourier modes below $e^{\psi}/\lambda_{\rm s}$. When the Fourier mode expansion is possible, the condition $\epsilon \ll 1$ implies that all the Fourier modes included in the smeared fields $\varphi^a$ should satisfy $k/e^{\psi} \ll K/c_{\rm max}$, ensuring also $\epsilon_{\rm LMS} \ll 1$.

\subsection{Homogeneity and smoothing}  \label{SSec:homogeneitysmoothing}
If non-vanishing gradient of some field is necessary to describe the configuration, the smoothing condition $\epsilon \ll 1$ is violated. As an example, let us consider the case where the coarse-grained spatial metric in each causal patch, $\bar{g}_{ij}$, is approximately given by the Friedmann‐Lema\^{i}tre‐Robertson-Walker (FLRW) metric with a non-zero spatial curvature as
\begin{align}
    g_{ij}(t,\, \bm{x}) \simeq a^2(t) \left[ \delta_{ij} + \frac{{\cal K} x_i x_j}{1 - {\cal K} \delta_{kl} x^k x^l}  \right]\,, \label{Eq:hatgnK}
\end{align}
where ${\cal K} = \pm 1$ and $a(t)$ denotes the scale factor. The condition, $\epsilon \ll 1$, requires that the curvature radius $a/\sqrt{|{\cal K}|}$ should be much larger than the size of the causal patch $c_{\rm max}/K$, indicating that the spatial curvature within each patch is effectively zero. This requirement is safely satisfied in computing the superhorizon evolution of the wavelengths below the present horizon scale, because the spatial curvature of the present universe $|\Omega_K|$ is bounded by ${\cal O}(10^{-3})$~\cite{Aghanim:2018eyx}. Meanwhile, the coarse-grained spatial metric needs to be almost flat but it is not necessarily isotropic, including Bianchi type I spacetime (recall that the spatial curvature is not accepted).

Another example is the solid inflation, which was proposed in Ref.~\cite{Endlich:2012pz}. The matter content in the solid inflation consists of $d$ scalar fields $\phi^I(x)$ with $I=1,\,2,\,\cdots, d$ that preserve the internal shift and rotational symmetries. The solid action which is compatible with the required symmetry is given by
\begin{align}
 & S = \int d^4 x F (X,\, Y,\, Z) + \cdots \,, \label{Exp:actionsolid}
\end{align} 
where $X$, $Y$, and $Z$ are given by $X= [B]$, $Y= [B^2]/[B]^2$, and $Z= [B^3]/[B]^3$ with $B^{IJ} \equiv \partial_\mu \phi^I \partial^\mu \phi^J$. Here, the dots denote the trace over the matrix indices $I,\, J,\, \cdots$. In Eq.~(\ref{Exp:actionsolid}), higher derivative terms, which do not contribute to low-energy physics, are abbreviated. The background configuration of the solid is space dependent, $ \phi^I= x^I$, while the translation and rotation symmetries are preserved if simultaneous change of the internal field space coordinates is performed. In this model with the spatial configuration, a term with $\partial_i \phi^I$ can contribute as ${\cal O}(\epsilon^0)$ without being suppressed by small quantities. A Lagrangian density which includes $B^{IJ}$ in denominator can yield a negative power of $\epsilon$, when $B^{IJ}$ is dominated by the spatial gradient term. Meanwhile, as long as $B^{IJ}$ is dominated by the time derivative term as a consequence of the smoothing, the same theory does not give rise to any negative powers of $\epsilon$.

Since we postulate the smoothing which ensures that operating the spatial gradient always yields the suppression with a positive power of the small parameter $\epsilon$, the following discussions on the IR physics do not apply to the cases mentioned above. In fact, it is known that the WAM does not exist both for the perturbed non-flat FLRW spacetime~\cite{Garriga:1998he, Linde:1999wv} and solid inflation~\cite{Endlich:2012pz} (see also \cite{Bordin:2017ozj}).

\subsection{Basic conditions}
The gradient expansion developed in Refs.~\cite{Salopek:1990jq, Shibata:1999zs} assumes general relativity where any acausal propagation of information is prohibited. The causality ensures that two different patches which are separated much more than $c_{\rm max}/K$ evolve almost independently, allowing only the tiny mutual influence among them, which is mediated by the spatial gradient terms suppressed by $\epsilon \ll 1$. This effect can be taken into account order by order through the perturbative expansion with respect to $\epsilon$. In particular, in the limit $\epsilon \to 0$, the evolution of each causal patch can be determined completely independently of other patches for scalar field systems~\cite{Salopek:1990jq, Sasaki:1998ug}.

Having considered that the inflationary universe provides a natural laboratory to explore new physics at an extremely high energy, it is crucially important to establish a systematic tool to compute the primordial perturbations for a general model of inflation. For this purpose, in this paper, we derive the general condition that ensures the validity of the separate universe evolution, $(\star)$, and the $\delta N$ formalism. It turns out that $(\star)$ can be verified generically, including extended models of gravity and also non-zero spin fields as matter contents.

\subsubsection{Spatial Diff invariance} 
First, we require 
\begin{itemize}
    \item \sDif : The theory which describes the coarse-grained fields $\{\varphi^a (t,\, \bm{x})\}$ should remain invariant under the $d$-dim spatial diffeomorphism (Diff),  
\begin{align}
     x^i \to \tilde{x}^i(t,\, x^i).
\end{align}
\end{itemize}
The $d$-dim spatial Diff invariance is a sub group of $(d+1)$-dim Diff invariance. Variations of the spatial Diff invariant theories are summarized, e.g., in Ref.~\cite{Blas:2010hb}. The foliation preserving Diff invariance is satisfied in the Horava-Lifshitz (HL) gravity~\cite{Horava:2009uw}, which is known to be power-counting renormalizable. Lately, in Refs.~\cite{Barvinsky:2015kil, Barvinsky:2017zlx}, the perturbative renormalizability was shown for the projectable subclass of the HL gravity. The violation of the $(d+1)$-dim Diff invariance leads to the appearance of an additional scalar degree of freedom in the gravity sector, dubbed Khronon.

One may want to consider a theory whose action does not include the lapse function. Nevertheless, since the action for most theories, including a $(d+1)$-dim Diff invariant theory and the HL gravity, is naturally written down using the lapse function, in what follows we include $N$, when we write down an explicit example of the action. Let us, however, emphasize that the Hamiltonian constraint, which can be derived by the functional derivative of the action with respect to $N$, does not play any important role in the following discussion, enabling an extension to a theory without containing the lapse function straightforwardly.

In a $(d+1)$-dim Diff invariant theory, using the energy conservation, which is ensured by the time Diff, we can show the existence of the constant solution for the curvature perturbation in the uniform density slicing~\cite{Wands:2000dp, Lyth:2004gb}. It is intuitively natural that the time Diff implies the existence of a conserved quantity. However, this argument cannot apply to a theory where the $(d+1)$-dim Diff invariance is violated. Meanwhile, it is known that the curvature perturbation is also conserved at large scales, e.g., in non-projectable version of the HL gravity~\cite{ArmendarizPicon:2010rs, Arai:2018whk} (see also Ref.~\cite{Kobayashi:2010eh}). We show that the constant solution also exists, even when the $(d+1)$-dim Diff invariance is broken down to the $d$-dim Diff invariance.

\subsubsection{Locality condition} \label{SSSec:GE}
As emphasized above, in general relativity, the causality is crucial to verify the separate universe approach and the gradient expansion. In the absence of the $(d+1)$-dim Diff invariance, there is no clear bound on the propagation speed from the causality. Therefore, instead, as the second condition we require
\begin{itemize}
    \item \locality : After solving all the constraint equations except for gauge constraints ${\cal G}$, the effective dynamics of the coarse-grained fields is described by the Lagrangian density ${\cal L}(t, x^i)$ given locally by a function of fields at $x^\mu=(t, x^i)$. Taking variation of the corresponding action gives local field equations for the coarse-grained fields.
\end{itemize}
We call a constraint which generates a gauge symmetry a gauge constraint, expressing it as ${\cal G}$. Gauge constraints can be derived by taking derivative with respect to Lagrange multipliers associated with the gauge symmetry. The gauge constraint, which is first class, can transform into a second class constraint by imposing a gauge condition.  The variation of the action for ${\cal L}(t, x^i)$ gives field equations for the set of the fields $\{ \varphi^a \}$ which remain after all the non-gauge constraints are solved and the gauge conditions are employed. Since the locality is imposed for the coarse-grained fields, our \locality condition does not exclude a theory in which non-locality appears only in the dynamics of the fine-grained fields but not in that of the coarse-grained fields (a recent attempt to construct such theories can be found e.g., in Ref.~\cite{Buoninfante:2018lnh}). We implicitly assume that all the evolution equations do not vanish in the limit $\epsilon\to 0$ and can be expanded in powers of spatial gradient.

Eliminating a Lagrange multiplier can transform a local Lagrangian density to a non-local one, especially when the Lagrange multiplier is accompanied by a spatial gradient. For example, let us consider a system with a single dynamical scalar field $\phi$ and a Lagrange multiplier $\chi$ whose Lagrangian density is given by 
\begin{align}
    {\cal L}(x) = \cdots  + g^{ij} \partial_i \chi \partial_j \chi/2 + \chi\, g^{ij} \partial_i \phi \partial_j \phi
    + \cdots \,,  \label{Example}
\end{align}
where we have written only the terms which include $\chi$. 
Taking the functional derivative of the action with respect to $\chi$, we obtain the constraint equation as
\begin{align}
   \frac{1}{N \sqrt{g}} \partial_i ( g^{ij} N \sqrt{g} \,\partial_j \chi(x)) =  g^{ij} \partial_i \phi \partial_j \phi
   \,.  
   \label{eq:elliptic}
\end{align}
Once we eliminate $\chi$ by solving this constraint equation, ${\cal L}$ becomes non-local because of the appearance of the inverse Laplacian. Such models do not satisfy the \locality condition, even though the original Lagrangian does not contain non-local operators manifestly.

The presence of non-local terms can cause a serious problem for the validity of the separate universe approach, since two patches whose physical distance is much larger than $c_{\rm max}/K$ can interact through non-local terms. Especially, if the Lagrangian density includes terms with the inverse Laplacian $\partial^{-2}$, a term with the spatial derivative operator is no longer necessarily suppressed by $\epsilon$, since it may be compensated by $\partial^{-2}$. The \locality condition requires that after eliminating the Lagrange multipliers by solving all the constraints, ${\cal C}$,  except for the gauge constraints ${\cal G}$, the action becomes local with an appropriate choice of the gauge fixing conditions. Solving a gauge constraint such as the momentum constraints can also give rise to  non-local terms in the action. Nevertheless, as will be discussed in the next subsection, unlike non-gauge constraints, the existence of gauge constraints does not disturb the validity of the separate universe evolution.

The Hamiltonian constraint ${\cal H}$ for a $(d+1)$-dim Diff invariant theory and momentum constraints ${\cal H}_i$ are one of the gauge constraints ${\cal G}$. When the $(d+1)$-dim Diff is broken down to the foliation preserving Diff, e.g., in the HL gravity, ${\cal H}$ is not necessarily ${\cal G}$. For the projectable version of the HL gravity, since the lapse function is not allowed to depend on the spatial coordinates, ${\cal H}$ only gives a global constraint, which is integrated over a time constant hypersurface. Therefore, precisely speaking the projectable version does not satisfy the \locality condition. Instead, one may want to consider a theory where the time reparametrization invariance is also explicitly broken, e.g., by simply setting $N$ to 1 in the projectable version. In this case, since the (global) Hamiltonian constraint does not exist, the \locality condition can be fulfilled. Meanwhile, for the non-projectable version, since $N$ also depends on the spatial coordinates, we obtain ${\cal H}$ at each spacetime point. Since the time coordinate transformation is limited to $t \to \tilde{t}(t)$, unlike the global Hamiltonian constraint, the Hamiltonian constraint for each patch is not ${\cal G}$. As long as the latter can be solved without giving rise to non-local contributions, the \locality condition is satisfied. This is the situation discussed in Refs.~\cite{ArmendarizPicon:2010rs, Arai:2018whk}.

Strictly speaking, to figure out when the \locality condition holds, we need to understand how the short modes below the smoothing scale, $\lambda_{\rm s}$, affect the dynamics of $\{ \varphi^a \}$. As is widely known in the context of stochastic inflation~\cite{Starobinsky:1986fx, Nambu:1987ef, Nambu:1988je, Starobinsky:1994bd} (see e.g., Refs.~\cite{Tokuda:2017fdh, Gorbenko:2019rza, Baumgart:2019clc} for a recent progress), the effective action obtained by integrating out the degrees of freedom of short modes is, in general, given by a non-local effective Lagrangian density, which does not satisfy the \locality condition defined above. 
Therefore, in our forthcoming paper~\cite{YUquantum}, we will relax the \locality condition. The refined \locality condition restricts the smoothing scheme and also the quantum system, particularly the quantum state of the short modes to be traced out. In Ref.~\cite{YUquantum}, we will show that under the refined \locality condition, the discussion proceeds almost in the same way as in a classical theory. In this paper, focusing on a classical theory, we simply require the \locality condition defined above.

\subsection{Validity of separate universe evolution}
In this subsection, we show that the separate universe evolution $(\star)$ indeed holds under the \locality and \sDif conditions.

\subsubsection{Gauge constraints}  \label{SSSec:GC}
When all the terms in a gauge constraint are suppressed by the spatial gradient, solving the constraint gives rise to non-local terms in the action (recall the discussion in the previous subsection). In what follows, we express a set of gauge constraints as ${\cal G}_{\epsilon}$, when all the terms therein are suppressed by at least one spatial gradient, i.e., 
$$
{\rm a~constraint~} \chi \in  {\cal G}_{\epsilon} \qquad \quad {\rm if~} \chi \in {\cal G} {\rm ~and~} \chi = {\cal O}(\epsilon)\,. 
$$

For example, the momentum constraints ${\cal H}_i=0$, which are obtained by taking the variation of the action with respect to $N_i$, is one of ${\cal G}_\epsilon$. We can confirm that all the terms in ${\cal H}_i$ are indeed accompanied by at least one spatial gradient as follows. Using ${\cal H}_i$, which is the generator of the spatial Diff, the change of a field $\varphi^a$ under a spatial coordinate transformation (\ref{Exp:sDiff}) is given by
\begin{align}
   \left\{ \varphi^a(x), \int d^d y \sqrt{g(y)} \xi^i {\cal H}_i(y) \right\} = \Lee_\xi  \varphi^a(x)\,.  \label{Eq:Higenerator}
\end{align}
Here, $\{\cdot,\, \cdot \}$ denotes the Poisson brackets and $\Lee_\xi$ denotes the Lie derivative. The generator, which satisfies Eq.~(\ref{Eq:Higenerator}), is expressed as
\begin{align}
    \int d^d x \sqrt{g(x)} \xi^i {\cal H}_i(x) = \sum_a{}' \int d^d x\, \pi_a(x)  \Lee_\xi  \varphi^a(x)\,, \label{Exp:Higenerator}
\end{align}
where $\sum{}'$ sums over all the fields in $\{ \varphi^a \}$ other than $N$ and $N^i$. Tensor indices of $\{ \varphi^a \}$ are abbreviated, and $\pi_a(x)\equiv \partial(N\sqrt{g}{\cal L})/\partial \dot\varphi^a$ is the conjugate momentum of $\varphi^a$.

One can easily understand that ${\cal H}_i$ in Eq.~\eqref{Exp:Higenerator} is 
equivalent to the usual momentum constraints defined by 
$\tilde {\cal H}_i\equiv N \partial {\cal L}/\partial N^i$, when we use the equations of motion $\delta S/\delta \varphi^a=0$. The variation of the action caused by the spatial Diff transformation, 
\begin{align}
 x^i \to e^{\xi^\mu \partial_\mu} x^i = x^i + \xi^i + \cdots  \label{Exp:sDiff}   
\end{align}
with $\xi^\mu = (0,\, \xi^i(t,\, \bm{x}))$, which should vanish, can be calculated as 
\begin{align} \label{eq:spatialtranslation}
    0 = \delta_\xi S =  &\, \int d^{d+1} x \left\{ \xi^i \partial_t\left(\sqrt{g}{\cal \tilde H}_i\right)+ \sum_{a}{}' \frac{\delta S}{\delta \varphi^a} \delta_\xi \varphi^a + {\cal O}(\epsilon^2)\right\}\cr
      & +\left[-\int d^d x \sqrt{g(x)} \xi^i \tilde {\cal H}_i(x) + \sum_a{}' \int d^d x\, \pi_a(x)  \Lee_\xi  \varphi^a(x)\ \right]_{t_i}^{t_f}\,, 
\end{align}
where we used $\delta_\xi N=0$ and $\delta_\xi N^i = - \partial_t \xi^i + {\cal O}(\epsilon)$. 
Here $\delta_\xi$ denotes the variation under the spatial Diff transformation (\ref{Exp:sDiff}). The variation of $\varphi^a$ is given by $\delta_\xi\varphi^a= \Lee_\xi \varphi^a$. Using the equations of motion $\delta S/\delta \varphi^a=0$, we find $\tilde{\cal H}_i = {\cal H}_i$ and 
\begin{align}
    \frac{\partial (\sqrt{g}{\cal H}_i)}{\partial t} =0\,,\label{Exp:deltaS_sDiff2}
\end{align}
at the leading order of $\epsilon$. 

Performing the integration by parts in the left hand side of (\ref{Exp:Higenerator}), we can derive the explicit form of ${\cal H}_i$. Since all the terms in $\Lee_\xi  \varphi^a(x)$ are accompanied with at least one spatial derivative, so are the terms in ${\cal H}_i$, i.e., ${\cal H}_i \in {\cal G}_\epsilon$. Using 
\begin{align}
  \Lee_\xi g_{ij} = - \xi^k \partial_k g_{ij} - \partial_i \xi^k g_{kj} - \partial_j \xi^k g_{ik}\,,
\end{align}
and performing the integration by parts, the contribution of the variation of the spatial metric $\delta_\xi g_{ij}$ in the right hand side of Eq.~(\ref{Exp:Higenerator}) can be further rewritten into a more widely used form as 
\begin{align}
     \int d^d x\, \pi^{ij}(x)  \Lee_\xi  g_{ij}(x) =  2 \int d^d x\, \sqrt{g(x)}\, \xi^i \nabla_j\! \left(\frac{{\pi^j}_i(x)}{\sqrt{g(x)}} \right)\,,  \label{Exp:Hi}
\end{align}
where $\pi^{ij}$ is the conjugate momentum of $g_{ij}$, defined by
\begin{align}
\label{eq:piij}
    \pi^{ij} \equiv \frac{\partial (N\!\sqrt{g} {\cal L})}{\partial \dot{g}_{ij}} \,,
\end{align}
and 
${\pi^j}_i \equiv g_{ik} \pi^{jk}$. Here, $\nabla_i$ is the covariant derivative for the spatial metric $g_{ij}$. The conjugate momentum divided by $\!\sqrt{g}$, which transforms as a rank $(1,\, 1)$ tensor, satisfies $\sqrt{g}\, \nabla_j ({\pi^j}_i/\sqrt{g}) = \partial_j {\pi^j}_i - \pi^{jk} g_{jk, i}$. Unlike the momentum constraints, the Hamiltonian constraint is not ${\cal G}_{\epsilon}$.

Another example of ${\cal G}_\epsilon$ is the gauge constraint for a U(1) gauge field. Let us consider a matter Lagrangian density ${\cal L}_{\rm matter}$, given by   
\begin{align}
    {\cal L}_{\rm matter} =  {\cal F} (\phi^I,\,X^{IJ},\, Y,\, \tilde{Y})\,, \label{Exp:Lmatter_U1}
\end{align}
where $\phi^I$ with $I=1,\, 2,\, \cdots$ denotes neutral or charged scalar fields with $X^{IJ}$, $Y$ and $\tilde{Y}$ being
\begin{align}
    X^{IJ} \equiv - \frac{1}{4} (D^{(I)}_\mu \phi^I D^{(J)\mu} \phi^{J*} + {\rm c.c.})\,,
\end{align}
and
\begin{align}
    Y \equiv - \frac{1}{4} F_{\mu \nu} F^{\mu \nu}\,, \qquad \tilde{Y} \equiv -\frac{1}{4} \tilde{F}_{\mu \nu} F^{\mu \nu}\,.
\end{align}
Here $F_{\mu \nu}$ and $\tilde{F}_{\mu \nu}$ are the field strength and its dual for a U(1) gauge field $A_\mu$, respectively, and $D^{(I)}_\mu$ is given by
\begin{align}
    D^{(I)}_\mu \equiv \partial_\mu - i e^{(I)} A_\mu \,, 
\end{align}
where $e^{(I)}$ is the U(1)-charge of $\phi^I$. 
For a neutral scalar field, the charge $e^{(I)}$ should be set to 0. The Lagrangian density, (\ref{Exp:Lmatter_U1}), includes the typical Lagrangian for the axion and also the Euler-Heisenberg Lagrangian as a special case.

In what follows (unless stated), we impose the gauge condition 
\begin{align}
    N_i=0 \,, \label{GC}
\end{align}
which is local in the sense that it does not vanish in the limit $\epsilon\to 0$. 
As is known, $N_i=0$ does not completely remove the degrees of freedom for the spatial small gauge transformations. In fact, at linear perturbation, $N_i$ transforms under the spatial coordinate transformation, $x^i \to \tilde{x}^i = x^i + \xi^i$ as 
\begin{align}
    \tilde{N}_i = N_i - \dot{\xi}^j g_{ji}\,.\label{rGC}
\end{align}
This indicates that we can still perform a time-independent spatial coordinate transformation, maintaining the gauge condition $N_i=0$. We will discuss an additional gauge condition to remove this residual gauge degree of freedom at the end of Sec.~\ref{SSec:proofGE}.

Taking derivative with respect to $A^0$, we obtain the gauge constraint, which corresponds to the Gauss law, as 
\begin{align}
  {\cal H}_{\rm U(1)} \equiv \partial_i\pi^i_A + J^0
   = 0 \label{Exp:GC_U1}\,,
\end{align}
where $\pi_A^i$ is the conjugate momentum given by
\begin{align}
    \pi_A^i \equiv \frac{\partial (N \sqrt{g} {\cal L}_{\rm matter}) }{\partial \dot{A}_i} =   \sqrt{g} {\cal F}_Y g^{ij} \frac{\dot{A}_j}{N}  + {\cal O}(\epsilon) \,, \label{Exp:piAex_U1}
\end{align}
with ${\cal F}_Y \equiv \partial {\cal F}/\partial Y$, and $J^0$ is the 0-th component of the current, which corresponds to the electric charge density (multiplied by $\sqrt{g}$) and is given by
\begin{align}
    J^0 = \frac{\partial (N \sqrt{g} {\cal L}_{\rm matter}) }{\partial A_0} = \frac{ \sqrt{g} {\cal F}_{IJ}}{4N} \left\{ ie^{(I)} \phi^I (D_0^{(J)} \phi^J)^* - i D_0 \phi^I e^{(J)} \phi^{J*} + {\rm c.c.}  \right\} \,,
\end{align}
with ${\cal F}_{IJ} \equiv \partial {\cal F}/\partial X^{IJ}$. In the presence of the charged scalar fields, since the current $J^0$ includes terms which are not suppressed by $\epsilon$, the gauge constraint does not become ${\cal G}_\epsilon$. To satisfy consistently the gauge constraint (\ref{Exp:GC_U1}), which we denote as ${\cal H}_{\rm U(1)}$, $J^0 = {\cal O}(\epsilon)$ should hold, requiring the fall-off of the current $J^0$ in the large scale limit. In the temporal gauge with $A_0=0$, the gauge constraint reads\footnote{As an example, let us consider the case when only one of the scalar fields $I=\bar{I}$ is charged and the other fields are neutral. Then, expressing $\phi_{\bar{I}} = \bar{\phi} e^{i \theta}$, we can rewrite Eq.~(\ref{Exp:current}) as
$$
 \sqrt{g} {\cal F}_{\bar{I} \bar{I}} \bar{\phi}^2 \frac{\dot{\theta}}{N} = {\cal O}(\epsilon)\,,   
$$
which requires the absence of the azimuth motion in the large scale limit. This is consistent with the known background evolution (see, e.g., Ref.~\cite{Emami:2009vd} and references therein). The constant phase $\theta$ is canceled out from all the field equations at the leading order of the gradient expansion.  
}
\begin{align}
    \frac{\sqrt{g} {\cal F}_{IJ}}{N} \left\{ i e^{(I)} (\phi^I \dot{\phi}^{J*} - {\rm c.c.}) + (I \leftrightarrow J )  \right\} = {\cal O} (\epsilon)\,.  \label{Exp:current}
\end{align}
This is nothing but the charge neutrality on large scales. 
In the absence of charged scalar fields, since $J^0$ vanishes, the gauge constraint ${\cal H}_{\rm U(1)}$ becomes ${\cal G}_\epsilon$.

Similarly to the gauge constraints, the gauge conditions also can vanish in the limit $\epsilon \to 0$. We express the whole set of gauge conditions as ${\cal X}$ and those which vanish in the limit $\epsilon \to 0$ as ${\cal X}_{\epsilon}$.  A typical example of ${\cal X}_{\epsilon}$ is the transverse condition, $\partial_i \gamma^{ij}=0$. Meanwhile, our gauge condition $N_i=0$ is not ${\cal X}_\epsilon$. When we remove gauge degrees of freedom by imposing ${\cal X}_{\epsilon}$, the \locality condition can fail to hold due to non-locality caused by employing ${\cal X}_{\epsilon}$.

\subsubsection{Proof of separate universe evolution}  \label{SSec:proofGE}
Let us classify all the field equations which remain after solving the non-gauge constraints, ${\cal E}_{\rm all}$, into the evolution equations ${\cal E}_{\rm ev}$ and the gauge constraints ${\cal G}$, i.e., ${\cal E}_{\rm all}= {\cal E}_{\rm ev} \cup {\cal G}$. For our convenience, we introduce ${\cal E} \equiv  {\cal E}_{\rm ev} \cup ({\cal G} \cap {\cal G}_\epsilon^c)$, where the gauge constraints whose leading terms are suppressed by $\epsilon$, ${\cal G}_\epsilon$, are excluded. Here, the superscript $c$ denotes a complementary set. The \locality condition ensures that solving the non-gauge constraints ${\cal C} \cap {\cal G}^c$ does not give rise to non-local terms in the action. In addition, solving the gauge constraints whose leading terms are not suppressed by $\epsilon$, i.e., ${\cal G} \cap {\cal G}^c_{\epsilon}$ does not yield non-local contributions, either. Therefore, the solution of ${\cal E}$ remains local, verifying the basic condition of the separate universe $(\star)$ and allowing us to systematically solve equations ${\cal E}$ expanded by $\epsilon$. As discussed in Sec.~\ref{SSSec:GC}, we solve ${\cal E}$, choosing the gauge conditions which are not ${\cal X}_\epsilon$.

We express the set of the fields which remain after solving ${\cal C} \cap {\cal G}^c$ and employing the gauge conditions which are not ${\cal X}_\epsilon$ as $\{ \varphi^a \}$. This set of the fields, $\{\varphi^a\}$, includes redundant variables, since the constraints ${\cal G}$ are not yet solved. In addition, the gauge conditions ${\cal X} \cap {\cal X}^c_\epsilon$ may still leave residual gauge degrees of freedom. Distinguishing it from $\{ \varphi^a \}$, we denote the variables which are necessary and sufficient to specify a physical initial configuration by $\{\varphi^{a_{\rm phys}}\}_{\rm phys}$ with the indices $a_{\rm phys}$.

Obviously, the solution space of ${\cal E}$ becomes broader than the physical solution space of the system, since ${\cal G}_\epsilon$ is not yet solved. To address this aspect, let us emphasize the following two properties of ${\cal G}_\epsilon$. First, solving only ${\cal E}$ without ${\cal G}_\epsilon$ completely determines the time evolution of $\{\varphi^{a}\}$ for a given initial condition. Second, when ${\cal G}_\epsilon$ hold at a certain time, say at the initial time, they continue to hold at any time. For example, 
under the gauge transformation,
\begin{equation}\label{eq:U1gauga}
    A_\mu \to A_\mu +  \partial_\mu \theta\,,
\end{equation}
and spatial Diff transformation \eqref{Exp:sDiff}, the variation of the action whose Lagrangian density is given by Eq.~(\ref{Exp:Lmatter_U1}) can be evaluated as
\begin{align}
    0 &\, = \left(\delta_\theta+\delta_\xi \right) S_{\rm matter} \cr
     &\, = \int d^{d+1} x\,
      \left[  -\theta \partial_t\left(\sqrt{g} {\cal H}_{\rm U(1)}\right) 
           +\xi^i \partial_t\left(\sqrt{g} {\cal H}_{\rm U(1)}A_i\right)
            + \xi^i \partial_t\left(\sqrt{g} {\cal H}_i\right)
      + {\cal O}(\epsilon^2) \right]\,,
\end{align}
where we have used $\delta_\xi A_0= -\dot\xi^i A_i + {\cal O}(\epsilon)$, and we have abbreviated the boundary terms and the terms which vanish by using the evolution equations ${\cal E}_{\rm ev}$. 
From this expression, we find 
\begin{align}
 & {\partial \over \partial t}\left(\sqrt{g} \,{\cal H}_{\rm U(1)}A_i + \sqrt{g}{\cal H}_i\right)= 0\,,\\
 & {\partial\sqrt{g} \, {\cal H}_{\rm U(1)}\over \partial t}=0\,,  \label{Eq:consU1g}
\end{align}
holds when all the evolution equations ${\cal E}_{\rm ev}$ are satisfied, 
at the leading order of the expansion with respect to $\epsilon$. In particular, to show the conservation of U(1) gauge constraint (\ref{Eq:consU1g}), we only need to use the evolution equations for the matter fields that have non-zero U(1) charges, for which $\delta_\theta$ does not vanish. When ${\cal H}_{\rm U(1)}$ holds at an initial time, it continues to hold at an arbitrary time, i.e., being conserved in time. Meanwhile, only the linear combination of ${\cal H}_{\rm U(1)}$ and ${\cal H}_i$ is conserved. This was not explicit in Eq.~(\ref{Exp:deltaS_sDiff2}), since all the matter field equations, including ${\cal H}_{\rm U(1)}$, was used to derive it. For $\dot{A}_i \neq 0$, only when both ${\cal H}_i$ and ${\cal H}_{\rm U(1)}$ hold at an initial time, ${\cal H}_i$ continues to hold at an arbitrary time.

Therefore, we can reduce the solution space of $\{ \varphi^a \}$ to the one of $\{ \varphi^{a_{\rm phys}} \}_{\rm phys}$ merely by choosing a proper initial condition that satisfies ${\cal G}$ (and also by removing residual gauge degrees of freedom). The right hand side of Eq.~(\ref{Cond:asymp}) follows a local time evolution determined by solving the equations ${\cal E}$ in the limit $\epsilon \to 0$ at each Hubble patch for a given initial condition $\{\varphi_*^{a'}\}'$. We can specify the initial condition of the separate universe by using $\{ \varphi^{a'} \}'$ which consist of the remaining fields of $\{ \varphi^a \}$ after solving the gauge constraints which do not vanish in the limit $\epsilon \to 0$, i.e., ${\cal G} \cap {\cal G}^c_\epsilon$ such as the Hamiltonian constraint for a $(d+1)$-dim Diff invariant theory.

In summary, at the leading order of the gradient expansion, using Eq.~(\ref{Cond:asymp}), correlation functions of $\varphi^a$ can be computed as follows 
\begin{align} \label{Exp:correlation}
    \langle \varphi^{a_1} (t_1,\, \bm{x}_1) \cdots  \varphi^{a_n} (t_n,\, \bm{x}_n) \rangle 
    &= \int \prod_{i=1}^n D\bar{\varphi}^{a'_i}_*\,  P(\bar{\varphi}^{a'_1}_*,\, \cdots,\, \bar{\varphi}^{a'_n}_*) \big|_{\bar{\varphi}^{a'}_*= \varphi^{a'}(t_*,\, \sbm{x})} \nonumber \\
    & \qquad \qquad \times \bar{\varphi}^{a_1}(t_1;\{\bar{\varphi}^{a'_1}_* \}') \cdots \bar{\varphi}^{a_n} (t_n; \{\bar{\varphi}^{a'_n}_*\}')\,.
\end{align}
The probability distribution of the initial condition, $ P(\bar{\varphi}^{a'_1}_*,\, \cdots,\, \bar{\varphi}^{a'_n}_*) $, should be determined by solving all the equations ${\cal E}_{\rm all}$, which also include ${\cal G}_\epsilon$, until $t_*$. As a result, in general, $ P(\bar{\varphi}^{a'_1}_*,\, \cdots,\, \bar{\varphi}^{a'_n}_*) $ has correlations among different patches. Meanwhile, as we have shown, the quantities in the second line of Eq.~\eqref {Exp:correlation} can be determined by solving ${\cal E}$, the local field equations for $\{\varphi^{a}\}$, correspondingly to the independent evolution of each local patch in the limit $\epsilon \to 0$. This verifies the separate universe evolution $(\star)$. The separate universe evolution $(\star)$ only ensures the locality of its evolution for a given initial condition, but not the locality of the initial distribution. This point becomes important in discussing the existence of WAM, the second arrow in Fig.~\ref{fig:chart}. At higher orders of the gradient expansion, an influence from other patches can be taken into account order by order in the expansion with respect to $\epsilon$.

We have introduced three different sets of the fields, whose mutual relation can be summarized as follows,
\begin{align*}
     & \{ \varphi^a \}: \mbox{All metric and matter fields after solving}~ {\cal C} \cap {\cal G}^c~ \&~ \mbox{imposing gauge conditions} \cr
     &\xrightarrow[ 
 \mbox{solving}~({\cal G} \cap {{\cal G}}_{\epsilon}^c)]{} ~~~ 
 \{\varphi^{a'}\}': \mbox{variables that specify initial conditions of separate universe}~~~ \cr
 & \xrightarrow[\mbox{solving}~
 {\cal G}_{\epsilon}~\mbox{and removing residual gauge DOFs}]{}~~~  \{\varphi^{a_{\rm phys}}\}_{\rm phys}\,.
\end{align*}
If there still exist residual gauge degrees of freedom after imposing gauge conditions that are not ${\cal X}_\epsilon$, we also need to remove them to obtain the set of the physical degrees of freedom. As discussed around Eq.~(\ref{rGC}), to remove the residual gauge degrees of freedom which remain after employing $N_i=0$, one may want to impose the transverse condition at $t=t_*$ as
\begin{align}
    \partial_i \gamma^{ij} (t_*,\, \bm{x}) = 0\,, \label{Exp:Tad}
\end{align}
which is ${\cal X}_\epsilon$. Unlike the gauge condition $\partial_i \gamma^{ij} (t,\, \bm{x}) = 0$ for an arbitrary $t$, whether we employ Eq.~(\ref{Exp:Tad}) or not is irrelevant in computing the evolution of each patch for a given initial condition, i.e., in obtaining the quantities in the second line of Eq.~(\ref{Exp:correlation}). It only affects the initial probability distribution, the first line. Alternatively, one may want to employ the axial gauge conditions, $\gamma_{iz}=0$ at $t=t_*$, avoiding ${\cal X}_\epsilon$. The condition $(\star)$ holds for both gauge conditions.

\subsubsection{Previous works}
Before closing this section, let us summarize the previous works, highlighting the difference from our argument. In the conventional gradient expansion, the asymptotic solution in the limit $\epsilon \to 0$ is restricted by imposing the additional conditions (see, e.g., Ref.~\cite{Lyth:2004gb}),
\begin{align}
 [{\rm removed~assumptions}]:\qquad   N_i = \oep{}\,, \qquad \dot{\gamma}_{ij} = \oep{2}\,.   \label{Cond:conventional}
\end{align}
Since $N_i$ can be set to 0 by performing the gauge transformation (even at the linear order in perturbation), the first condition is 
just a choice of the gauge. By contrast, it is obvious that the second condition does not hold when we consider general models of gravity, e.g., the massive gravity \cite{deRham:2010ik, deRham:2010kj}, in which the graviton is gapped, or in the presence of non-zero spin fields which can source the shear on a large scale. Our argument, which is solely based on the \sDif and \locality conditions, does not restrict the asymptotic behaviour in the limit $\epsilon \ll 1$. In particular, the condition (\ref{Cond:conventional}) requires that the shear (${A^i}_j$, whose definition is given in the next section) should be suppressed by $\epsilon$. However, as will be shown, the decaying solution of $\zeta$, which is also known as the Weinberg's second mode, cannot be properly reproduced without keeping the shear at the leading order of the gradient expansion.

The gradient expansion has been applied only to a scalar field system, with a few exceptions~\cite{Abolhasani:2013zya, Talebian-Ashkezari:2016llx}, which addressed a system with a U(1) gauge field and inflaton as matter components. In these papers, several non-trivial conditions such as $\gamma_{ij} = {\cal O}(\epsilon)$ for $i\ne j$, which restrict large scale anisotropy, were employed. Our argument shows that the separate universe $(\star)$ and the $\delta N$ formalism can apply to general matter contents, as long as the \sDif and \locality conditions are fulfilled, including models with non-zero spin fields, such as anisotropic inflation models \cite{Watanabe:2009ct, Watanabe:2010fh, Kanno:2010nr, Soda:2012zm}. This generalization becomes possible by solving the evolution for the set of fields $\{ \varphi^{a} \}$ not for the independent degrees of freedom $\{ \varphi^{a_{\rm phys}} \}_{\rm phys}$. This enables us to confine the potential non-locality which emerges by solving ${\cal G}_\epsilon$ only in the choice of the initial condition, ensuring the locality of the evolution for a given initial condition.

Specifically for a scalar field system, where the momentum constraints ${\cal H}_i$ are the only 
${\cal G}_\epsilon$ in the system, it is not very important whether we solve the evolution for $\{ \varphi^{a_{\rm phys}} \}_{\rm phys}$ or $\{ \varphi^{a} \}$. Sugiyama et al. showed that, when the other equations of motion than ${\cal H}_i$ are fulfilled, the violation of the momentum constraints, even if it exists, decays with the inverse of the physical spatial volume under the slow-roll approximation~\cite{Sugiyama:2012tj}. This result was extended to a more general time evolution by removing the slow-roll approximation in Ref.~\cite{Garriga:2015tea} (see also Ref.~\cite{Naruko:2012fe}). A lesson from Refs.~\cite{Sugiyama:2012tj, Garriga:2015tea, Naruko:2012fe} is that as far as all the other equations are properly solved, the momentum constraints are also satisfied, leaving aside the error that decays in an expanding universe as $1/\sqrt{g}$, i.e., $\sqrt{g} {\cal H}_i=\,$const.

In Refs.~\cite{Sugiyama:2012tj, Garriga:2015tea}, the approximate validity of the momentum constraints was shown based on a brute force computation in a specific model. This can be also shown generically by using \sDif condition as follows\footnote{In a $d+1$ Diff invariant theory, the Bianchi identity and the conservation law of the energy momentum tensor relate the divergence of the momentum constraints to the time derivative of the Hamiltonian constraint. Therefore, one projected component of the momentum constraints is automatically satisfied (without the error) as long as the Hamiltonian constraint is satisfied. This single relation is sufficient for a system composed of scalar fields at linear order, but obviously not so when one wants to consider non-linear orders in perturbation or a model with a non-zero spin field, e.g., a vector field, which does not immediately decay at large scales. 
}. Evaluating the variation of the action under the spatial coordinate transformation, 
we obtained Eq.~\eqref{eq:spatialtranslation}. 
From this relation, we derived
Eq.~(\ref{Exp:deltaS_sDiff2}), using the equations $\{ {\cal E} \}$. Namely, when the momentum constraints are the only ${\cal G}_\epsilon$ in the system, as long as the equations ${\cal E}$ are solved properly, 
Eq.~(\ref{Exp:deltaS_sDiff2}) holds. This implies that the momentum constraints ${\cal H}_i$ are automatically satisfied, if we accept the "error" which decays as $\propto 1/\sqrt{g}$ in an expanding universe. 
In fact, as will be discussed in the next section, the non-local components that appear in the shear by solving ${\cal H}_i$ decay as $\propto 1/\sqrt{g}$, which implies exponential decay, especially, during inflation. This fact becomes important when we consider the evolution of the physical degrees of freedom $\{ \varphi^{a_{\rm phys}} \}_{\rm phys}$, which is determined by solving all the field equations in the system including ${\cal G}_{\epsilon}$. Meanwhile, if there is another ${\cal G}_\epsilon$, e.g., for a system with neutral scalar fields and U(1) gauge field, we need to solve the evolution for $\{ \varphi^{a} \}$ (see Sec.~\ref{SSec:deltaN}).

\section{Weinberg's adiabatic mode}   \label{Sec:WAM}
The approximate validity of ${\cal G}_\epsilon$, discussed at the end of the previous section, is closely related to the existence of WAM\footnote{In cosmology, the matter content is called adiabatic (even without
referring to their thermodynamic property), when the pressure $P$ is
non-perturbatively given by a function of the energy density $\rho$ without
depending on any other quantities.
Moving to perturbation theory this leads to
\begin{align} 
 & \frac{\delta P}{\dot{\bar{P}}} = \frac{\delta \rho}{\dot{\bar{\rho}}}\,,  \label{Cond:Ad}
\end{align}
where the dot denotes the time derivative with respect to the
cosmological time. For a scalar field system, this condition should be imposed only in the long wavelength limit. Here, $\delta P$ and $\delta \rho$ denote the perturbations and $\bar{P}$ and
$\bar{\rho}$ denote their background values. A solution which satisfies
the adiabatic condition is called the adiabatic mode. The
adiabatic mode can still exist, even when the non-adiabatic mode
also exists. Equation (\ref{Cond:Ad}) states that the time
step measured by the change of the pressure coincides with the one
measured by the change of the energy density, i.e., there is only one clock.
If there is only a scalar field, this can be rephrased as being on an
attractor solution. As is discussed in this section, Weinberg proposed one way to
single out the adiabatic mode by extending the change under the
dilatation into soft modes $\bm{k} \neq 0$~\cite{Weinberg:2003sw}. The WAM can be absorbed by a coordinate transformation in a local patch~\cite{Tanaka:2011aj}. However, this property is not exclusive of the  WAM. Following the convention, we call a mode which is locally identical to a coordinate transformation an adiabatic mode, even if Eq.~(\ref{Cond:Ad}) is not satisfied. Other types of adiabatic modes were reported in Refs.~\cite{Finelli:2017fml, Pajer:2019jhb}.}. In this section, we show that when a system under consideration satisfies \locality and \sDif conditions and also a condition \approxGe$\!$, there approximately exists the WAM as a solution beyond the horizon scale. Here, we define 
\begin{itemize}
    \item \approxGe: When the equations for $\{ \varphi^{a} \}$, ${\cal E}$, are satisfied, the gauge constraints that trivially vanish in the limit $\epsilon \to 0$, ${\cal G}_\epsilon$, are approximately satisfied at the leading order of $\epsilon$.  
\end{itemize}
As discussed in the previous section, a system which only contains scalar fields satisfies the \approxGe condition~\cite{Sugiyama:2012tj, Garriga:2015tea}. 
By contrast, a model with a U(1) gauge field requires a more careful consideration, particularly when the kinetic term factor ${\cal F}_Y$ rapidly decreases in time, making the approximate validity of ${\cal H}_{\rm U(1)}$ non-trivial (see Eq.~\eqref{Exp:piAex_U1}).

As discussed in Sec.~\ref{SSec:smoothing}, when the Fourier mode expansion is possible, the regime with $\epsilon \ll 1$ directly corresponds to considering the superhorizon limit. In this section, in the regime with $\epsilon \ll 1$ we show the approximate existence of the Weinberg's adiabatic mode (WAM), the constant solution of $\psi$, under the three conditions. Unlike the argument in Ref.~\cite{Lyth:2004gb}, which relies on the energy conservation, our argument also applies to the case in which the $(d+1)$-dim Diff invariance is broken down to the $d$ dim spatial Diff invariance. For the non-projectable HL gravity, in Refs.~\cite{ArmendarizPicon:2010rs, Arai:2018whk}, the existence of the constant solution was manifestly shown in linear perturbation. (See Refs.~\cite{Kobayashi:2009hh,Izumi:2011eh,Gumrukcuoglu:2011ef} about the existence of the constant solution in  the projectable HL gravity).

The existence of the constant solution should be clearly distinguished from the  conservation in time. Even if the constant solution of $\psi$, the WAM, exists, if it is dominated by other modes, $\psi$ is not conserved.

\subsection{WAM and [approx\texorpdfstring{${\cal G}_\epsilon$}{Gepsilon}] condition}  \label{SSec:RelationW}
When all the equations to be solved remain invariant under the shift of $\psi$, 
\begin{align}
    \psi(t,\, \bm{x}) \to \psi(t,\, \bm{x}) + c({\bm x}) \,,  \label{Exp:psishift}
\end{align}
where $c({\bm x})$ satisfies
$$ c_{\rm max} |\partial_i c(\bm{x})|/( e^{\psi} K) \ll |c(\bm{x})|\,,$$
there exists a time independent solution of $\psi$ at the leading order of the $\epsilon$ expansion. In the Weinberg's argument~\cite{Weinberg:2003sw}, the invariance under Eq.~(\ref{Exp:psishift}) is ensured by extending the transformation of $\psi$ under the dilatation, $\bm{x} \to \bm{x}^c \equiv e^{-c} \bm{x}$, which is one of the large gauge transformations\footnote{In cosmological perturbation theory, we usually consider only the perturbed variables around an FLRW spacetime which can be expanded in terms of the Fourier modes with finite wavenumbers. Notice that the change due to the dilatation does not satisfy this condition. In fact, the Fourier
transformation of $\delta x^i  \propto  c x^i$ reads $\partial \delta (\bm{k})/(\partial \bm{k})$, 
which includes the delta function for the $\bm{k}=0$ mode. Therefore, the dilatation is out of consideration in the conventional cosmological
perturbation theory.}.

Under the dilatation $\bm{x} \to \bm{x}_c \equiv e^{-c} \bm{x}$, the spatial metric $g_{ij}$ transforms as
 \begin{align}
     g^c_{ij}(t,\, \bm{x}_c) = e^{ 2 c} g_{ij} (t,\, \bm{x})\,, 
 \end{align}
 and $\psi(x) \equiv \ln g(x)/(2d)$ as 
 \begin{align}
     \psi^c (t,\, \bm{x}_c) = \psi(t,\, \bm{x}) + c\,.  \label{Exp:psi_shift}
 \end{align}
 If only $\psi$ changes under the dilatation at ${\cal O}(\epsilon^0)$, the dilatation invariance implies the invariance of the action under the introduction of the additive constant shift of $\psi$ as $\psi_{\sbm{k} =0} \to \psi_{\sbm{k} =0} +c  + \cdots$. Since this is a (large) spatial gauge transformation, with the \sDif condition, the system remains invariant without leaving any change in the physical
configuration. Meanwhile, let us extend this dilatation transformation to the one with a
time-independent but inhomogeneous parameter $c(\bm{x})$, which has comoving wave number $\bm{k}_L\neq 0$. 
Here, the change of the $k=0$ mode under the dilatation with the constant parameter $c$ is extended to that of the soft mode $k_L \neq 0$, just by replacing the parameter $c$ with an inhomogeneous one $c_{\sbm{k}_L}$. Namely, 
$$
\psi_{\sbm{k}_L} \to \psi_{\sbm{k}_L}+ c_{\sbm{k}_L}  + {\cal O}(\epsilon)\,.
$$
Since this extended transformation is no longer a gauge transformation, it alters the physical configuration. The continuity at $\bm{k}=0$ ensures that the additive shift induced by this inhomogeneous dilatation still solves the field equations at the leading order of $\epsilon$, although the solution at higher orders is modified. This solution is called the WAM. In Ref.~\cite{Tanaka:2017nff}, this argument was extended to a non-linear quantum theory.

Since the \locality condition ensures that the equations ${\cal E}$ are local, 
${\cal E}$ are all continuous at $k=0$. Therefore, the WAM should be a solution of ${\cal E}$. 
However, this does not immediately ensure the continuity of physical solutions at $k=0$. In fact, the configuration after the transformation does not satisfy ${\cal G}_\epsilon$ in general at the leading order of $\epsilon$. 
As a result, computing the physical solutions $\{ \varphi^{a_{\rm phys}} \}_{\rm phys}$ by solving ${\cal G}_\epsilon$ can give rise to a singular pole at $k=0$.  When the \approxGe condition is satisfied in addition to the \locality and \sDif conditions, all the solutions of ${\cal E}$ should approximately satisfy ${\cal G}_\epsilon$ at the leading order of the $\epsilon$ expansion. Therefore, the WAM approximately solves the whole set of equations, ${\cal E}_{\rm all}$ in the limit $\epsilon \to 0$.

\subsection{Momentum constraints}  \label{Sec:Eq}
In the rest of this section, we address when the \approxGe condition holds, considering models that satisfy the \locality and \sDif conditions. Let us express the total action as
\begin{align}
    S =\int d^{d+1} x N \sqrt{g} {\cal L} =\int d^{d+1} x N \sqrt{g} \left[  {\cal L}_{\rm g} + \Lma \right] \,,  \label{Exp:S}
\end{align}
where ${\cal L}_{\rm g}$ is the Lagrangian density for the gravitational field and $\Lma$ denotes the matter Lagrangian density. The time derivative of the spatial metric, $g_{ij}$ which remains invariant under the spatial coordinate transformation $x^i \to \tilde{x}^i(t, \bm{x})$ is given by the extrinsic curvature and its trace part, defined as 
\begin{align}
    K_{ij} = \frac{1}{2N} (\dot{g}_{ij} - \nabla_i N_j - \nabla_j N_i)\,, \qquad K \equiv g^{ij} K_{ij}.
\end{align}
Using the extrinsic curvature, the kinetic term of gravity is given by~\footnote{Since the shift vector $N_i$ is related to the $(d+1)$-dim metric component, $g_{0i}$, as $g_{0i} = N_i$, it does not transform as a vector field under $x^i \to \tilde{x}^i(t,  \bm{x})$ (while it does under $x^i \to \tilde{x}^i( \bm{x})$). Therefore, a contraction of $N_i$, $N^i N_i$, is not invariant under the time dependent spatial coordinate transformation.}
\begin{align}
  {\cal L}_{\rm g} &= \frac{F}{16 \pi G} \biggl[K^i\!_j K^j\!_i - \lHL K^2 + \lambda_1 K^3 - \lambda_2 K K^i\!_j K^j\!_i + \lambda_3 K^i\!_j K^j\!_l K^l\!_i + \cdots  +{\cal O}(\epsilon)   \biggr] \,, \label{Def:Lg}
\end{align}
where the coefficients $F$, $\lHL$, $\lambda_i$ with $i=1,\, 2,\, 3$ are arbitrary functions of the matter fields\footnote{The action (\ref{Def:Lg}) only includes the terms which preserve the time reparametrization invariance, i.e., $t \to \tilde{t}(t)$. However, as argued in Ref.~\cite{Blas:2010hb}, without this invariance, more terms can appear. For instance, the coefficients can depend on the lapse function $N$.}. Potential terms of gravity such as the spatial Ricci scalar ${^sR}$ are ${\cal O}(\epsilon^2)$, since they have at least two more spatial gradient compared to the terms shown in Eq.~(\ref{Def:Lg}). With the \locality condition, the non-local contributions are absent in the above expression.

For an illustrative purpose, we have written down several possible terms of ${\cal L}_{\rm g}$ in Eq.~(\ref{Def:Lg}). However, as far as the \sDif and \locality conditions hold, the explicit form is not important. The above Lagrangian density includes general relativity $(\lHL =1,\, \lambda_1= \lambda_2= \lambda_3=0)$, the projectable and non-projectable versions of the HL gravity\footnote{In the HL gravity, ${\cal L}_{\rm g}$ can include the spatial Riemann tensor, the spatial Ricci tensor, and the spatial Ricci scalar with the spatial covariant derivative operators~(see, e.g., \cite{Arai:2018whk}). In the non-projectable version of the HL gravity, the gradient instability in the IR can be avoided by introducing a term proportional to $g^{ij} \partial_i \ln N \partial_j \ln N$~\cite{Blas:2009qj}. This term is also included in ${\cal O}(\epsilon^2)$.} ($\lambda_1= \lambda_2= \lambda_3=0$), and beyond Horndeski ($\lHL =1, \lambda_2= 3 \lambda_1,\, \lambda_3= 2 \lambda_1$ for $d=3$)~\cite{Gleyzes:2014dya}.

Taking the functional derivative of the total action with respect to the Lagrange multipliers $N_i$, we obtain the momentum constraints as
\begin{align}
  & \nabla_j \left[F \left\{ (1 - \lambda_2 K) K^j\!_i + \frac{3}{2} \lambda_3 K^j\!_l K^l\!_i -  \delta^j\!_i \left(\lHL K - \frac{3}{2} \lambda_1 K^2 + \frac{1}{2} \lambda_2 K^i\!_j K^j\!_i  \right) \right\} \right]   \cr
  & \qquad \qquad \qquad \qquad + \cdots + 8 \pi G N \frac{\partial \Lma}{\partial N^i}  =  {\cal O} \left(\epsilon^2\right)\,. \label{Eq:Mc}
\end{align}
Here, the terms expressed by $\cdots$ denote contributions that potentially appear from other kinetic terms of gravity, e.g., $K^4$ and ${K^i}_j{K^j}_l{K^l}_m {K^m}_i$. As discussed in the previous section, not only the geometrical contributions, the matter contribution in the second line should be also suppressed by ${\cal O}(\epsilon)$.

Following Refs.~\cite{Deruelle:1994iz, Shibata:1999zs, Shibata:1995dg}, let us introduce the tracelsss part of $K_{ij}$, the shear $A_{ij}$, as
\begin{align}
   K_{ij} = \frac{K}{d} g_{ij} + A_{ij}\,,  \qquad g^{ij} A_{ij} = 0\,.   \label{Exp:Kdec}
\end{align}
The tensor indices of $A_{ij}$ are raised and lowered by $g_{ij}$. Using the expansion $K$ and the shear ${A^i}_j$, the momentum constraints are rewritten as
\begin{align}
  & 
 \partial_i \left[ F \left\{ \left(  \frac{1}{d} - \lHL \right) K + \frac{3}{2} \left(\lambda_1  - \frac{\lambda_2}{d} +  \frac{\lambda_3}{ d^2}  \right) K^2 - \frac{\lambda_2}{2} {A^k}_l {A^l}_k \right\}   \right] \cr
  &\! +\! \nabla_j \!\left[ F\! \left\{\! {A^j}_i + \! \left( \frac{3\lambda_3}{d}  - \lambda_2 \right)\! K {A^j}_i + \frac{3\lambda_3}{2} A^j\!_l A^l\!_i \right\}\!   \right] \! 
  + \cdots + 8 \pi G N \frac{\partial \Lma}{\partial N^i}\!  =  {\cal O} \left(\epsilon^2  \right)\!.\,\,\,\,\,\label{Eq:dKts}
\end{align}

In $N_i=0$ gauge, $K$ and ${A^i}_j$ are given by 
\begin{align}
    & K = \frac{d}{N} \dot{\psi}\,, \label{Exp:KS} \\
    & {A^i}_j = \frac{1}{2N} \left( \delta^i_k \delta^l_j - \frac{1}{d} \delta^i_j \delta^l_k \right) \gamma^{km} \dot{\gamma}_{ml} = \frac{1}{2N} \gamma^{im} \dot{\gamma}_{mj} = - \frac{1}{2N} \dot{\gamma}^{im} \gamma_{mj}\,. \label{Exp:AijS}
\end{align}
In this gauge, $K$ and ${A^i}_j$ directly correspond to the derivative of $\psi$ and $\gamma_{ij}$ with respect to the proper time $\tau$, which is related to the cosmological time as $d \tau|_{N_i=0} = N dt$. In this gauge, by integrating Eq.~(\ref{Exp:KS}), $\psi$ directly corresponds to the $e$-folding number as
\begin{align}
    \psi(t,\, \bm{x}) = \frac{1}{d} \int^{\tau(t)} d \tau' K(\tau',\, \bm{x})= \frac{1}{d} \int^t dt' N(t',\, \bm{x}) K(t',\, \bm{x}) \,. \label{Exp:deltaN}
\end{align}
 This is the basic equation in the $\delta N$ formalism, which will be discussed in Sec.~\ref{SSec:deltaN}.

\subsection{Evolution of shear on large scales}\label{SSec:shear}
In the next subsection, we point out that the shear, ${A^i}_j$, can be schematically solved as
\begin{align}
    F {A^i}_j(t,\, \bm{x})  \sim \frac{{D^i}_j(\bm{x})}{\sqrt{g}} + ({\rm contributions~sourced~by~matter~fields})^i\!_j \,, \label{Exp:shearKEY}
\end{align}
where ${D^i}_j(\bm{x})$ is time-independent. 
In the absence of the source, Eq.~(\ref{Exp:shearKEY}) simply states that the shear decays in an expanding universe. This property of the shear has been confirmed in many examples~\cite{Deruelle:1994iz}, but it is not (immediately) clear what ensures it. In the next subsection, we will show that the precise version of Eq.~(\ref{Exp:shearKEY}) can be derived, making use of the Noether charges of the large gauge transformations.

The first term of Eq.~(\ref{Exp:shearKEY}), ${D^i}_j(\bm{x})/\sqrt{g}$, corresponds to the geometrical degrees of freedom, whose evolution can be determined without referring to the detail of matter fields. Inserting this expression into the momentum constraints (\ref{Eq:dKts}), we can determine ${D^i}_j(\bm{x})$ so that the momentum constraints are properly solved at a certain time, e.g., at the horizon crossing time of the scales of interest, $t=t_{\rm hoc}$. The symmetric and traceless tensor ${D^i}_j(\bm{x})$ has $(d/2+1)(d-1)$ degrees of freedom, among which $d$ components are determined by solving the momentum constraints and the remaining $(d/2-1)(d+1)$ components correspond to the decaying modes of gravitational waves. Once ${D^i}_j(\bm{x})$ is chosen to satisfy the momentum constraints at one time, as discussed in the previous section, we can verify that the momentum constraints hold at any time. As is clear from Eq.~(\ref{Eq:dKts}), the resultant expression of ${D^i}_j(\bm{x})$ in terms of matter fields becomes non-local\footnote{If ${\cal H}_i=0$ were to be written in the longitudinal type form $\partial_i (\cdots)=0$, they can be solved without introducing non-local integral expression. At linear perturbation around the FLRW spacetime, whose spatial metric is maximally symmetric, the momentum constraints for the scalar type perturbation can be written in the form, $\partial_i (\cdots)=0$~\cite{Kodama:1985bj} (see the momentum constrains derived in Sec.~\ref{SSSec:USR}). Nevertheless, in general, the transverse part of ${\cal H}_i=0$ does not vanish.}. However, since the non-local terms are multiplied by the inverse of the physical volume, $1/\sqrt{g}$, which decays in an expanding universe, they die off after a while. After that, whether the momentum constraints are solved or not becomes insignificant. In particular, if the momentum constraints are the only ${\cal G}_\epsilon$ in the system, the \approxGe condition holds, since all the solutions of ${\cal E}$ satisfy ${\cal H}_i$ approximately, just allowing the errors which decay with $1/\sqrt{g}$. In this approximation, solving the momentum constraints is equivalent to solving the manifestly local equations obtained by setting ${D^i}_j(\bm{x})=0$ in Eq.~\eqref{Exp:shearKEY}.

Before deriving the precise version, taking a detour, let us derive Eq.~(\ref{Exp:shearKEY}) in a specific example also by solving the field equations straightforwardly. As an example, we consider a theory of gravity whose action is given by
\begin{align}
 ({\rm ex}) \qquad \quad {\cal L}_{\rm g} &= \frac{F}{16 \pi G} \biggl[K^i\!_j K^j\!_i - \lHL K^2  +{\cal O}(\epsilon)   \biggr] \,, \label{Def:Lgex}
\end{align}
which includes, e.g., general relativity and the Horava-Lifshitz gravity. Taking the derivative of the action with respect to the spatial metric, $g_{ij}$, we obtain~\footnote{With our definition of the extrinsic curvature $K_{ij}$, $K$ becomes positive for an expanding universe. In Ref.~\cite{Shibata:1999zs}, the definition of $K_{ij}$ has the opposite signature, which has made Eq.~(\ref{Eq:EijTLgen}) slightly different from theirs.} 
\begin{align}
    &\frac{1}{N \sqrt{g} F} \partial_t \left( \sqrt{g} F P^{ij} \right) + 2 (K^{il} {K^j}_{l} - \lHL K^{ij} K) - \frac{1}{2} g^{ij} ({K^l}_m {K^m}_l - \lHL K^2 ) \cr
    & \quad - \frac{1}{N} D_k(P^{ij} N^k ) + \frac{1}{N} D_k (P^{ik} N^j)  + \frac{1}{N} D_k (P^{kj} N^i) + {\cal O} (\epsilon^2 )   = \frac{8 \pi G}{F} T^{ij}\,,  \label{Eq:EijTLgen}
\end{align}
where $T^{ij}$ is (the spatial component of) the energy-momentum tensor and $P^{ij}$ is given by
\begin{align}
    P^{ij} \equiv K^{ij} - \lHL K g^{ij} = A^{ij} + (1 - \lHL d) \frac{K}{d} g^{ij}\,.
\end{align}

Using Eqs.~(\ref{Exp:KS}) and (\ref{Exp:AijS}), the traceless part of the field equation (\ref{Eq:EijTLgen}) can be given by a compact expression as~\footnote{The trace part of Eq.~(\ref{Eq:EijTLgen}) reads
\begin{align}
    \frac{1}{N \sqrt{g} F} \partial_t (F \sqrt{g} K) - \frac{1}{2} K^2  - \frac{d}{2 (1 - \lHL d)} {A^i}_j {A^j}_i =  \frac{d}{1 - \lHL d} \frac{8 \pi G}{F} P + {\cal O}(\epsilon)\,,  \label{Eq:EijT}
\end{align}
where the isotropic pressure $P$ is defined as $P \equiv {T^i}_i/d$. In deriving Eq.~(\ref{Eq:EijT}), we have eliminated the time derivative of ${A^i}_j$ by using Eq.~(\ref{Eq:ijS}). Using the Hamiltonian constraint, we can rewrite Eq.~(\ref{Eq:EijT}) as
\begin{align}
    \frac{1}{N  F} \partial_t (F K)  - \frac{d}{1 - \lHL d} {A^i}_j {A^j}_i = \frac{d}{1 - \lHL d} \frac{8 \pi G}{F} (\rho + P) +  {\cal O}(\epsilon)\,. \label{Eq:EijT2}
\end{align}}
\begin{align}
    \frac{1}{N \sqrt{g}} \partial_t \left( \sqrt{g} F {A^i}_j \right) = 8 \pi G\,{\Pi^i}_j  + {\cal O}(\epsilon)\,, \label{Eq:ijS}
\end{align}
where ${\Pi^i}_j$ is the anisotropic pressure, defined by
\begin{align}
    \Pi^{ij} \equiv \left( \delta^i\!_k \delta^j\!_l - \frac{1}{d} g^{ij} g_{kl} \right) T^{kl}\,.
\end{align}
The quadratic term of $A_{ij}$ does not appear in Eq.\eqref{Eq:EijT2}, when we express the left hand side as the time derivative of ${A^i}_j$ instead of $A_{ij}$. This is the fully non-perturbative equation for the leading order of the gradient expansion. Integrating Eq.~(\ref{Eq:ijS}) over time from $t_{\rm hoc}$, we obtain
\begin{align}
 F {A^i}_j(t,\, \bm{x}) = \frac{\sqrt{g_{\rm hoc}(\bm{x})}}{\sqrt{g(t,\, \bm{x})}} \left[ F_{\rm hoc}  {A^i}_{j\,{\rm hoc}}(\bm{x}) + \int^{\tau(t)}_{\tau_{\rm hoc}} d\tau' \frac{\left(\sqrt{g}  \, 8 \pi G {\Pi^i}_j \right)_{(\tau',\, \sbm{x})}}{(\sqrt{g})_{(\tau_{\rm hoc},\, \sbm{x})}} + {\cal O}(\epsilon) \right], \label{Sol:AijS}
\end{align}
where the $\tau$ integral is performed for a given $\bm{x}$. 
The variables with the subscript ``hoc'' denote those evaluated at $t= t_{\rm hoc}$. The first term in the right hand side generically decays in an expanding universe, being inversely proportional to the physical spatial volume as $\propto 1/(\sqrt{g})$ \cite{Deruelle:1994iz}. Integrating Eq.~(\ref{Sol:AijS}) over time, we obtain
\begin{align}
    & \gamma_{ij} (t,\, \bm{x}) = \gamma_{ij} (t_{\rm hoc},\, \bm{x})  + 2 \int^{\tau(t)}_{\tau_{\rm hoc}} d \tau' \gamma_{il}(\tau',\, \bm{x})   \frac{(F \sqrt{g})_{(\tau_{\rm hoc},\, \sbm{x})}}{(F \sqrt{g})_{(\tau',\, \sbm{x})}} \cr
    & \qquad \qquad \qquad  \qquad  \qquad  \qquad \times \left[{A^i}_{j\,{\rm hoc}}(\bm{x}) + \int^{\tau'}_{\tau_{\rm hoc}} d\tau'' \frac{\left(\sqrt{g}  \, 8 \pi G {\Pi^i}_j \right)_{(\tau'',\, \sbm{x})}}{(F \sqrt{g})_{(\tau_{\rm hoc},\, \sbm{x})}} \right] + {\cal O}(\epsilon) \,, \cr  \label{Sol:gammaij}  
\end{align}
which is a formal solution of $\gamma_{ij}$ at the leading order of the gradient expansion for the theory of gravity, (\ref{Def:Lgex}).

When the large scale anisotropic pressure is suppressed as 
\begin{align}
    |8 \pi G \,{\Pi^i}_j| = {\cal O} (\epsilon)\,, \label{Cond:Pi}
\end{align}
e.g., in a scalar field system, Eq.~(\ref{Sol:AijS}) immediately provides the key expression (\ref{Exp:shearKEY}) we need to show. By choosing ${D^i}_j(\bm{x})$ or equivalently $F_{\rm hoc} {A^i}_{j\,{\rm hoc}}(\bm{x})$ properly, the momentum constraints ${\cal H}_i$ can be satisfied at $t= t_{\rm hoc}$. Although ${D^i}_j(\bm{x})$ determined by solving ${\cal H}_i$ becomes non-local, it soon decays as $\propto 1/\sqrt{g}$ in an expanding universe. Since the shear decays exponentially for general relativity in an inflationary spacetime, Salopek and Bond set ${A^i}_j$ to 0~\cite{Salopek:1990jq}. As was pointed out in Ref.~\cite{Sasaki:1998ug}, while the shear ${A^i}_j$ decays in time, we need to take into account the contribution of ${A^i}_j \propto \sqrt{g}_*/\sqrt{g}$ to reproduce the decaying solution of the curvature perturbation $\zeta$, a.k.a., the Weinberg's second mode, at the leading order of the gradient expansion (see also Refs.~\cite{Sugiyama:2012tj, Garriga:2015tea} and the discussion in Sec.~\ref{SSec:Wseconed}). Meanwhile, when the anisotropic pressure is not suppressed by $\epsilon$ in the limit $\epsilon \ll 1$, the relation between Eq.~(\ref{Exp:shearKEY}) and Eq.~(\ref{Sol:AijS}) may be less clear, since ${{\Pi}^i}_j$ also depends on $ {A^i}_{j\,{\rm hoc}}(\bm{x})$ through a matter source, e.g., when their equations of motion depend on the shear.

\subsection{Noether charge and fall-off of non-local contributions} \label{SSSec:Noether}
In this subsection, we show the key expression, (\ref{Exp:shearKEY}), using a set of the Noether charge (densities) for the spatial large gauge transformations. 

\subsubsection{Noether charge of large gauge transformations}
Gauge transformations, which are parametrized by functions of the spacetime, are in general classified into two categories: small gauge transformations and large gauge transformations. In Ref.~\cite{Avery:2015rga}, the former are defined as local transformations with bounded support and the latter as local transformations with support (as well) on the boundary. When
the support of the theory extends to the infinity, the change under the
former falls off in the limit $|x| \to \infty$, while the change
under the latter does not.

Among the infinite number of large spatial gauge transformations~\cite{Urakawa:2010it, Urakawa:2010kr} (see also Ref.~\cite{Hinterbichler:2013dpa}), we focus on a global transformation, 
\begin{align}
    x^i \to \tilde{x}^i = x^i + {M^i}_j \, x^j \,, \label{Exp:LGT}
\end{align}
where ${M^i}_j$ denotes an infinitesimal constant rank $(1,\, 1)$ tensor with $d^2$ independent components. The trace part of ${M^i}_j$ describes the scale transformation and the traceless part describes the shear transformation and the rotation.

Following the textbook argument, the Noether charges can be derived by writing down the action evaluated at two coordinates related by the large gauge transformations, (\ref{Exp:LGT}) and setting the difference to 0 as 
\begin{align}
     0 &= 
     \int d^{d+1} x 
         \left[\det\left({\partial \tilde x^i\over \partial x^j}\right) \tilde{N}\sqrt{\tilde g} \tilde {\cal L}
           -N \sqrt{g} {\cal L} \right]\cr
    & =  \int d^{d+1} x \biggl[ {M^i}_i N \sqrt{g} {\cal L}  
     +\sqrt{g}{\cal H}_i \delta_M N^i + \frac{\partial (N\!\sqrt{g} {\cal L})}{\partial g_{ij}} \delta_M g_{ij} + \pi^{ij} \delta_M \dot{g}_{ij}  \cr
    & \qquad \qquad \qquad + \sum_\alpha \frac{\partial (N\!\sqrt{g} {\cal L})}{\partial \varphi_{\rm matter}^{\alpha}} \delta_M \varphi_{\rm matter}^{\alpha} + \sum_\alpha \pi^{\alpha}_{\rm matter} \delta_M \dot{\varphi}_{\rm matter}^{\alpha} \biggr] \,, \label{Eq:Noether00}
\end{align}
where $\alpha$ labels different species of matter fields and $\delta_M$ denotes the change of each field under the transformation, (\ref{Exp:LGT}), e.g.,
\begin{align}
    &\delta_M g_{ij} = \tilde{g}_{ij} (t,\, \tilde{x}^i) - g_{ij}(t,\, x^i) = - {M^k}_i g_{kj}(t,\, x^i) - {M^k}_j g_{ik}(t,\, x^i) + {\cal O}(M^2)\,.  \label{Exp:transgij}
\end{align}
The first term in the right hand side of Eq.~(\ref{Eq:Noether00}) appears from the Jacobian for the large gauge transformation, i.e., $\det(\partial\tilde{x}^i/\partial x^j) = 1 + {M^k}_k$. For a notational brevity, tensor indices of matter fields (even if any) are ignored. The conjugate momenta are defined as
\begin{align}
  \pi^{\alpha}_{\rm matter} \equiv   \frac{\partial (N\!\sqrt{g} {\cal L})}{\partial \dot{\varphi}_{\rm matter}^{\alpha}}\,.  \label{Exp:piij} 
\end{align}

Using the equations of motion and also Eq.~(\ref{Exp:transgij}), we can rewrite Eq.~(\ref{Eq:Noether00}) as
\begin{align}
    0 = {M^i}_j \int d^{d+1} x  \left[ N \sqrt{g} {\cal L} \, {\delta^j}_i + \partial_t \left( -  2 {\pi^j}_i +  \sum_\alpha  \pi^{\alpha}_{\rm matter}  \frac{\partial  \delta_M \varphi_{\rm matter}^{\alpha}}{\partial {M^i}_j}  \right) + {\cal O}(\epsilon) \right],  \label{Eq:Noether1}
\end{align}
with ${\pi^j}_i=g_{ik} \pi^{jk}$ defined in Eq.~\eqref{eq:piij}. Since 
$\delta_M N^i=M^i_{~ j} N^j$, we eliminate the term with the momentum constraints in rewriting Eq.~(\ref{Eq:Noether00}) into Eq.~(\ref{Eq:Noether1}), employing the gauge $N^i=0$. To determine the matter contributions, we need to specify the tensor structure of the existing matter fields. For instance, a scalar field $\phi$ and a vector field $A_\mu$ transform under Eq.~(\ref{Exp:LGT}) as
\begin{align}
    & \delta_M \phi = \delta_M A_0 = 0\,, \qquad \delta_M A_i = -  {M^k}_i A_k + {\cal O}(M^2,\, \epsilon)\,.
\end{align}
Therefore, $\phi$ does not contribute to Eq.~(\ref{Eq:Noether1}), while $A_\mu$ contributes as $- \pi^j_A A_i$, where $\pi^i_A$ denotes the conjugate momentum of $A_i$. Here, higher order terms of ${M^i}_j$ are ignored, considering  infinitesimally small $|{M^i}_j|$. Since ${\cal G}_\epsilon$ is higher order in $\epsilon$, Eq.~(\ref{Eq:Noether1}) does not require ${\cal G}_\epsilon$ to hold at all.

Since ${M^i}_j$ is an arbitrary constant matrix, we obtain $d^2-1$ Noether charge densities for the shear transformation and the rotation in Eq.~(\ref{Exp:LGT}) as 
\begin{align}
  {Q^i}_{j}\equiv 8 \pi G \left( \left[- 2 {\pi^i}_j  + \sum_{\alpha} \pi^{\alpha}_{\rm matter}  \frac{\partial  \delta_{M} \varphi_{\rm matter}^{\alpha}}{\partial {M^j}_i} \right]^{\rm TL}+ {\cal O}(\epsilon) \right)\,, 
   \label{Eq:Noether}
\end{align}
that satisfy
\begin{align}
    \frac{d}{dt} \int d^d\bm{x}\,  {Q^i}_j = 0\,.  \label{Exp:Noetherglobal}
\end{align}
Here and hereafter, the superscript TL indicates the operation to pick up the traceless part. For our later convenience, we have inserted the constant factor $8 \pi G$ in the definition of ${Q^i}_j$. The trace part, which corresponds to the scale transformation, does not give a conservation because of the  contribution of the Jacobian factor. Instead, the trace part of Eq.~(\ref{Eq:Noether1}) gives the equation which corresponds to the trace part of Eq.~(\ref{Eq:EijTLgen}).

The global conserved charges, given in Eq.~(\ref{Exp:Noetherglobal}), do not give any relation to the local quantity directly. 
Nevertheless, together with the \locality condition, Eq.~(\ref{Exp:Noetherglobal}) ensures the approximate time conservation of the Noether charge densities ${Q^i}_j$ at the leading order of $\epsilon$. To show this, let us first consider a specific example where all the fields in $\{ \varphi^{a} \}$ become globally homogeneous. In this case, since the spatial integral in Eq.~(\ref{Exp:Noetherglobal}) operates trivially, we obtain  $d{Q^i}_j/dt=0$. This can be also derived by using the equations of motion for the homogeneous fields $\{\bar{\varphi}^{a} (t) \}$, i.e.,
\begin{align}
 \frac{d}{dt} {Q^i}_j=0 \quad  \Longleftarrow \quad \mbox{set of equations for}~ \{\bar{\varphi}^{a} (t)\},\,{\rm i.e.}, {\cal E}\,. \label{Exp:flow_homogeneous}
\end{align}
The equations used to show the conservation of ${Q^i}_j$ are nothing but ${\cal E}$, which do not include ${\cal G}_\epsilon$,  since ${\cal G}_\epsilon$ become trivial for homogeneous fields.

Next, we turn our attention to the coarse-grained fields $\{\bar{\varphi}^a_{\sbm{x}}(t)\}$, which are approximately homogeneous within each causal patch. The \locality condition ensures that $\{ \bar{\varphi}^a_{\sbm{x}}(t)\}$ satisfies the same set of the equations as Eq.~(\ref{Exp:flow_homogeneous}), allowing ${\cal O}(\epsilon)$ corrections. Therefore, using these equations, we can also show 
\begin{align}
\label{eq:conservationQ}
 \frac{d}{dt} {Q^i}_{j\,{\sbm{x}}}={\cal O}(\epsilon) \quad  \Longleftarrow \quad \mbox{set of the equations for}~ \{\bar{\varphi}_{\sbm{x}}^{a} (t)\}\,,{\rm i.e.}, {\cal E}\,,
\end{align}
where ${Q^i}_{j\,{\sbm{x}}}$ denotes the Noether charge density computed by using $\{ \bar{\varphi}^a_{\sbm{x}}(t)\}$. 
Likewise $\{ \bar{\varphi}^a_{\sbm{x}}(t)\}$, ${Q^i}_{j\,{\sbm{x}}}$ is homogeneous within each patch in the limit $\epsilon \to 0$. 
Since $\{ \bar{\varphi}^a_{\sbm{x}}(t)\}$ is not exactly homogeneous, the constraints ${\cal G}_\epsilon$ are no longer trivially satisfied. However, the approximate conservation of ${Q^i}_{j\,{\sbm{x}}}$ \eqref{eq:conservationQ} can be shown without using ${\cal G}_\epsilon$, as one can understand from the fact that Eq.~(\ref{Exp:flow_homogeneous}) is derived by using only ${\cal E}$. 

In the absence of charged matter fields, repeating the same argument for the U(1) gauge transformation (\ref{eq:U1gauga}) with $\theta= M_i\,x^i$ ($M_i$: a constant vector), which is still allowed in the gauge $A_0=0$, we simply obtain the Maxwell equation, given by
\begin{align}
    \partial_t \pi^i_A = {\cal O}(\epsilon)\,, \label{Eq:Maxwell_neutral}
\end{align}
where the conjugate momentum $\pi^i_A$ corresponds to the Noether charge density. Meanwhile, in the presence of charged matter fields, the same argument does not apply, since the change of the matter fields depends on the spatial coordinates even for a constant $M_i$. 
Remember that since $\pi^i_A$ contains the factor $\sqrt{g}$, the time derivative of $A_i$ decrease in time unless there is some enhancement factor which compensates $1/\sqrt{g}$.

When the Lagrangian density ${\cal L}_{\rm g}$ is given by Eq.~(\ref{Def:Lgex}), we obtain\footnote{For a theory of gravity with the Lagrangian density, (\ref{Def:Lg}), ${\pi^j}_i$ is given by
\begin{align*}
   {\pi^j}_i &= \frac{F}{16 \pi G} \sqrt{g} \Biggl[ \left\{ 1 + \left( \frac{3\lambda_3}{d} - \lambda_2 \right) K \right\} {A^j}_i + \frac{3\lambda_3}{2}  \left[ {A^j}_l {A^l}_i \right]^{\rm TL}  \cr
   & \qquad \qquad \qquad \quad + \frac{{\delta^j}_i}{d} \left\{ (1- d \lHL) K + \frac{3}{2d} (d^2 \lambda_1 - \lambda_2 + \lambda_3 ) K^2 + \frac{1}{2} (3 \lambda_3 - d \lambda_2) {A^k}_l{A^l}_k \right\}   + \cdots   \Biggr] \,. 
\end{align*}
}
\begin{align}
     ({\rm ex}) \qquad \quad {\pi^j}_i = \frac{F}{16 \pi G} \sqrt{g} \left[ {A^j}_i + \frac{{\delta^j}_i}{d} (1- d \lHL) K  \right]\,.  \label{Exp:piex}
\end{align}
As one can see from this example, the symmetric part of Eq.~(\ref{Eq:Noether}) provides the precise version of Eq.~(\ref{Exp:shearKEY}). Here, the symmetric part of ${Q^i}_j(\bm{x})$ roughly corresponds to ${D^i}_j (\bm{x})$. In the next subsection, we will show that the symmetric part of ${Q^i}_j$, which is accompanied by (a positive power of) $(1/\sqrt{g})$, can be generically used to solve the momentum constraints at $t= t_{\rm hoc}$, just accepting the non-local contributions that decay rapidly in an expanding universe.

\subsubsection{Decay of non-local contributions}
As argued around Eq.~(\ref{Exp:Hi}), by using the conjugate momentum $\pi^{ij}$, the momentum constraints can be expressed in the following compact form,
\begin{align}
    2 \nabla_j \left( \frac{{\pi^j}_i}{\sqrt{g}} \right) + N \left(\frac{\partial \Lma}{\partial N^i} \right)_{\!\!K_{ij}:{\rm fixed}}  =  {\cal O} \left(\epsilon^2  \right)\,,  \label{Exp:MCcompact}
\end{align}
where the second term is given by taking the derivative of the matter Lagrangian with respect to $N_i$, while keeping $K_{ij}$ fixed. The variation with respect to $N^i$ through $K_{ij}$ contained in $\Lma$ (if exists) is included in the conjugate momentum $\pi^{ij}$. Eliminating the traceless part of ${\pi^j}_i$ in Eq.~(\ref{Exp:MCcompact}) by using the symmetric part of Eq.~(\ref{Eq:Noether}), we can further rewrite the momentum constraints as 
\begin{align}
    &\frac{1}{\sqrt{g}}{\nabla_j} {Q^{j}}^{~{\rm sym}}_i - \frac{16 \pi G}{d \sqrt{g}} \partial_i {\pi^k}_k \cr
    &= 8 \pi G \left[ \frac{1}{\sqrt{g}} { \nabla_{\!j}} \sum_{\alpha} \pi^{\alpha}_{\rm matter}  \left[ \frac{\partial  \delta_{M} \varphi_{\rm matter}^{\alpha}}{\partial {M^i}_j} \right]^{\rm symTL} \!\!\!\!\!\!\!+ N \left( \frac{\partial \Lma}{\partial N^i} \right)_{\!\!K_{ij}:{\rm fixed}} \right] + {\cal O} \left(\epsilon^2  \right)\!, \label{Eq:MCtobetuned}
\end{align}
where $Q^{{\rm sym}}_{ij}$ is the symmetric part of $Q_{ij}$. We put the superscript ``{\rm symTL}'' to indicate that only the symmetric and traceless part is extracted.

The momentum constraints (\ref{Eq:MCtobetuned}) relate the symmetric part of the Noether charge densities ${Q^i}_j{}^{\rm sym}$ to matter fields in the right hand side. As discussed in Sec.~\ref{SSec:proofGE}, when all other equations are satisfied and Eq.~(\ref{Eq:MCtobetuned}) are properly solved for ${Q^i}_j{}^{\rm sym}$, e.g., at $t=t_{\rm hoc}$, the momentum constraints (\ref{Eq:MCtobetuned}) hold also at an arbitrary time. Since Eq.~(\ref{Eq:MCtobetuned}) has $d$ components, we can solve it to determine $d$ components among the symmetric part of ${Q^i}_j$. Since we need to integrate the spatial derivative to determine the $d$ components, the obtained expression becomes non-local. Nevertheless, the non-local terms ${Q^i}_j{}^{\rm sym}$ in the expression of ${A^i}_j$ become smaller and smaller as the universe expands, being suppressed by $1/\sqrt{g}$. 
By looking at the expression, say, in \eqref{Exp:piex}, one may wonder what if $F$ decreases too rapidly to compensate the growth of $\sqrt{g}$. A temporal decrease of $F$ is not a problem, since we just have to wait further for the non-local terms to start decaying. Meanwhile, a long-lasting decrease of $F$ can lead to a strong coupling regime. (As will be discussed below Eq.~(\ref{Exp:epsilon}), a rapid change of $F$ also violates the slow-roll condition.)
Once these terms become negligible, whether ${\cal H}_i$ is solved or not should be irrelevant. Therefore, when ${\cal H}_i$ are the only ${\cal G}_\epsilon$ in the system, the \approxGe condition holds at a late time after these non-local terms become negligible.

The momentum constraints determine how to glue the neighboring local patches at each time, resulting in the appearance of non-local contributions. As discussed in Sec.~\ref{SSec:proofGE}, as far as the neighboring patches are glued properly at a certain time, satisfying the momentum constraints, they remain so also at later times. The decay of the non-local contributions indicates that even if the neighboring patches are not properly glued, violating the momentum constraints or choosing wrong values for the symmetric part of ${Q^i}_j$, the error becomes less and less important as time goes on.

One may wonder if the same argument applies for the U(1) gauge constraint ${\cal H}_{U(1)}$ in the absence of charged matter fields, by choosing $\pi^i_A$, which corresponds to the Noether charge, in such a way that ${\cal H}_{U(1)}$ is satisfied at a certain time. This is the case when the gauge fields become negligible at large scales. Nevertheless, when they do not decay at large scales, e.g., being sourced by a decreasing ${\cal F}_Y$, the non-local terms with $\pi^i_A$ are not negligible.

\subsubsection{Usage of conserved charges}
Equation (\ref{Eq:Noether}) can be useful also in solving the shear. To illustrate this, let us consider an example where ${\cal L}_{\rm g}$ is given by Eq.~(\ref{Def:Lgex}) and the matter Lagrangian density ${\cal L}_{\rm matter}$ is given by Eq.~(\ref{Exp:Lmatter_U1}). Inserting Eqs.~(\ref{Exp:piAex_U1}) and (\ref{Exp:piex}) into the symmetric part of Eq.~(\ref{Eq:Noether}), we obtain a similar expression to Eq.~(\ref{Sol:AijS}) as
\begin{align}
    F {A^j}_i =  - \frac{{Q^{j}}^{~{\rm sym}}_i}{\sqrt{g}}  - 8 \pi G {\cal F}_Y \left[ g^{jl} \frac{\dot{A}_l}{N} A_i   
    \right]^{\rm symTL} + {\cal O}(\epsilon)\,.  \label{Eq:Noetherex}
\end{align}

The main difference between Eqs.~(\ref{Sol:AijS}) and (\ref{Eq:Noetherex}) is in that the matter contributions in the latter are not in the form of the time integral. Integrating Eq.~(\ref{Eq:Noetherex}) formally, we obtain
\begin{align}
    \gamma_{ij} (t,\, \bm{x}) &=  \gamma_{ij} (t_{\rm hoc},\, \bm{x}) - 2  \int^{\tau(t)}_{\tau_{\rm hoc}} d \tau' \frac{\gamma_{il} (\tau',\, \bm{x})}{(F \sqrt{g})_{(\tau',\, \sbm{x})}}  {Q^{l}}^{~{\rm sym}}_j(\bm{x}) \cr
    & \qquad - 8 \pi G \int^{\tau(t)}_{\tau_{\rm hoc}} d \tau' e^{- 2 \psi(\tau',\, \sbm{x})} \frac{{\cal F}_Y}{F} \left[ (\partial_{\tau} A_i)  A_j 
    \right]^{\rm symTL}_{(\tau',\, \sbm{x})}  + {\cal O}(\epsilon).\,\,  \label{Sol:gammaij_Noether}
\end{align}
The second term corresponds to the decaying solution and the term in the second line is the sourced contributions. Compared to Eq.~(\ref{Sol:gammaij}), one time integral is already performed in Eq.~(\ref{Sol:gammaij_Noether}).

\subsection{Approximate existence of WAM}
In this section, we have shown that the Weinberg's argument can apply much more generically, extending it to an anisotropic universe with a large scale anisotropic pressure and modified theories of gravity. In fact, when the \approxGe condition holds in addition to the \locality and \sDif conditions, the WAM exists as an approximate solution in the large scale limit. Based on the argument in the previous subsection, let us summarize cases when the additional condition \approxGe holds.

When the momentum constraints are the only ${\cal G}_\epsilon$ in the system, as discussed in the previous subsection, the \approxGe holds since the non-local contributions inevitably become negligible at late times. This includes the cases when the system only includes scalar fields or includes also a gauge field with a charged field.

The WAM is not the exact solution even at the leading order of the gradient expansion, because it does not fulfill the leading order of ${\cal G}_\epsilon$, which is ${\cal O}(\epsilon)$, due to the decaying contribution. This can be seen explicitly by focusing on the first term in the second line of Eq.~(\ref{Eq:dKts}), which includes $(\partial_j \psi) {A^j}_i$. This term contributes as the leading order in the gradient expansion since ${A^i}_j = {\cal O}(\epsilon^0)$ (see the discussion in the next subsection), while it does not remain invariant under Eq.~(\ref{Exp:psishift}) for an inhomogeneous $c(\bm{x})$.

Let us consider an anisotropic universe with a $d$-dimensional rank $(m,\, n)$ tensor field with $m$ upper indices and $n$ lower indices, ${\varphi_{i_1 \cdots i_n}}^{j_1 \cdots j_m}$, which transforms under the dilatation as 
\begin{align}
    \varphi_{i_1 \cdots i_n}^{c~~~~~j_1 \cdots j_m} (t,\, \bm{x}^c) = e^{(n-m)c} {\varphi_{i_1 \cdots i_n}}^{j_1 \cdots j_m} (t,\, \bm{x})\,,
\end{align}
satisfying the \approxGe condition in addition to the other two conditions. Since the system remains approximately invariant under Eq.~(\ref{Exp:psishift}), when $\varphi^{a}(t,\, \bm{x}) = f^{a}(t,\, \bm{x})$ is a solution for $\{ \varphi^{a} \}$,
\[
  \varphi^{a}(t,\, \bm{x}) = \begin{cases}
    f^{a}(t,\, \bm{x}) + c(\bm{x})  & (a= \psi), \\
    e^{(n-m)c(\sbm{x})} f^{a}(t,\, \bm{x})  & (a={\rm a~rank~}(m,\, n) {\rm~tensor}), 
  \end{cases}
\]
also satisfies the equations {${\cal E}$} at the leading order of $\epsilon$ and ${\cal G}_\epsilon$ approximately as well, owing to \approxGe\hspace{-2pt}\footnote{When one wants to consider a solution where only $\psi$ is shifted, the rank-$(m,\, n)$ tensor fields should be projected by the tetrad in $d$-dim. space ${e^{(\alpha)}}_i$ as
\begin{align}
  \varphi_{(\alpha_1) \cdots (\alpha_n) (\beta_1) \cdots (\beta_m)} (t,\, \bm{x}) \equiv {e_{(\alpha_1)}}^{i_1} \cdots  {e_{(\alpha_n)}}^{i_n} e_{(\beta_1)\,j_1} \cdots  e_{(\beta_n)\, j_m}\,  {\varphi_{i_1 \cdots i_n}}^{j_1 \cdots j_m} (t,\, \bm{x}) \,, 
\end{align}
where the indices with the parentheses are the tetrad basis coordinates. For example, in the projected tetrad bases, the kinetic term for a gauge field, $ g^{ij}\dot{A}_i \dot{A}_i$, is given as 
$$
 g^{ij}\dot{A}_i \dot{A}_i = (\dot{A}_\alpha - {\dot{e}_\alpha\!}^i A_i) (\dot{A}^\alpha - \dot{e}^{\alpha j} A_j)\,,
$$
where we have used $g^{ij} = {e_\alpha}^i e^{\alpha j}$. } . As it is clear from the definition, $\gamma_{ij}$ does not change under the dilatation. We express the spatial metric perturbation as  $\delta\gamma^i_{~j}=\log (\delta^{ik} \gamma_{kj})$, which transforms like a (1,1) tensor under the scale
transformation. When there exists a non-decaying tensor field with $n \neq m$, the change of these fields also should be taken into account in the definition of WAM.

Meanwhile, when a system includes gauge fields, the gauge constraint can become ${\cal G}_\epsilon$. As one can see from Eq.~(\ref{Exp:GC_U1}), the gauge constraint ${\cal H}_{\rm U(1)}$ does not remain invariant under the dilatation with an inhomogeneous parameter $c(\bm{x})$\footnote{Recall that the dilatation with an inhomogeneous parameter $c(\bm{x})$ should be understood as simply replacing a constant $c$ with $c(\bm{x})$ in the transformation law of the dilatation with a constant $c$, which is definitely not a coordinate transformation. This point was emphasized in Ref.~\cite{Matarrese:2020why}.}, since there appears a term with $\partial_i c(\bm{x}) A^i(t,\, \bm{x})$. While this term is suppressed by the spatial gradient, this is a leading order contribution in the gradient expansion for ${\cal H}_{\rm U(1)}$, which is ${\cal G}_\epsilon$, similarly to $\partial_j \psi {A^j}_i$ in ${\cal H}_i$. Therefore, when the gauge fields do not decay at large scales, e.g., being enhanced by a coupling with the inflaton, the gauge constraint ${\cal H}_{\rm U(1)}$ does not hold even approximately, resulting in the absence of the WAM. Exceptionally, when the background value of $A^i$ vanishes, ${\cal H}_{\rm U(1)}$ remains to be valid also after the dilatation at the linear order of perturbation, since $\partial_i c(\bm{x}) A^i(t,\, \bm{x})$ becomes second order in perturbation. Then, the WAM approximately exists at the linear perturbation.

Let us emphasize again that when $\psi$ is dominated by other modes, the (approximate) existence of the constant solution is not particularly useful to shortcut the time evolution on super-horizon scales. The constant solution becomes more important as an asymptotic solution after the period of all the non-trivial dynamics, which potentially generates growing modes of $\psi$.

In this paper, we have focused on the adiabatic solution of the scalar perturbation, but the same argument also applies to the tensor perturbations~\cite{Tanaka:2017nff}. The conservation of the tensor perturbations in a scalar field system was first pointed out in the historic paper by Starobinsky~\cite{Starobinsky:1986fx}. Under the shear transformation, 
\begin{align}
    x^i \to x^i_C \equiv \bigl[ e^{- \frac{C}{2}} \bigr]{}^i_{\, j}\,  x^j\,,
\end{align}
with $C^i_{\, j}$ being a constant traceless and symmetric tensor, the anisotropic spatial metric transforms as
\begin{align}
    \gamma_{ij\,C}(t,\, \bm{x}_C) = \bigl[e^{ \frac{C}{2}} \bigr]{}^k_{~ i}\, \bigl[e^{ \frac{C}{2}}\bigr]{}^l_{~ j}\,
    \gamma_{kl}(t,\, \sbm{x})\,. 
\end{align}
Similarly to Eq.~(\ref{Exp:psi_shift}), this transformation shifts $\delta {\gamma^i}_j$, defined as $\gamma_{ij} \equiv \bar{\gamma}_{ik} {[e^{\delta \gamma}]^k}_j$ by using the (background) homogeneous contribution $\bar{\gamma}_{ij}$, as follows
\begin{align}
    \delta {\gamma^i}_{j\, C} (t,\, \bm{x}_C) = \delta {\gamma^i}_j (t,\, \bm{x}) + {C^i}_j 
     + {1\over 2} \left( {C^i}_k \delta {\gamma^k}_{j}+\delta {\gamma^i}_{k}{C^k}_j\right) +\cdots\,. 
\end{align}
Unlike the dilatation, the change under the shear transformation is not a simple additive one, but is given by the above expression. When we promote the shear transformation by replacing ${C^i}_j$ with an inhomogeneous parameter $C^i\!_j(\bm{x})$, repeating the same argument as the one for dilatation, we find that the change under the promoted transformation provides a solution of ${\cal E}$ at the leading order of the gradient expansion. When the \approxGe condition holds in addition to the \locality and \sDif conditions, the solutions of ${\cal E}$ approximately satisfy all the equations in the system, ${\cal E}_{\rm all}$. If one imposes the transverse condition (\ref{Exp:Tad}) at $t=t_*$, which is ${\cal X}_{\epsilon}$, to remove the residual degrees of freedom in the spatial coordinates, ${C^i}_j(\bm{x})$ should be chosen so that Eq.~(\ref{Exp:Tad}) is satisfied both before and after the transformation. At the linear perturbation, this requires $\partial_i {C^i}_j(\bm{x})=0$. The remaining degrees of freedom in ${C^i}_j(\bm{x})$ correspond to the decaying two solutions of the tensor perturbation.

As an example, let us consider a model with a U(1) gauge field and neutral fields. 
When we perform the shear transformation with an inhomogeneous parameter $C^i\!_j(\bm{x})$, there appears a term with $[e^{C/2}]_{~i}^j (\partial_j [e^{-C/2}]^i_{~k})\, \pi_A^k$ in the gauge constraint ${\cal H}_{\rm U(1)}$. With $\partial_i {C^i}_j(\bm{x})=0$, we find that the violation of ${\cal H}_{\rm U(1)}$ vanishes exceptionally at the linear perturbation even when $\pi_A^k$ has a non-zero background value. For the WAM to exist approximately also at higher orders of perturbation, the violation of ${\cal H}_{\rm U(1)}$ should be approximately negligible, i.e, satisfying the \approxGe condition. When the gauge fields survive also at the large scales, the \approxGe condition can be violated at the higher orders of perturbation. Meanwhile, when the U(1) gauge constraint is not ${\cal G}_\epsilon$ owing to the presence of charged fields, once ${\cal H}_i$, which is the only ${\cal G}_\epsilon$, start to hold approximately, the WAM exists as an approximate solution. Earlier studies of the gravitational waves in Bianchi background can be found, e.g., in Refs.~\cite{Gumrukcuoglu:2007bx, Gumrukcuoglu:2008gi, Pereira:2007yy}.

\subsection{Weinberg's second solution}  \label{SSec:Wseconed}
In previous works, the shear has been treated as a sub-leading contribution in the gradient expansion. However, as we will see in this subsection, if we assume ${A^i}_j = {\cal O}(\epsilon)$, the other solution of $\psi$, which corresponds to the Weinberg's second mode, does not appear at the leading order of the gradient expansion. This was shown in Ref.~\cite{Sasaki:1998ug}, considering a scalar field system (see also Ref.~\cite{Garriga:2015tea}, where a more general class of scalar field models was discussed).

In this subsection, for simplicity, let us consider a single field model of inflation, whose matter Lagrangian density is given by
\begin{align}
    {\cal L}_{\rm matter} = P(X,\, \phi) \,, \qquad X \equiv - \frac{1}{2} \partial_\mu \phi  \partial^\mu \phi\,,   \label{Exp:L_kinf}
\end{align}
in general relativity (with $F=1$). In this case, the energy density, defined as 
\begin{align}
    \rho \equiv - \frac{\partial}{\partial N} \left(  N \Lma \right)\,,
\end{align}
is given by 
\begin{align}
    \rho =  2 P_X X - P + {\cal O}(\epsilon^2)
    \,, \label{Exp:rho_single}
\end{align}
with $P_X \equiv \partial P/\partial X$
and the shear simply decays as ${A^i}_j = - Q^i_{\,j}/\sqrt{g}$, since there is no anisotropic pressure on large scales.

In general relativity, the Hamiltonian and momentum constraints are given by 
\begin{align}
    & \frac{1-d}{d} K^2 +  A^i\!_j A^j\!_i + 16 \pi G \rho  = {\cal O} (\epsilon)\,,  \label{Eq:HC_single}
\end{align}
and
\begin{align}
  & 
 \left(  \frac{1}{d} - 1 \right)  \partial_i K + \nabla_j {A^j}_i - {8 \pi G} P_X  \frac{\dot{\phi}}{N} \partial_i \phi = {\cal O}(\epsilon^2)\,.  \label{Eq:MC_single}
\end{align}
For our convenience, let us introduce the flat FLRW spacetime as the background and consider the deviation from there. Once we ignore ${A^i}_j$, setting ${A^i}_j = {\cal O}(\epsilon)$, the constraint equations imply that the uniform Hubble slicing with
\begin{align}
    K(t,\, \bm{x}) = d H(t)\,,  \label{Def:UHgauge}
\end{align}
the uniform density slicing, and the uniform field slicing all agree with each other. Here, $H(t)$ is the Hubble parameter for the FLRW background and we have noticed $P_X \dot{\phi} \neq 0$ in a sensible model of inflation. Evaluating $\delta \rho = {\cal O}(\epsilon)$ with $\delta \phi= {\cal O}(\epsilon)$, we obtain $\delta N = \delta (\dot{\psi}/H) = {\cal O}(\epsilon)$, i.e., the fluctuation of $\psi$ is time independent at the leading order of $\epsilon$.

Therefore, in order to properly reproduce the Weinberg's second mode, we need to take into account the shear as the leading order of the gradient expansion, i.e., ${A^i}_j = {\cal O}(\epsilon^0)$. In Ref.~\cite{Garriga:2015tea}, this was pointed out for the curvature perturbation $\zeta$ at linear perturbation, taking the transverse gauge. It is natural that the same story follows for $\psi$ in our gauge with Eqs.~(\ref{GC}) and (\ref{Exp:Tad}), because our $\psi$ and $\zeta$ agree {when ${A^i}_j$ is negligible all the time}\footnote{At linear perturbation, ${A^i}_j \sim 0$ can be recast into $\dot{\gamma}_{ij} \sim 0$, which in turn implies that the transverse condition (\ref{Exp:Tad}) also holds at an arbitrary time.}.

In particular, in a scalar field system, one may think that the local spacetime metric approaches the FLRW metric if we take the limit $\epsilon \to 0$. As we have argued here, this is not correct, {in contrast to the observation made from the definition of the Bardeen potential in Ref.~\cite{Lyth:2003im}}. Instead, only after we take both the $\epsilon \to 0$ limit and the late time limit, the local spacetime metric approaches the FLRW metric because of the shear contribution.

\section{Generalized \texorpdfstring{$\delta N$}{delta N} formalism}    \label{SSec:deltaN}
As is widely known, the $\delta N$ formalism \cite{Starobinsky:1982ee, Starobinsky:1986fxa, Sasaki:1995aw, Lyth:2004gb, Sasaki:1998ug} provides a powerful tool to calculate the large scale evolution of the perturbations generated during inflation (for a review, see Refs.~\cite{Tanaka:2010km, Abolhasani:2019qcn}). In this paper, we have shown that the gradient expansion can be verified for a general theory which satisfies the \sDif and \locality conditions. This enables us to apply the $\delta N$ formalism also to models of anisotropic inflation. Furthermore, we can also calculate non-zero spin fields such as the gravitational waves in this generalized formalism, which we dub the generalized $\delta N$ formalism (the \dNex).

\subsection{User's guide of g\texorpdfstring{$\delta N$}{delta N} formalism}
In this subsection, we summarize the basic procedure of the \dNex. Let us consider a theory which satisfies the \sDif and \locality conditions, including scalar fields $\phi^I~(I=1,\, 2,\, \cdots)$ and a vector field $A^\mu$ and so on, i.e., $\{\varphi_{\rm mat}^a\} = \phi^I,\, A^\mu,\, \cdots $.

\subsubsection{With redundant fields}
As discussed in Sec.~\ref{SSec:proofGE}, the separate universe evolution $(\star)$ can be verified by considering the evolution of $\{\varphi^{a} \}$, which still contains redundant fields to be removed. The \dNex~applies as long as the \locality and \sDif conditions hold. The initial conditions can be set around the horizon crossing time, i.e., $t_\star = t_{\rm hoc}$. By solving the evolution of homogeneous cosmological models with various initial conditions, we obtain the map from $\{\varphi^{a}(t_\star) \}$ to $\{\varphi^{a}(t_f) \}$, where $t_f$ is the final time. We dub this mapping, given by solving ${\cal E}$, as \dNex. The initial condition can be set before the \approxGe condition starts to hold. In order to apply this \dNex, the initial distribution needs to be specified also for the redundant fields. (When we have only the initial conditions for physical degrees of freedom, we need to determine the initial conditions for the remaining variables, using ${\cal G}_\epsilon$.)

For a theory with the $(d+1)$-dim Diff 
invariance, the equations to be solved, ${\cal E}$, are the Hamiltonian constraint, Eq.~(\ref{Eq:EijT2}), the evolution equations for matter fields, and the expression of the shear, Eq.~(\ref{Eq:Noether}) or alternatively Eq.~(\ref{Eq:ijS}). 
For non-projectable version of the HL gravity, the Hamiltonian constraint is a non-gauge constraint ${\cal C}\cup {\cal G}^c$ that reduces the variables in the original Lagrangian to $\{\varphi^a\}$. When the action is given by Eq.~(\ref{Def:Lg}), the Hamiltonian constraint reads 
\begin{align}
    &\frac{1- d \lHL }{d} K^2 +2 \left( \lambda_1 - \frac{\lambda_2}{d} + \frac{\lambda_3}{d^2} \right) K^3+ \left\{ 1 + 2 \left( - \lambda_2 + \frac{3 \lambda_3}{d}  \right) K \right\} A^i\!_j A^j\!_i  \cr
  & \qquad \qquad  \quad + 2 \lambda_3 A^i\!_j A^j\!_l A^l\!_i  + \cdots + \frac{16 \pi G}{F} \rho  = {\cal O} (\epsilon)\,.  \label{Eq:Hc2}
\end{align}

As shown in Eq.~(\ref{Exp:deltaN}), in $N_i=0$ gauge, $\psi$ directly corresponds to the non-perturbative $e$-folding number. Equation (\ref{Exp:deltaN}) states that the the difference of $\psi$ between $t_*$ and the final time $t_f$ is computed by integrating the expansion between these two time slices as
\begin{align}
    \psi(t_f,\, \bm{x}) - \psi(t_*,\, \bm{x}) = \frac{1}{d} \int^{t_f}_{t_*} dt' N(t',\, \bm{x}) K(t',\, \bm{x}) \,. \label{Exp:deltaN2}
\end{align}
In this gauge, $\psi$ does not directly correspond to the curvature perturbation $\zeta$\footnote{Here, $\zeta$ is the curvature perturbation in $\phi=$const. time slice, which is at linear order given by
\begin{align}
    \zeta \equiv {\cal R}- \frac{H}{\dot{\phi}} \delta \phi = \delta \psi - \frac{1}{2d} \chi - \frac{H}{\dot{\phi}} \delta \phi\,, \label{Def:zeta}
\end{align}
where we have introduced $\dot{\chi}$ as the scalar part of the shear, given by
\begin{align}
    {A^i}_j \sim \frac{1}{2} \delta^{im} \dot{\gamma}_{mj} \sim \frac{1}{2} \delta^{im} \left( \partial_m \partial_j \partial^{-2} - \frac{1}{d} \delta_{mj} \right) \dot{\chi}\,. 
\end{align}}, since we also have to include the longitudinal part of $\gamma_{ij}$ (see Eq.~(\ref{Exp:zeta2})).

Let us summarize a couple of key differences between the conventional $\delta N$ formalism and the \dNex. First, in the former we can only calculate the curvature perturbation, i.e., a scalar perturbation, while, in the latter, we can calculate an arbitrary field in the theory under consideration, $\{\varphi^{a}\}$, unless it trivially vanishes in the limit $\epsilon \to 0$. For example, we can also compute gravitational waves, evaluating $\gamma_{ij}$ for different values of $\{\varphi^{a'}_* \}'$. Second, all the fields are systematically incorporated through the initial condition $\{\varphi_{\sbm{x}\,*}^{a'}\}'$, including non-zero integer spin fields and also purely geometrical degrees of freedom, such as the gravitational waves $\gamma_{ij}$ and the Khronon in the Horava-Lifshitz gravity. This has become possible by getting rid of unnecessary assumptions such as the absence of anisotropic pressure at the leading order of $\epsilon$.

\subsubsection{Without redundant fields}
When the \approxGe condition is satisfied and the residual gauge degrees of freedom are removed without employing ${\cal X}_\epsilon$ or become trivially negligible, we can also apply the \dNex~ even when we give the initial conditions only for the physical degrees of freedom $\{\varphi^{a_{\rm phys}} \}_{\rm phys}$, excluding all the redundant fields. In this case, $t_*$ should be set by requiring that ${Q^i}_j$ is negligible in all the equations in ${\cal E}$. Typically, $t_*$ can be set to several $e$-foldings after the horizon crossing in an inflationary universe.  In this case, the initial conditions at $t=t_*$, can be given by solely specifying $\{\varphi^{a_{\rm phys}} \}_{\rm phys}$.

\subsection{Implications on anisotropic inflation}    \label{SSec:anisotropic}
In this subsection, we apply the \dNex~to a model of anisotropic inflation~\cite{Watanabe:2009ct, Watanabe:2010fh, Kanno:2010nr, Soda:2012zm}, considering the gravity Lagrangian density (\ref{Def:Lgex}) with $\lHL=1$ and the matter Lagrangian density (\ref{Exp:Lmatter_U1}). The Hamiltonian constraint and the trace part of the $(i,\, j)$ component of Einstein's equation give
\begin{align}
    &\frac{1- d }{d} K^2 + A^i\!_j A^j\!_i  + \frac{16 \pi G}{F} \rho  = {\cal O} (\epsilon^2)\,, \label{Eq:Hc_GR} \\
    &  \frac{1}{N  F} \partial_t (F K)  - \frac{d}{1 -  d} {A^i}_j {A^j}_i = \frac{d}{1 -  d} \frac{8 \pi G}{F} (\rho + P) +  {\cal O}(\epsilon^2)\,. \label{Eq:ijT_GR}
\end{align}
The shear is given by Eq.~(\ref{Eq:Noetherex}). In our gauge with $N_i=0$ and $A_0=0$, $Y$, $\tilde{Y}$, and $X^{IJ}$ are given by
\begin{align}
    & Y = \frac{1}{2N^2} g^{ij} \dot{A}_i \dot{A}_j + {\cal O}(\epsilon^2)\,, \qquad \tilde{Y} =  - \frac{1}{2 N \sqrt{g}} \epsilon^{ijk} \dot{A}_i F_{jk}\,, \nonumber \\
    & X^{IJ} = \frac{1}{4N^2} \left( \dot{\phi}^I \dot{\phi}^{J*} + {\rm c.c.}  \right) - \frac{1}{4} g^{ij} A_i A_j e^{(I)} e^{(J)} \left( \phi^I \phi^{J*} + {\rm c.c.}  \right) + {\cal O}(\epsilon)\,. 
\end{align}
Using these expressions and Eq.~(\ref{Exp:Lmatter_U1}), the energy density $\rho$ and the isotropic and asisotropic pressures are given by 
\begin{align}
   &  \rho = 2 X^{IJ} {\cal F}_{IJ}  + 2 Y {\cal F}_Y  - {\cal F} + \frac{1}{2} g^{ij} A_i A_j |\phi|^2 + {\cal O}(\epsilon) \,, \label{Exp:rho_U1} \\
   &  P = {\cal F} - \frac{2}{d}Y {\cal F}_Y  +  \frac{1}{2d} g^{ij} A_i A_j |\phi|^2 + {\cal O}(\epsilon)\,, \label{Exp:P_U1} \\
   & {\Pi}_{ij} = - \frac{1}{N^2} {\cal F}_Y  \left[ \dot{A}_i \dot{A}_j \right]^{\rm TL}+ \frac{1}{2} |\phi|^2 \left[ A_i A_j \right]^{\rm TL}   + {\cal O}(\epsilon) \,,
\end{align}
where we have used $T_{ij} = g_{ij} {\cal L}_{\rm matter} - 2 \partial {\cal L}_{\rm matter}/\partial g^{ij}$. Here, we have introduced the amplitude of the charged scalar fields, given by
\begin{align}
  |\phi|^2 \equiv  {\cal F}_{IJ}e^{(I)} e^{(J)} (\phi^I \phi^{J*} + {\rm c.c.}) \,. 
\end{align}
The Maxwell equation gives
\begin{align}
    \partial_t \pi^i_A = - J^i\,,  \label{Eq:Maxwell}
\end{align}
where the current $J^i$ is given by
\begin{align}
    J^i \equiv \frac{\partial N \sqrt{g} {\cal L}_{\rm matter}}{\partial A_i} = - \frac{1}{2} N \sqrt{g} |\phi|^2 A^i + {\cal O} (\epsilon) \,.  \label{Exp:Ji}
\end{align}
The current $J^\mu$ satisfies the conservation $\partial_\mu J^\mu = 0$. Sending the charge $e^{(I)}$ to 0, we can also consider neutral scalar fields.

\subsubsection{Neutral scalar fields}
First, let us consider the case where all the scalar fields are neutral. The Klein-Gordon equation for neutral scalar fields is given by
\begin{align}
    \frac{1}{N \sqrt{g}} \partial_t \left( {\cal F}_{IJ} \sqrt{g} \frac{\dot{\phi}^J}{N} \right) -  \frac{\partial {\cal F}}{\partial \phi^{I}} = {\cal O}(\epsilon)\,. 
\end{align}
In this case, using Eq.~(\ref{Eq:Maxwell}), we find Eq.~(\ref{Eq:Maxwell_neutral}). Using this solution, $Y$ can be recast into
\begin{align}
    Y(t,\, \bm{x}) = \frac{1}{2 ({\cal F}_Y)^2} \frac{g_{ij}}{g} \pi_A^i(\bm{x}) \pi_A^j(\bm{x})  + {\cal O}(\epsilon)\,. \label{Exp:Y_U1_neutral}
\end{align}
Suppose that ${\cal F}$ is dominated by a term which is proportional to $Y^n$ with an almost time independent coefficient. 
Then, we have 
$$ Y \propto (g_{ij}/g)^{1/(2n-1)} \propto e^{- \frac{2(d-1)}{2n-1} \psi}.$$ Since $Y$, ${\cal F}$, and ${\cal F}_Y$ all decrease in time for $n> 1/2$, if ${\cal F}$ is given by a polynomial of $Y$ with almost time independent coefficients, the lowest power term starts to dominate ${\cal F}$ as time goes on. Therefore, in the following, we focus on the linear term of $Y$, assuming that $F_Y$ is a function of the scalar fields.

Integrating Eq.~\eqref{Exp:piAex_U1}, we obtain the solution of $A_i$ as
\begin{align}
    A_i (t,\, \bm{x}) =  A_{i*} (\bm{x}) + \pi_A^j(\bm{x}) \int^{\tau(t)}_{\tau_*} \frac{d \tau'}{\sqrt{g} {\cal F}_Y} g_{ij} (\tau',\, \bm{x}) + {\cal O}(\epsilon)\,.   \label{Sol:Ai}  
\end{align}
Inserting this solution (\ref{Sol:Ai}) into Eq.~(\ref{Eq:Noetherex}), we obtain the shear as
\begin{align}
     {A^j}_i &=  - \frac{1}{F\sqrt{g}} \left[ {
    Q^j}_i + 4 \pi G \Biggl[ {\delta^j}_k {\delta_i}^l +  \gamma^{jl} {\gamma}_{ik} - \frac{2}{d} {\delta^j}_i {\delta^l}_k   \right] \pi^k_A \cr
    & \qquad \qquad \qquad \qquad  \times \left\{ A_{l *} + d \pi^m_A  \int^{\psi(t)}_{\psi_*} \frac{d \psi'}{\sqrt{g} K {\cal F}_Y} g_{ml} (\psi',\, \bm{x})   \right\} \Biggr] + {\cal O}(\epsilon).\,   \label{Eq:shear_U1neutral}
\end{align}
In Refs.~\cite{Watanabe:2009ct, Kanno:2009ei, Kanno:2010nr}, considering a single canonical scalar field coupled with a U(1) gauge field in general relativity ($F=1$) with $d=3$, Watanabe, Kanno, and Soda constructed a model in which the shear survives without being diluted by the exponential expansion. As argued in these papers, when the coupling ${\cal F}_Y$ is proportional to $e^{-4 \hat{c} \psi}$ with $\hat{c}=1$ or $\hat{c}>1$, the energy density of the gauge field becomes almost constant or grows exponentially during a slow-roll inflation. Especially for $\hat{c}>1$, it enters the anisotropic inflation phase, where the gauge fields slows down the inflaton.

In the following, using the formulae derived in this paper, let us consider the anisotropic inflation model more closely for a more general matter Lagrangian density (\ref{Exp:Lmatter_U1}) with ${\cal F}_Y(\phi^I,\, X^{IJ})$ in a general spacetime dimension. Considering an inflationary solution, we assume that the expansion $K$ remains almost constant, i.e.,
\begin{align}
    \varepsilon \equiv - d \frac{\dot{K}}{N K^2} \ll 1 \,, 
\end{align}
which corresponds to the usual slow-roll parameter, when we take the homogeneous and isotropic limit. This condition does not restrict quantities that correspond to other slow-roll parameters. Using Eqs.~(\ref{Eq:Hc_GR}) and (\ref{Eq:ijT_GR}), we can compute $\varepsilon$ as
\begin{align}
 \varepsilon = d \frac{\partial_t \ln F}{N K} + \frac{d^2}{d-1} \frac{{A^i}_j {A^j}_i}{K^2} +  \frac{d}{2} \frac{\rho + p}{\rho} \frac{1}{1 + F {A^i}_j {A^j}_i/(16 \pi G \rho)} + {\cal O}(\epsilon)\,. \label{Exp:epsilon}
\end{align}
Unless we accept a fine tuning, the variation of $F$ in the time scale of the cosmic expansion violates the slow-roll condition. 

When we ignore the time variation of $K$ and $F$, the time integral in the shear, i.e., the second term in the second line of Eq.~(\ref{Eq:shear_U1neutral}), evolves as
$$
\frac{g_{ij}}{g {\cal F}_Y} \propto \frac{e^{-2(d-1)\psi}}{{\cal F}_Y}\,.  
$$
Therefore, when ${\cal F}_Y$ evolves as ${\cal F}_Y \propto e^{- 2 \hat{c} (d-1) \psi}$ with $\hat{c} > 1$ or $\hat{c} =1$, ${A^i}_j {A^j}_i/K^2$ grows or approaches to a constant. Meanwhile, the energy density of the gauge fields can be expressed as $Y {\cal F}_Y$, which also scales in the same way as the above shear contribution. This is just a hand-waving argument to look for a model where the shear can survive in an expanding universe. To build a concrete model, one needs to start with ${\cal F}$ which is a function of $\phi^I$ and/or $X^{IJ}$ as in Refs.~\cite{Watanabe:2009ct, Kanno:2009ei, Kanno:2010nr}.

Equation (\ref{Exp:epsilon}) immediately gives an upper bound on the shear as $\sqrt{{A^i}_j {A^j}_i/K^2} \leq {\cal O}(\sqrt{\varepsilon})$ when the null energy condition is satisfied and the time variation of $F$ is negligible~\cite{Maleknejad:2012as}. In Ref.~\cite{Watanabe:2009ct}, a more stringent upper bound, $\sqrt{{A^i}_j {A^j}_i/K^2} \leq {\cal O}(\varepsilon)$, was obtained under the slow-roll approximation. In what follows, considering a more general matter Lagrangian density (\ref{Exp:Lmatter_U1}), we show that this inequality holds at the leading order of the gradient expansion, when $\varepsilon \ll 1$, $F$ satisfies $\partial_t F \geq 0$, and the scalar fields are not ghosts. While we assume $\varepsilon \ll 1$, other slow-roll parameters can be large likewise in the ultra slow-roll model, which will be discussed in the next subsection.

Using Eq.~(\ref{Eq:shear_U1neutral}) and ignoring the decaying contributions, which are suppressed by $1/(F \sqrt{g})$, we obtain
\begin{align}
    \sqrt{\frac{{A^i}_j {A^j}_i}{K^2}} = \frac{8 \pi G}{F \sqrt{g} K} \sqrt{d(d-1)}  \pi_A^i \pi_A^j  \int^{\psi(t)}_{\psi_*} \frac{d \psi'}{\sqrt{g} K {\cal F}_Y} g_{ij} (\psi',\, \bm{x}) \,, \label{Exp:shear_U1}
\end{align}
at the leading order of the gradient expansion, where we have ignored the time variation of $\gamma_{ij} = e^{ -2 \psi} g_{ij}$ during the time scale of the expansion, ${\cal O}(1)/K$, based on ${A^i}_j {A^j}_i/K^2 < {\cal O}(\varepsilon) \ll 1$. Using Eqs.~(\ref{Eq:Hc_GR}), (\ref{Exp:rho_U1}), (\ref{Exp:P_U1}), and (\ref{Exp:epsilon}), we obtain
\begin{align}
    \frac{16 \pi G}{Fd} \frac{Y {\cal F}_Y}{K^2} \leq \frac{d}{2} \frac{\rho + P}{\rho}  \leq  \varepsilon\,. \label{Exp:Upper_U1}
\end{align}
To show the first $\leq$, we have used $X^{IJ} {\cal F}_{IJ} \geq 0$, which is ensured from the absence of the ghost instability. The equality holds, in the absence of the scalar fields, while in such cases there is no enhancement of the gauge fields either. Using 
\begin{align}
    \frac{8 \pi G}{F} \frac{g_{ij} \pi^i_A \pi^j_A}{d K^2 g {\cal F}_Y} \leq \varepsilon\,,
\end{align}
which can be obtained from Eqs.~(\ref{Exp:Y_U1_neutral}) and (\ref{Exp:Upper_U1}), in Eq.~(\ref{Exp:shear_U1}), we arrive at 
\begin{align}
    \sqrt{\frac{{A^i}_j {A^j}_i}{K^2}} \leq  \sqrt{d^3 (d-1)}\, \varepsilon \int^{\psi}_{\psi_*} d \psi' \frac{(F \sqrt{g} K \varepsilon)_{\psi'}}{(F \sqrt{g} K \varepsilon)_{\psi}} 
    \,,
\end{align}
where the numerator and the denominator are evaluated for $\psi'$ and $\psi$, respectively. When $(F \sqrt{g} K \varepsilon)$ increases in time, the integral is dominated by the upper end, giving the upper bound of ${\cal O}(\varepsilon)$. 
In particular, for a slow-roll inflation, using $F \sqrt{g} K \varepsilon \propto e^{d \psi}$, we can compute the upper bound on the shear more accurately as
\begin{align}
     \sqrt{\frac{{A^i}_j {A^j}_i}{K^2}} \leq \sqrt{d (d-1)}\, \varepsilon \,. 
\end{align}
Meanwhile, when $(F \sqrt{g} K \varepsilon)$ decreases in time likewise in the ultra slow-roll model, the integral is dominated by the lower end, giving the upper bound of ${\cal O}(\varepsilon_*) \times (F\! \sqrt{g} K)_*/(F\! \sqrt{g} K)$.

Finally, let us discuss an application of the \dNex. As has been argued, when the \approxGe condition does not hold, e.g., in the absence of charged matter fields, the initial condition for the \dNex~should be given by using $\{ \varphi_*^{a'} \}'$. Perturbing Eq.~(\ref{Exp:deltaN2}), we can compute $\delta \psi(t_f,\, \bm{x})$ as
\begin{align}
    \delta \psi (t_f,\, \bm{x}) = \sum_{a'} \frac{\partial N}{\partial \bar{\varphi}^{a'}_*} \delta \varphi^{a'}_* + \frac{1}{2} \sum_{a',\, b'} \frac{\partial^2 N}{\partial \bar{\varphi}^{a'}_* \partial \bar{\varphi}^{b'}_*} \delta \varphi^{a'}_*  \delta \varphi^{b'}_* + \cdots \,.   
\end{align}
Since the gauge constraint ${\cal H}_{\rm U(1)}$ should be left unsolved, the longitudinal mode of $\pi^i_{A*}$ should be also included in computing the coefficients of the expansion, $\partial^n N/\partial \bar{\varphi}^{a'_1} \cdots \partial \bar{\varphi}^{a'_n}$. However, whether the longitudinal mode is kept or not does not affect the computation of the coefficients, since there is no distinction between the longitudinal and transverse modes in the homogeneous limit. Meanwhile, the initial fluctuation $\delta \varphi^{a'}_*$ should be given, satisfying also ${\cal G}_\epsilon$ properly.

To compute the curvature perturbation, we also need to include the traceless part of $\gamma_{ij}$  (see Eq.~(\ref{Def:zeta})) by solving the equations ${\cal E}$ in the homogeneous limit. This can yield a difference from the result in Ref.~\cite{Abolhasani:2013zya}, especially when the shear is not negligible. A more detailed analysis is left elsewhere.

\subsubsection{Charged scalar fields}
Next, let us consider a model with charged scalar fields, whose equations of motion are given by
\begin{align}
    \frac{1}{N \sqrt{g}} \partial_t \left( {\cal F}_{IJ} \sqrt{g} \frac{\dot{\phi}^J}{N} \right) + g^{ij} A_i A_j {\cal F}_{IJ} e^{(I)} e^{(J)} \phi^J - 2 \frac{\partial {\cal F}}{\partial \phi^{I*}} = {\cal O}(\epsilon)\,. 
\end{align}
As is known and also as one can confirm by using Eq.~(\ref{Eq:Maxwell}), through the Higgs mechanism, a non-zero vacuum expectation value of the charged scalar fields can induce the time dependent mass term of $A^i$, breaking the conformal symmetry~\cite{Turner:1987bw}. Equation (\ref{Exp:Ji}) indicates that as long as the field space metric ${\cal F}_{IJ}$ is positive definite, the acquired mass of the gauge field is positive. Once the gauge symmetry is spontaneously broken, the gauge constraint ${\cal H}_{\rm U(1)}$ can be solved locally, likewise the corresponding equation for a massive vector field. Namely, the U(1) gauge constraint no longer belongs to ${\cal G}_\epsilon$. Naively, this is because with the mass gap, the gauge fields are screened in the large scale limit. Therefore, when there exist non-vanishing charged scalar fields in the limit $\epsilon \to 0$, the \approxGe condition holds even in the presence of gauge fields. As a result, the initial conditions for the \dNex~can be also given by using only the physical degrees of freedom $\{ \varphi_*^{a_{\rm phys}} \}_{\rm phys}$ as long as $t_*$ is set after all the non-local contributions which appear by solving ${\cal H}_i$ become negligibly small. Then, the approximate existence of the WAM is also guaranteed.

\subsection{Remark on ultra slow-roll model} \label{SSSec:USR}
For the \approxGe condition to be valid, the Noether charge density ${Q^i}_j$, whose expression determined by solving ${\cal H}_i$ is non-local, should become negligible in ${\cal E}$. Meanwhile, as discussed in Sec.~\ref{SSec:Wseconed}, when we set ${Q^i}_j$ to 0 in the whole set of the equations, ${\cal E}_{\rm all}$, we only find the WAM but not the second mode, obtaining $\delta(\dot{\psi}/H) = {\cal O}(\epsilon)$. Having considered these aspects, one may think that when the second mode is important\footnote{For example, in an inflection-type model or the ultra slow-roll model~\cite{Tsamis:2003px, Kinney:2005vj}, the approximate shift symmetry of the potential leads to the rapid decrease of the velocity of the inflaton field $|\dot{\phi}|$, resulting in the dominance of Weinberg's second mode.}, the \approxGe condition may not hold, indicating the absence of the WAM. This is not the case, because the \approxGe condition requires that ${Q^i}_j$ should be negligible in all equations of ${\cal E}$, while we find the absence of the second mode for ${Q^i}_j=0$ by solving also ${\cal H}_i$\footnote{To see this point, the Hamiltonian-Jacobi formalism~\cite{Salopek:1990jq} is useful. Once the shear becomes negligible in ${\cal E}$, the second order KG equation (\ref{Eq:KG_single}) can be replaced with the first order equation
\begin{align}
    P_X \dot{\phi} = - \frac{d-1}{8 \pi G d} \partial_\phi K(\phi,\, p) + {\cal O}(\epsilon^2)\,,  \label{Exp:HJ}
\end{align}
where $p$ characterizes the initial condition of the conjugate momentum $\pi_{\phi *}$. The Weinberg's second mode corresponds to the fluctuation of $p_{\sbm{x}}$~\cite{Garriga:2016poh}. In fact, the second mode is described as
\begin{align}
     \zeta_2 (t,\, \bm{x}) \sim \Delta p_{\sbm{x}} \int^t_{t_*} \frac{c_s^2 d\psi}{K \varepsilon e^{\psi}}\,,  \label{Exp:zeta2}
\end{align}
with $c_s^2 = P_X/(P_X + 2X P_{XX})$, $\varepsilon = 4 \pi G P_X (d \phi/d \psi)^2$, and $\Delta p_{\sbm{x}}$ being the spatial variation of $p$ among different patches. Equation (\ref{Exp:zeta2}) indicates that when the trajectory in the field space is an attractor, the dependence on the initial conjugate momentum, $\pi_{\phi *}$, is negligible, so is the Weinberg's second mode. In this case, $\zeta$ is conserved in time, since the WAM becomes the dominant mode. 
Meanwhile, the momentum constraints (\ref{Eq:MC_single}) relate $p_{\sbm{x}}$ to ${Q^i}_{j\,\sbm{x}}$ as~\cite{Garriga:2015tea}
\begin{align}
     \nabla_j \left( e^{-d \psi_{\sbm{x}}} {Q^j}_{i\,\sbm{x}} \right) = \frac{d-1}{d} \frac{\partial K}{\partial p} 
     \partial_i p_{\sbm{x}} + {\cal O}(\epsilon^2)\,. \label{Exp:RelcQ}
\end{align}
While Eq.~(\ref{Exp:zeta2}) is the expression for the linear perturbation, Eq.~(\ref{Exp:RelcQ}) holds non-linearly. Using Eq.~(\ref{Exp:zeta2}) and the momentum constraints (\ref{Exp:RelcQ}), we find that for ${Q^j}_{i\,\sbm{x}}=0$, the Weinberg's second mode vanishes, reproducing the result in Sec.~\ref{SSec:Wseconed}.}, in which the contributions of ${Q^i}_j$ need not be negligible for the validity of the \approxGe condition.

In what follows, we consider a single scalar inflation, which is the same setup as in Sec.~\ref{SSec:Wseconed}. The Klein-Gordon (KG) equation is given by
\begin{align}
    \frac{e^{-d\psi}}{N} \partial_t \left( e^{d \psi} \frac{P_X \dot{\phi}}{N} \right) -  P_\phi = {\cal O} (\epsilon^2)\,. \label{Eq:KG_single}
\end{align}
Among ${\cal E}$, Eqs.~(\ref{Eq:KG_single}) and (\ref{Eq:HC_single}) are sufficient to describe the evolution of the scalar field at the leading order of $\epsilon$, if the influence of the shear ${A^i}_j$ is negligible.  
In the two equations mentioned above, ${Q^i}_j$ appears only through the second term of Eq.~(\ref{Eq:HC_single}), ${A^i}_j {A^j}_i$, at the leading order of $\epsilon$. Therefore, when this second term becomes negligibly small compared to the first, i.e.,
\begin{align}
     \frac{{A^i}_j {A^j}_i}{K^2} = \frac{{Q^i}_j {Q^j}_i}{K^2} e^{- 2d \psi} \ll 1 \,, \label{Eq:shear_negligible}
\end{align}
one would be able to safely neglect ${Q^i}_j$. 
Once this situation is realized, whether ${Q^i}_j$ is chosen so that ${\cal H}_i$ is satisfied or not is not important, ensuring the approximate validity of ${\cal H}_i$ and subsequently the validity of the \approxGe condition. Let us emphasize that the validity of the \approxGe condition or Eq.~(\ref{Eq:shear_negligible}) does not imply that the Weinberg's second mode, written as $\zeta_2$, should be much smaller than the WAM. In fact, when the time integral in $\zeta_2$ is largely enhanced, this can happen, satisfying Eq.~(\ref{Eq:shear_negligible}). This is the situation when the scalar field goes through an almost flat potential region, leading to $\varepsilon \propto e^{-2 d\psi}$~\cite{Kinney:2005vj, Namjoo:2012aa}.

Manifestly, the equations ${\cal E}$, corresponding to Eqs.~(\ref{Eq:HC_single}) and (\ref{Eq:KG_single}), have two solutions also for ${Q^i}_j=0$, because the momentum constraints ${\cal H}_i$ are not solved. One corresponds to the WAM and the other decays roughly with $1/\sqrt{g}$. As long as the shear becomes negligible compared to the expansion, satisfying Eq.~(\ref{Eq:shear_negligible}),  these two solutions approximately solve the whole equations ${\cal
E}_{\rm all}$. On the flat slicing or $N=1$ slicing (the synchronous gauge), it becomes manifest that the well-known enhancement of the super-horizon curvature perturbation $\zeta$ in the ultra slow-roll limit $|\dot\phi|\to 0$ is merely caused by the time coordinate transformation to the $\phi=$const. gauge. In fact, the necessary coordinate shift to evaluate $\zeta$ amounts to $\delta t=\delta\phi/\dot\phi$. Even when the Weinberg's second mode $\zeta_2$ becomes much larger than the WAM, the perturbed variables in the present gauge can be computed without encountering the breakdown of perturbative expansion. This is because the dominance of $\zeta_2$ is simply caused by choosing $\phi$ as a clock which is about to stop in the limit $\dot{\phi} \to 0$. We also encounter a similar situation in the uniform Hubble slicing and the uniform density slicing, since the time variation of the expansion and the energy density rapidly decreases, being proportional to $\varepsilon \propto e^{-2 d\psi}$.

\section{Remaining issues: Quantization}  \label{Sec:futureissues}
In this paper, we have considered only the classical evolution, but for completeness, we also need to take into account quantum effects. It is widely known that an influence of the soft modes of $\zeta$ and $\gamma_{ij}$ are rather generically characterized by the so-called consistency relation~\cite{Maldacena:2002vr, Creminelli:2004yq, Hinterbichler:2013dpa}, which is an example of the soft theorem in cosmology. Meanwhile, in single field models of inflation, as was first pointed out in Ref.~\cite{Urakawa:2009gb} based on the approach of Ref.~\cite{Geshnizjani:2002wp}, the IR divergence \cite{Tsamis:1993ub, Tsamis:1996qm, Sloth:2006az, Sloth:2006nu, Seery:2007we, Seery:2007wf, Urakawa:2008rb, Kitamoto:2012vj, Kitamoto:2013rea} of the adiabatic perturbation can be attributed to residual gauge degrees of freedom. Therefore, the IR divergence is cancelled out in a quantity which remains invariant under these residual gauge transformations~\cite{Byrnes:2010yc, Gerstenlauer:2011ti, Urakawa:2010it, Urakawa:2010kr, Giddings:2010nc, Giddings:2011zd, Senatore:2012nq, Tanaka:2012wi}. (See Refs.~\cite{Geshnizjani:2003cn, Urakawa:2009gb, Urakawa:2011fg, Tokuda:2018eqs, Gorbenko:2019rza} for IR issues of isocurvature modes.) The residual gauge degrees of freedom also require attention in discussing the local type non-Gaussianity~\cite{Tanaka:2011aj, Pajer:2013ana}. This argument was extended to higher loops in Ref.~\cite{Tanaka:2013caa} and also to gravitational waves in Ref.~\cite{Tanaka:2014ina}. The IR regularity of $\zeta$ also has a connection to the conservation of $\zeta$ in the presence of loop corrections~\cite{Senatore:2012ya, Assassi:2012et}. A review of IR issues can be found in Refs.~\cite{Seery:2010kh, Tanaka:2013xe, Hu:2018nxy}.

In Weinberg's paper~\cite{Weinberg:2003sw}, the existence of the constant solution was shown under the requirement of the continuity at $k=0$. In Ref.~\cite{Tanaka:2017nff}, extending the argument of Ref.~\cite{Weinberg:2003sw} so that it also applies to a quantum system, we showed that the continuity at $k=0$ ensures the existence of the WAM, the validity of the consistency relation and the cancellation of the infrared (IR) divergence (see also Refs.~\citep{Goldberger:2013rsa, Berezhiani:2013ewa, Garriga:2016poh}). In particular, we can show the conservation of $\zeta$ also in the presence of radiative corrections of massive fields~\cite{Tanaka:2015aza, Tanaka:2017nff}. More recently, based on the power counting of $\epsilon$, the evolution of the adiabatic curvature perturbation under the influence of loops was discussed in Ref.~\cite{Cohen:2020php}.

In this paper, considering classical theories, we have rephrased the requirement of the continuity at $k=0$ with physically more tractable conditions, the \locality and \approxGe conditions. While the former is usually satisfied for a healthy theory, the latter can be violated, e.g., in a system with a gauge field and neutral fields. In our forthcoming paper~\cite{YUquantum}, extending the discussion in this paper to a quantized system, we show that the \locality\hspace{-4pt}, \sDif\hspace{-4pt}, and \approxGe conditions are also closely related to the validity of the consistency relation and the cancellation of the IR divergence.

\acknowledgments
We would like to thank Lorenzo Bordin, Atsushi Naruko, and Jiro Soda for their fruitful comments on the draft. Y.~U. would like to thank Dick Bond and Jonathan Braden for insightful discussions. T.~T. is supported by JSPS Grant-in-Aid for Scientific Research JP17H06358 (and also JP17H06357), as a part of the innovative research area, ``Gravitational wave physics and astronomy: Genesis'', and also by JP20K03928.
Y.~U. is supported by JSPS Grant-in-Aid for Scientific Research on
Innovative Areas under Contract No.~18H04349, Grant-in-Aid for Scientific Research (B) under Contract No. 19H01894, and the Deutsche Forschungsgemeinschaft (DFG, German Research Foundation) - Project number 315477589 - TRR 211.

\bibliography{refst}

\providecommand{\href}[2]{#2}\begingroup\raggedright\begin{thebibliography}{100}

\bibitem{Salopek:1990jq}
D.~S. Salopek and J.~R. Bond, {\it {Nonlinear evolution of long wavelength
  metric fluctuations in inflationary models}},  {\em Phys. Rev.} {\bf D42}
  (1990) 3936--3962.

\bibitem{Shibata:1999zs}
M.~Shibata and M.~Sasaki, {\it {Black hole formation in the Friedmann universe:
  Formulation and computation in numerical relativity}},  {\em Phys. Rev.} {\bf
  D60} (1999) 084002, [\href{http://arxiv.org/abs/gr-qc/9905064}{{\tt
  gr-qc/9905064}}].

\bibitem{Deruelle:1994iz}
N.~Deruelle and D.~Langlois, {\it {Long wavelength iteration of Einstein's
  equations near a space-time singularity}},  {\em Phys. Rev.} {\bf D52} (1995)
  2007--2019, [\href{http://arxiv.org/abs/gr-qc/9411040}{{\tt gr-qc/9411040}}].

\bibitem{Wands:2000dp}
D.~Wands, K.~A. Malik, D.~H. Lyth, and A.~R. Liddle, {\it {A New approach to
  the evolution of cosmological perturbations on large scales}},  {\em Phys.
  Rev.} {\bf D62} (2000) 043527,
  [\href{http://arxiv.org/abs/astro-ph/0003278}{{\tt astro-ph/0003278}}].

\bibitem{Lyth:2004gb}
D.~H. Lyth, K.~A. Malik, and M.~Sasaki, {\it {A General proof of the
  conservation of the curvature perturbation}},  {\em JCAP} {\bf 0505} (2005)
  004, [\href{http://arxiv.org/abs/astro-ph/0411220}{{\tt astro-ph/0411220}}].

\bibitem{Starobinsky:1982ee}
A.~A. Starobinsky, {\it {Dynamics of Phase Transition in the New Inflationary
  Universe Scenario and Generation of Perturbations}},  {\em Phys. Lett.} {\bf
  117B} (1982) 175--178.

\bibitem{Starobinsky:1986fxa}
A.~A. Starobinsky, {\it {Multicomponent de Sitter (Inflationary) Stages and the
  Generation of Perturbations}},  {\em JETP Lett.} {\bf 42} (1985) 152--155.
  [Pisma Zh. Eksp. Teor. Fiz.42,124(1985)].

\bibitem{Sasaki:1995aw}
M.~Sasaki and E.~D. Stewart, {\it {A General analytic formula for the spectral
  index of the density perturbations produced during inflation}},  {\em Prog.
  Theor. Phys.} {\bf 95} (1996) 71--78,
  [\href{http://arxiv.org/abs/astro-ph/9507001}{{\tt astro-ph/9507001}}].

\bibitem{Sasaki:1998ug}
M.~Sasaki and T.~Tanaka, {\it {Superhorizon scale dynamics of multiscalar
  inflation}},  {\em Prog. Theor. Phys.} {\bf 99} (1998) 763--782,
  [\href{http://arxiv.org/abs/gr-qc/9801017}{{\tt gr-qc/9801017}}].

\bibitem{Tanaka:2006zp}
Y.~Tanaka and M.~Sasaki, {\it {Gradient expansion approach to nonlinear
  superhorizon perturbations}},  {\em Prog. Theor. Phys.} {\bf 117} (2007)
  633--654, [\href{http://arxiv.org/abs/gr-qc/0612191}{{\tt gr-qc/0612191}}].

\bibitem{Weinberg:2008nf}
S.~Weinberg, {\it {Non-Gaussian Correlations Outside the Horizon}},  {\em Phys.
  Rev. D} {\bf 78} (2008) 123521, [\href{http://arxiv.org/abs/0808.2909}{{\tt
  arXiv:0808.2909}}].

\bibitem{Weinberg:2008si}
S.~Weinberg, {\it {Non-Gaussian Correlations Outside the Horizon II: The
  General Case}},  {\em Phys. Rev. D} {\bf 79} (2009) 043504,
  [\href{http://arxiv.org/abs/0810.2831}{{\tt arXiv:0810.2831}}].

\bibitem{Takamizu:2010xy}
Y.-i. Takamizu, S.~Mukohyama, M.~Sasaki, and Y.~Tanaka, {\it {Non-Gaussianity
  of superhorizon curvature perturbations beyond $\delta$ N formalism}},  {\em
  JCAP} {\bf 06} (2010) 019, [\href{http://arxiv.org/abs/1004.1870}{{\tt
  arXiv:1004.1870}}].

\bibitem{Naruko:2012fe}
A.~Naruko, Y.-i. Takamizu, and M.~Sasaki, {\it {Beyond $\delta N$ formalism}},
  {\em PTEP} {\bf 2013} (2013) 043E01,
  [\href{http://arxiv.org/abs/1210.6525}{{\tt arXiv:1210.6525}}].

\bibitem{Ratra:1991bn}
B.~Ratra, {\it {Cosmological 'seed' magnetic field from inflation}},  {\em
  Astrophys. J. Lett.} {\bf 391} (1992) L1--L4.

\bibitem{Caldwell_2011}
R.~R. Caldwell, L.~Motta, and M.~Kamionkowski, {\it Correlation of
  inflation-produced magnetic fields with scalar fluctuations},  {\em Physical
  Review D} {\bf 84} (Dec, 2011).

\bibitem{Martin:2007ue}
J.~Martin and J.~Yokoyama, {\it {Generation of Large-Scale Magnetic Fields in
  Single-Field Inflation}},  {\em JCAP} {\bf 01} (2008) 025,
  [\href{http://arxiv.org/abs/0711.4307}{{\tt arXiv:0711.4307}}].

\bibitem{Turner:1987bw}
M.~S. Turner and L.~M. Widrow, {\it {Inflation Produced, Large Scale Magnetic
  Fields}},  {\em Phys. Rev. D} {\bf 37} (1988) 2743.

\bibitem{Demozzi:2009fu}
V.~Demozzi, V.~Mukhanov, and H.~Rubinstein, {\it {Magnetic fields from
  inflation?}},  {\em JCAP} {\bf 08} (2009) 025,
  [\href{http://arxiv.org/abs/0907.1030}{{\tt arXiv:0907.1030}}].

\bibitem{Watanabe:2009ct}
M.-a. Watanabe, S.~Kanno, and J.~Soda, {\it {Inflationary Universe with
  Anisotropic Hair}},  {\em Phys. Rev. Lett.} {\bf 102} (2009) 191302,
  [\href{http://arxiv.org/abs/0902.2833}{{\tt arXiv:0902.2833}}].

\bibitem{Watanabe:2010fh}
M.-a. Watanabe, S.~Kanno, and J.~Soda, {\it {The Nature of Primordial
  Fluctuations from Anisotropic Inflation}},  {\em Prog. Theor. Phys.} {\bf
  123} (2010) 1041--1068, [\href{http://arxiv.org/abs/1003.0056}{{\tt
  arXiv:1003.0056}}].

\bibitem{Kanno:2010nr}
S.~Kanno, J.~Soda, and M.-a. Watanabe, {\it {Anisotropic Power-law Inflation}},
   {\em JCAP} {\bf 12} (2010) 024, [\href{http://arxiv.org/abs/1010.5307}{{\tt
  arXiv:1010.5307}}].

\bibitem{Soda:2012zm}
J.~Soda, {\it {Statistical Anisotropy from Anisotropic Inflation}},  {\em
  Class. Quant. Grav.} {\bf 29} (2012) 083001,
  [\href{http://arxiv.org/abs/1201.6434}{{\tt arXiv:1201.6434}}].

\bibitem{Wald:1983ky}
R.~M. Wald, {\it {Asymptotic behavior of homogeneous cosmological models in the
  presence of a positive cosmological constant}},  {\em Phys. Rev. D} {\bf 28}
  (1983) 2118--2120.

\bibitem{Cheung:2007st}
C.~Cheung, P.~Creminelli, A.~Fitzpatrick, J.~Kaplan, and L.~Senatore, {\it {The
  Effective Field Theory of Inflation}},  {\em JHEP} {\bf 03} (2008) 014,
  [\href{http://arxiv.org/abs/0709.0293}{{\tt arXiv:0709.0293}}].

\bibitem{Gong:2019hwj}
J.-O. Gong, T.~Noumi, G.~Shiu, J.~Soda, K.~Takahashi, and M.~Yamaguchi, {\it
  {Effective Field Theory of Anisotropic Inflation and Beyond}},  {\em JCAP}
  {\bf 08} (2020) 027, [\href{http://arxiv.org/abs/1910.11533}{{\tt
  arXiv:1910.11533}}].

\bibitem{Graham:2015rva}
P.~W. Graham, J.~Mardon, and S.~Rajendran, {\it {Vector Dark Matter from
  Inflationary Fluctuations}},  {\em Phys. Rev. D} {\bf 93} (2016), no.~10
  103520, [\href{http://arxiv.org/abs/1504.02102}{{\tt arXiv:1504.02102}}].

\bibitem{Nakayama:2019rhg}
K.~Nakayama, {\it {Vector Coherent Oscillation Dark Matter}},  {\em JCAP} {\bf
  10} (2019) 019, [\href{http://arxiv.org/abs/1907.06243}{{\tt
  arXiv:1907.06243}}].

\bibitem{Kehagias:2017cym}
A.~Kehagias and A.~Riotto, {\it {On the Inflationary Perturbations of Massive
  Higher-Spin Fields}},  {\em JCAP} {\bf 07} (2017) 046,
  [\href{http://arxiv.org/abs/1705.05834}{{\tt arXiv:1705.05834}}].

\bibitem{Arkani-Hamed:2015bza}
N.~Arkani-Hamed and J.~Maldacena, {\it {Cosmological Collider Physics}},
  \href{http://arxiv.org/abs/1503.08043}{{\tt arXiv:1503.08043}}.

\bibitem{Lee:2016vti}
H.~Lee, D.~Baumann, and G.~L. Pimentel, {\it {Non-Gaussianity as a Particle
  Detector}},  {\em JHEP} {\bf 12} (2016) 040,
  [\href{http://arxiv.org/abs/1607.03735}{{\tt arXiv:1607.03735}}].

\bibitem{Ghosh:2014kba}
A.~Ghosh, N.~Kundu, S.~Raju, and S.~P. Trivedi, {\it {Conformal Invariance and
  the Four Point Scalar Correlator in Slow-Roll Inflation}},  {\em JHEP} {\bf
  07} (2014) 011, [\href{http://arxiv.org/abs/1401.1426}{{\tt
  arXiv:1401.1426}}].

\bibitem{Akrami:2019izv}
{\bf Planck} Collaboration, Y.~Akrami et~al., {\it {Planck 2018 results. IX.
  Constraints on primordial non-Gaussianity}},
  \href{http://arxiv.org/abs/1905.05697}{{\tt arXiv:1905.05697}}.

\bibitem{Bartolo:2017sbu}
N.~Bartolo, A.~Kehagias, M.~Liguori, A.~Riotto, M.~Shiraishi, and V.~Tansella,
  {\it {Detecting higher spin fields through statistical anisotropy in the CMB
  and galaxy power spectra}},  {\em Phys. Rev.} {\bf D97} (2018), no.~2 023503,
  [\href{http://arxiv.org/abs/1709.05695}{{\tt arXiv:1709.05695}}].

\bibitem{Bordin:2019tyb}
L.~Bordin and G.~Cabass, {\it {Probing higher-spin fields from inflation with
  higher-order statistics of the CMB}},  {\em JCAP} {\bf 06} (2019) 050,
  [\href{http://arxiv.org/abs/1902.09519}{{\tt arXiv:1902.09519}}].

\bibitem{Franciolini:2018eno}
G.~Franciolini, A.~Kehagias, A.~Riotto, and M.~Shiraishi, {\it {Detecting
  higher spin fields through statistical anisotropy in the CMB bispectrum}},
  {\em Phys. Rev.} {\bf D98} (2018), no.~4 043533,
  [\href{http://arxiv.org/abs/1803.03814}{{\tt arXiv:1803.03814}}].

\bibitem{MoradinezhadDizgah:2018pfo}
A.~Moradinezhad~Dizgah, G.~Franciolini, A.~Kehagias, and A.~Riotto, {\it
  {Constraints on long-lived, higher-spin particles from galaxy bispectrum}},
  {\em Phys. Rev.} {\bf D98} (2018), no.~6 063520,
  [\href{http://arxiv.org/abs/1805.10247}{{\tt arXiv:1805.10247}}].

\bibitem{MoradinezhadDizgah:2018ssw}
A.~Moradinezhad~Dizgah, H.~Lee, J.~B. Muñoz, and C.~Dvorkin, {\it {Galaxy
  Bispectrum from Massive Spinning Particles}},  {\em JCAP} {\bf 1805} (2018),
  no.~05 013, [\href{http://arxiv.org/abs/1801.07265}{{\tt arXiv:1801.07265}}].

\bibitem{Schmidt:2015xka}
F.~Schmidt, N.~E. Chisari, and C.~Dvorkin, {\it {Imprint of inflation on galaxy
  shape correlations}},  {\em JCAP} {\bf 1510} (2015), no.~10 032,
  [\href{http://arxiv.org/abs/1506.02671}{{\tt arXiv:1506.02671}}].

\bibitem{Chisari:2016xki}
N.~E. Chisari, C.~Dvorkin, F.~Schmidt, and D.~Spergel, {\it {Multitracing
  Anisotropic Non-Gaussianity with Galaxy Shapes}},  {\em Phys. Rev. D} {\bf
  94} (2016), no.~12 123507, [\href{http://arxiv.org/abs/1607.05232}{{\tt
  arXiv:1607.05232}}].

\bibitem{Kogai:2018nse}
K.~Kogai, T.~Matsubara, A.~J. Nishizawa, and Y.~Urakawa, {\it {Intrinsic galaxy
  alignment from angular dependent primordial non-Gaussianity}},  {\em JCAP}
  {\bf 08} (2018) 014, [\href{http://arxiv.org/abs/1804.06284}{{\tt
  arXiv:1804.06284}}].

\bibitem{Kogai:2020vzz}
K.~Kogai, K.~Akitsu, F.~Schmidt, and Y.~Urakawa, {\it {Galaxy imaging surveys
  as spin-sensitive detector for cosmological colliders}},
  \href{http://arxiv.org/abs/2009.05517}{{\tt arXiv:2009.05517}}.

\bibitem{Karciauskas:2008bc}
M.~Karciauskas, K.~Dimopoulos, and D.~H. Lyth, {\it {Anisotropic
  non-Gaussianity from vector field perturbations}},  {\em Phys. Rev. D} {\bf
  80} (2009) 023509, [\href{http://arxiv.org/abs/0812.0264}{{\tt
  arXiv:0812.0264}}]. [Erratum: Phys.Rev.D 85, 069905 (2012)].

\bibitem{Abolhasani:2013zya}
A.~A. Abolhasani, R.~Emami, J.~T. Firouzjaee, and H.~Firouzjahi, {\it {$\delta
  N$ formalism in anisotropic inflation and large anisotropic bispectrum and
  trispectrum}},  {\em JCAP} {\bf 1308} (2013) 016,
  [\href{http://arxiv.org/abs/1302.6986}{{\tt arXiv:1302.6986}}].

\bibitem{Sugiyama:2012tj}
N.~S. Sugiyama, E.~Komatsu, and T.~Futamase, {\it {$\delta$N formalism}},  {\em
  Phys. Rev.} {\bf D87} (2013), no.~2 023530,
  [\href{http://arxiv.org/abs/1208.1073}{{\tt arXiv:1208.1073}}].

\bibitem{Garriga:2015tea}
J.~Garriga, Y.~Urakawa, and F.~Vernizzi, {\it {$\delta N$ formalism from
  superpotential and holography}},  {\em JCAP} {\bf 1602} (2016), no.~02 036,
  [\href{http://arxiv.org/abs/1509.07339}{{\tt arXiv:1509.07339}}].

\bibitem{Izumi:2011eh}
K.~Izumi and S.~Mukohyama, {\it {Nonlinear superhorizon perturbations in
  Horava-Lifshitz gravity}},  {\em Phys. Rev.} {\bf D84} (2011) 064025,
  [\href{http://arxiv.org/abs/1105.0246}{{\tt arXiv:1105.0246}}].

\bibitem{Gumrukcuoglu:2011ef}
A.~E. Gumrukcuoglu, S.~Mukohyama, and A.~Wang, {\it {General relativity limit
  of Horava-Lifshitz gravity with a scalar field in gradient expansion}},  {\em
  Phys. Rev.} {\bf D85} (2012) 064042,
  [\href{http://arxiv.org/abs/1109.2609}{{\tt arXiv:1109.2609}}].

\bibitem{Weinberg:2003sw}
S.~Weinberg, {\it {Adiabatic modes in cosmology}},  {\em Phys. Rev.} {\bf D67}
  (2003) 123504, [\href{http://arxiv.org/abs/astro-ph/0302326}{{\tt
  astro-ph/0302326}}].

\bibitem{Horava:2009uw}
P.~Horava, {\it {Quantum Gravity at a Lifshitz Point}},  {\em Phys. Rev.} {\bf
  D79} (2009) 084008, [\href{http://arxiv.org/abs/0901.3775}{{\tt
  arXiv:0901.3775}}].

\bibitem{Aghanim:2018eyx}
{\bf Planck} Collaboration, N.~Aghanim et~al., {\it {Planck 2018 results. VI.
  Cosmological parameters}},  {\em Astron. Astrophys.} {\bf 641} (2020) A6,
  [\href{http://arxiv.org/abs/1807.06209}{{\tt arXiv:1807.06209}}].

\bibitem{Endlich:2012pz}
S.~Endlich, A.~Nicolis, and J.~Wang, {\it {Solid Inflation}},  {\em JCAP} {\bf
  1310} (2013) 011, [\href{http://arxiv.org/abs/1210.0569}{{\tt
  arXiv:1210.0569}}].

\bibitem{Garriga:1998he}
J.~Garriga, X.~Montes, M.~Sasaki, and T.~Tanaka, {\it {Spectrum of cosmological
  perturbations in the one bubble open universe}},  {\em Nucl. Phys. B} {\bf
  551} (1999) 317--373, [\href{http://arxiv.org/abs/astro-ph/9811257}{{\tt
  astro-ph/9811257}}].

\bibitem{Linde:1999wv}
A.~D. Linde, M.~Sasaki, and T.~Tanaka, {\it {CMB in open inflation}},  {\em
  Phys. Rev. D} {\bf 59} (1999) 123522,
  [\href{http://arxiv.org/abs/astro-ph/9901135}{{\tt astro-ph/9901135}}].

\bibitem{Bordin:2017ozj}
L.~Bordin, P.~Creminelli, M.~Mirbabayi, and J.~Nore\~na, {\it {Solid
  Consistency}},  {\em JCAP} {\bf 03} (2017) 004,
  [\href{http://arxiv.org/abs/1701.04382}{{\tt arXiv:1701.04382}}].

\bibitem{Blas:2010hb}
D.~Blas, O.~Pujolas, and S.~Sibiryakov, {\it {Models of non-relativistic
  quantum gravity: The Good, the bad and the healthy}},  {\em JHEP} {\bf 04}
  (2011) 018, [\href{http://arxiv.org/abs/1007.3503}{{\tt arXiv:1007.3503}}].

\bibitem{Barvinsky:2015kil}
A.~O. Barvinsky, D.~Blas, M.~Herrero-Valea, S.~M. Sibiryakov, and C.~F.
  Steinwachs, {\it {Renormalization of Ho\v{r}ava gravity}},  {\em Phys. Rev.
  D} {\bf 93} (2016), no.~6 064022,
  [\href{http://arxiv.org/abs/1512.02250}{{\tt arXiv:1512.02250}}].

\bibitem{Barvinsky:2017zlx}
A.~O. Barvinsky, D.~Blas, M.~Herrero-Valea, S.~M. Sibiryakov, and C.~F.
  Steinwachs, {\it {Renormalization of gauge theories in the background-field
  approach}},  {\em JHEP} {\bf 07} (2018) 035,
  [\href{http://arxiv.org/abs/1705.03480}{{\tt arXiv:1705.03480}}].

\bibitem{ArmendarizPicon:2010rs}
C.~Armendariz-Picon, N.~F. Sierra, and J.~Garriga, {\it {Primordial
  Perturbations in Einstein-Aether and BPSH Theories}},  {\em JCAP} {\bf 1007}
  (2010) 010, [\href{http://arxiv.org/abs/1003.1283}{{\tt arXiv:1003.1283}}].

\bibitem{Arai:2018whk}
S.~Arai, S.~Sibiryakov, and Y.~Urakawa, {\it {Inflationary perturbations with
  Lifshitz scaling}},  {\em JCAP} {\bf 1903} (2019), no.~03 034,
  [\href{http://arxiv.org/abs/1803.01352}{{\tt arXiv:1803.01352}}].

\bibitem{Kobayashi:2010eh}
T.~Kobayashi, Y.~Urakawa, and M.~Yamaguchi, {\it {Cosmological perturbations in
  a healthy extension of Horava gravity}},  {\em JCAP} {\bf 04} (2010) 025,
  [\href{http://arxiv.org/abs/1002.3101}{{\tt arXiv:1002.3101}}].

\bibitem{Buoninfante:2018lnh}
L.~Buoninfante, G.~Lambiase, and M.~Yamaguchi, {\it {Nonlocal generalization of
  Galilean theories and gravity}},  {\em Phys. Rev. D} {\bf 100} (2019), no.~2
  026019, [\href{http://arxiv.org/abs/1812.10105}{{\tt arXiv:1812.10105}}].

\bibitem{Starobinsky:1986fx}
A.~A. Starobinsky, {\it {STOCHASTIC DE SITTER (INFLATIONARY) STAGE IN THE EARLY
  UNIVERSE}},  {\em Lect. Notes Phys.} {\bf 246} (1986) 107--126.

\bibitem{Nambu:1987ef}
Y.~Nambu and M.~Sasaki, {\it {Stochastic Stage of an Inflationary Universe
  Model}},  {\em Phys. Lett.} {\bf B205} (1988) 441--446.

\bibitem{Nambu:1988je}
Y.~Nambu and M.~Sasaki, {\it {Stochastic Approach to Chaotic Inflation and the
  Distribution of Universes}},  {\em Phys. Lett.} {\bf B219} (1989) 240--246.

\bibitem{Starobinsky:1994bd}
A.~A. Starobinsky and J.~Yokoyama, {\it {Equilibrium state of a selfinteracting
  scalar field in the De Sitter background}},  {\em Phys. Rev.} {\bf D50}
  (1994) 6357--6368, [\href{http://arxiv.org/abs/astro-ph/9407016}{{\tt
  astro-ph/9407016}}].

\bibitem{Tokuda:2017fdh}
J.~Tokuda and T.~Tanaka, {\it {Statistical nature of infrared dynamics on de
  Sitter background}},  {\em JCAP} {\bf 02} (2018) 014,
  [\href{http://arxiv.org/abs/1708.01734}{{\tt arXiv:1708.01734}}].

\bibitem{Gorbenko:2019rza}
V.~Gorbenko and L.~Senatore, {\it {$\lambda \phi^4$ in dS}},
  \href{http://arxiv.org/abs/1911.00022}{{\tt arXiv:1911.00022}}.

\bibitem{Baumgart:2019clc}
M.~Baumgart and R.~Sundrum, {\it {De Sitter Diagrammar and the Resummation of
  Time}},  {\em JHEP} {\bf 07} (2020) 119,
  [\href{http://arxiv.org/abs/1912.09502}{{\tt arXiv:1912.09502}}].

\bibitem{YUquantum}
T.~Tanaka and Y.~Urakawa, {\it {in preparation}}, .

\bibitem{Emami:2009vd}
R.~Emami, H.~Firouzjahi, and M.~Movahed, {\it {Inflation from Charged Scalar
  and Primordial Magnetic Fields?}},  {\em Phys. Rev. D} {\bf 81} (2010)
  083526, [\href{http://arxiv.org/abs/0908.4161}{{\tt arXiv:0908.4161}}].

\bibitem{deRham:2010ik}
C.~de~Rham and G.~Gabadadze, {\it {Generalization of the Fierz-Pauli Action}},
  {\em Phys. Rev. D} {\bf 82} (2010) 044020,
  [\href{http://arxiv.org/abs/1007.0443}{{\tt arXiv:1007.0443}}].

\bibitem{deRham:2010kj}
C.~de~Rham, G.~Gabadadze, and A.~J. Tolley, {\it {Resummation of Massive
  Gravity}},  {\em Phys. Rev. Lett.} {\bf 106} (2011) 231101,
  [\href{http://arxiv.org/abs/1011.1232}{{\tt arXiv:1011.1232}}].

\bibitem{Talebian-Ashkezari:2016llx}
A.~Talebian-Ashkezari, N.~Ahmadi, and A.~A. Abolhasani, {\it {$\delta$ M
  formalism: a new approach to cosmological perturbation theory in anisotropic
  inflation}},  {\em JCAP} {\bf 03} (2018) 001,
  [\href{http://arxiv.org/abs/1609.05893}{{\tt arXiv:1609.05893}}].

\bibitem{Tanaka:2011aj}
T.~Tanaka and Y.~Urakawa, {\it {Dominance of gauge artifact in the consistency
  relation for the primordial bispectrum}},  {\em JCAP} {\bf 05} (2011) 014,
  [\href{http://arxiv.org/abs/1103.1251}{{\tt arXiv:1103.1251}}].

\bibitem{Finelli:2017fml}
B.~Finelli, G.~Goon, E.~Pajer, and L.~Santoni, {\it {Soft Theorems For
  Shift-Symmetric Cosmologies}},  {\em Phys. Rev. D} {\bf 97} (2018), no.~6
  063531, [\href{http://arxiv.org/abs/1711.03737}{{\tt arXiv:1711.03737}}].

\bibitem{Pajer:2019jhb}
S.~Jazayeri, E.~Pajer, and D.~van~der Woude, {\it {Solid Soft Theorems}},  {\em
  JCAP} {\bf 06} (2019) 011, [\href{http://arxiv.org/abs/1902.09020}{{\tt
  arXiv:1902.09020}}].

\bibitem{Kobayashi:2009hh}
T.~Kobayashi, Y.~Urakawa, and M.~Yamaguchi, {\it {Large scale evolution of the
  curvature perturbation in Horava-Lifshitz cosmology}},  {\em JCAP} {\bf 0911}
  (2009) 015, [\href{http://arxiv.org/abs/0908.1005}{{\tt arXiv:0908.1005}}].

\bibitem{Tanaka:2017nff}
T.~Tanaka and Y.~Urakawa, {\it {Large gauge transformation, Soft theorem, and
  Infrared divergence in inflationary spacetime}},  {\em JHEP} {\bf 10} (2017)
  127, [\href{http://arxiv.org/abs/1707.05485}{{\tt arXiv:1707.05485}}].

\bibitem{Blas:2009qj}
D.~Blas, O.~Pujolas, and S.~Sibiryakov, {\it {Consistent Extension of Horava
  Gravity}},  {\em Phys. Rev. Lett.} {\bf 104} (2010) 181302,
  [\href{http://arxiv.org/abs/0909.3525}{{\tt arXiv:0909.3525}}].

\bibitem{Gleyzes:2014dya}
J.~Gleyzes, D.~Langlois, F.~Piazza, and F.~Vernizzi, {\it {Healthy theories
  beyond Horndeski}},  {\em Phys. Rev. Lett.} {\bf 114} (2015), no.~21 211101,
  [\href{http://arxiv.org/abs/1404.6495}{{\tt arXiv:1404.6495}}].

\bibitem{Shibata:1995dg}
M.~Shibata and H.~Asada, {\it {PostNewtonian equations of motion in the flat
  universe}},  {\em Prog. Theor. Phys.} {\bf 94} (1995) 11--31.

\bibitem{Kodama:1985bj}
H.~Kodama and M.~Sasaki, {\it {Cosmological Perturbation Theory}},  {\em Prog.
  Theor. Phys. Suppl.} {\bf 78} (1984) 1--166.

\bibitem{Avery:2015rga}
S.~G. Avery and B.~U.~W. Schwab, {\it {Noether’s second theorem and Ward
  identities for gauge symmetries}},  {\em JHEP} {\bf 02} (2016) 031,
  [\href{http://arxiv.org/abs/1510.07038}{{\tt arXiv:1510.07038}}].

\bibitem{Urakawa:2010it}
Y.~Urakawa and T.~Tanaka, {\it {IR divergence does not affect the
  gauge-invariant curvature perturbation}},  {\em Phys. Rev. D} {\bf 82} (2010)
  121301, [\href{http://arxiv.org/abs/1007.0468}{{\tt arXiv:1007.0468}}].

\bibitem{Urakawa:2010kr}
Y.~Urakawa and T.~Tanaka, {\it {Natural selection of inflationary vacuum
  required by infra-red regularity and gauge-invariance}},  {\em Prog. Theor.
  Phys.} {\bf 125} (2011) 1067--1089,
  [\href{http://arxiv.org/abs/1009.2947}{{\tt arXiv:1009.2947}}].

\bibitem{Hinterbichler:2013dpa}
K.~Hinterbichler, L.~Hui, and J.~Khoury, {\it {An Infinite Set of Ward
  Identities for Adiabatic Modes in Cosmology}},  {\em JCAP} {\bf 01} (2014)
  039, [\href{http://arxiv.org/abs/1304.5527}{{\tt arXiv:1304.5527}}].

\bibitem{Matarrese:2020why}
S.~Matarrese, L.~Pilo, and R.~Rollo, {\it {Resilience of long modes in
  cosmological observables}},  \href{http://arxiv.org/abs/2007.08877}{{\tt
  arXiv:2007.08877}}.

\bibitem{Gumrukcuoglu:2007bx}
A.~Gumrukcuoglu, C.~R. Contaldi, and M.~Peloso, {\it {Inflationary
  perturbations in anisotropic backgrounds and their imprint on the CMB}},
  {\em JCAP} {\bf 11} (2007) 005, [\href{http://arxiv.org/abs/0707.4179}{{\tt
  arXiv:0707.4179}}].

\bibitem{Gumrukcuoglu:2008gi}
A.~Gumrukcuoglu, L.~Kofman, and M.~Peloso, {\it {Gravity Waves Signatures from
  Anisotropic pre-Inflation}},  {\em Phys. Rev. D} {\bf 78} (2008) 103525,
  [\href{http://arxiv.org/abs/0807.1335}{{\tt arXiv:0807.1335}}].

\bibitem{Pereira:2007yy}
T.~S. Pereira, C.~Pitrou, and J.-P. Uzan, {\it {Theory of cosmological
  perturbations in an anisotropic universe}},  {\em JCAP} {\bf 09} (2007) 006,
  [\href{http://arxiv.org/abs/0707.0736}{{\tt arXiv:0707.0736}}].

\bibitem{Lyth:2003im}
D.~H. Lyth and D.~Wands, {\it {Conserved cosmological perturbations}},  {\em
  Phys. Rev. D} {\bf 68} (2003) 103515,
  [\href{http://arxiv.org/abs/astro-ph/0306498}{{\tt astro-ph/0306498}}].

\bibitem{Tanaka:2010km}
T.~Tanaka, T.~Suyama, and S.~Yokoyama, {\it {Use of delta N formalism -
  Difficulties in generating large local-type non-Gaussianity during inflation
  -}},  {\em Class. Quant. Grav.} {\bf 27} (2010) 124003,
  [\href{http://arxiv.org/abs/1003.5057}{{\tt arXiv:1003.5057}}].

\bibitem{Abolhasani:2019qcn}
A.~A. Abolhasani, H.~Firouzjahi, A.~Naruko, and M.~Sasaki, {\it {Basic
  formulation of $\delta$N formalism}},  pp.~9--65.
\newblock 2019.

\bibitem{Kanno:2009ei}
S.~Kanno, J.~Soda, and M.-a. Watanabe, {\it {Cosmological Magnetic Fields from
  Inflation and Backreaction}},  {\em JCAP} {\bf 12} (2009) 009,
  [\href{http://arxiv.org/abs/0908.3509}{{\tt arXiv:0908.3509}}].

\bibitem{Maleknejad:2012as}
A.~Maleknejad and M.~Sheikh-Jabbari, {\it {Revisiting Cosmic No-Hair Theorem
  for Inflationary Settings}},  {\em Phys. Rev. D} {\bf 85} (2012) 123508,
  [\href{http://arxiv.org/abs/1203.0219}{{\tt arXiv:1203.0219}}].

\bibitem{Tsamis:2003px}
N.~Tsamis and R.~P. Woodard, {\it {Improved estimates of cosmological
  perturbations}},  {\em Phys. Rev. D} {\bf 69} (2004) 084005,
  [\href{http://arxiv.org/abs/astro-ph/0307463}{{\tt astro-ph/0307463}}].

\bibitem{Kinney:2005vj}
W.~H. Kinney, {\it {Horizon crossing and inflation with large eta}},  {\em
  Phys. Rev.} {\bf D72} (2005) 023515,
  [\href{http://arxiv.org/abs/gr-qc/0503017}{{\tt gr-qc/0503017}}].

\bibitem{Garriga:2016poh}
J.~Garriga and Y.~Urakawa, {\it {Consistency relations and conservation of
  $\zeta$ in holographic inflation}},  {\em JCAP} {\bf 10} (2016) 030,
  [\href{http://arxiv.org/abs/1606.04767}{{\tt arXiv:1606.04767}}].

\bibitem{Namjoo:2012aa}
M.~H. Namjoo, H.~Firouzjahi, and M.~Sasaki, {\it {Violation of non-Gaussianity
  consistency relation in a single field inflationary model}},  {\em EPL} {\bf
  101} (2013), no.~3 39001, [\href{http://arxiv.org/abs/1210.3692}{{\tt
  arXiv:1210.3692}}].

\bibitem{Maldacena:2002vr}
J.~M. Maldacena, {\it {Non-Gaussian features of primordial fluctuations in
  single field inflationary models}},  {\em JHEP} {\bf 05} (2003) 013,
  [\href{http://arxiv.org/abs/astro-ph/0210603}{{\tt astro-ph/0210603}}].

\bibitem{Creminelli:2004yq}
P.~Creminelli and M.~Zaldarriaga, {\it {Single field consistency relation for
  the 3-point function}},  {\em JCAP} {\bf 10} (2004) 006,
  [\href{http://arxiv.org/abs/astro-ph/0407059}{{\tt astro-ph/0407059}}].

\bibitem{Urakawa:2009gb}
Y.~Urakawa and T.~Tanaka, {\it {Influence on observation from IR divergence
  during inflation: Multi field inflation}},  {\em Prog. Theor. Phys.} {\bf
  122} (2010) 1207--1238, [\href{http://arxiv.org/abs/0904.4415}{{\tt
  arXiv:0904.4415}}].

\bibitem{Geshnizjani:2002wp}
G.~Geshnizjani and R.~Brandenberger, {\it {Back reaction and local cosmological
  expansion rate}},  {\em Phys. Rev. D} {\bf 66} (2002) 123507,
  [\href{http://arxiv.org/abs/gr-qc/0204074}{{\tt gr-qc/0204074}}].

\bibitem{Tsamis:1993ub}
N.~Tsamis and R.~Woodard, {\it {The Physical basis for infrared divergences in
  inflationary quantum gravity}},  {\em Class. Quant. Grav.} {\bf 11} (1994)
  2969--2990.

\bibitem{Tsamis:1996qm}
N.~Tsamis and R.~Woodard, {\it {The Quantum gravitational back reaction on
  inflation}},  {\em Annals Phys.} {\bf 253} (1997) 1--54,
  [\href{http://arxiv.org/abs/hep-ph/9602316}{{\tt hep-ph/9602316}}].

\bibitem{Sloth:2006az}
M.~S. Sloth, {\it {On the one loop corrections to inflation and the CMB
  anisotropies}},  {\em Nucl. Phys. B} {\bf 748} (2006) 149--169,
  [\href{http://arxiv.org/abs/astro-ph/0604488}{{\tt astro-ph/0604488}}].

\bibitem{Sloth:2006nu}
M.~S. Sloth, {\it {On the one loop corrections to inflation. II. The
  Consistency relation}},  {\em Nucl. Phys. B} {\bf 775} (2007) 78--94,
  [\href{http://arxiv.org/abs/hep-th/0612138}{{\tt hep-th/0612138}}].

\bibitem{Seery:2007we}
D.~Seery, {\it {One-loop corrections to a scalar field during inflation}},
  {\em JCAP} {\bf 11} (2007) 025, [\href{http://arxiv.org/abs/0707.3377}{{\tt
  arXiv:0707.3377}}].

\bibitem{Seery:2007wf}
D.~Seery, {\it {One-loop corrections to the curvature perturbation from
  inflation}},  {\em JCAP} {\bf 02} (2008) 006,
  [\href{http://arxiv.org/abs/0707.3378}{{\tt arXiv:0707.3378}}].

\bibitem{Urakawa:2008rb}
Y.~Urakawa and K.-i. Maeda, {\it {One-loop Corrections to Scalar and Tensor
  Perturbations during Inflation in Stochastic Gravity}},  {\em Phys. Rev. D}
  {\bf 78} (2008) 064004, [\href{http://arxiv.org/abs/0801.0126}{{\tt
  arXiv:0801.0126}}].

\bibitem{Kitamoto:2012vj}
H.~Kitamoto and Y.~Kitazawa, {\it {Soft Graviton effects on Gauge theories in
  de Sitter Space}},  {\em Phys. Rev. D} {\bf 87} (2013), no.~12 124004,
  [\href{http://arxiv.org/abs/1204.2876}{{\tt arXiv:1204.2876}}].

\bibitem{Kitamoto:2013rea}
H.~Kitamoto and Y.~Kitazawa, {\it {Soft gravitational effects in Kadanoff-Baym
  approach}},  {\em JHEP} {\bf 10} (2013) 145,
  [\href{http://arxiv.org/abs/1305.2029}{{\tt arXiv:1305.2029}}].

\bibitem{Byrnes:2010yc}
C.~T. Byrnes, M.~Gerstenlauer, A.~Hebecker, S.~Nurmi, and G.~Tasinato, {\it
  {Inflationary Infrared Divergences: Geometry of the Reheating Surface versus
  $\delta N$ Formalism}},  {\em JCAP} {\bf 08} (2010) 006,
  [\href{http://arxiv.org/abs/1005.3307}{{\tt arXiv:1005.3307}}].

\bibitem{Gerstenlauer:2011ti}
M.~Gerstenlauer, A.~Hebecker, and G.~Tasinato, {\it {Inflationary Correlation
  Functions without Infrared Divergences}},  {\em JCAP} {\bf 06} (2011) 021,
  [\href{http://arxiv.org/abs/1102.0560}{{\tt arXiv:1102.0560}}].

\bibitem{Giddings:2010nc}
S.~B. Giddings and M.~S. Sloth, {\it {Semiclassical relations and IR effects in
  de Sitter and slow-roll space-times}},  {\em JCAP} {\bf 01} (2011) 023,
  [\href{http://arxiv.org/abs/1005.1056}{{\tt arXiv:1005.1056}}].

\bibitem{Giddings:2011zd}
S.~B. Giddings and M.~S. Sloth, {\it {Cosmological observables, IR growth of
  fluctuations, and scale-dependent anisotropies}},  {\em Phys. Rev. D} {\bf
  84} (2011) 063528, [\href{http://arxiv.org/abs/1104.0002}{{\tt
  arXiv:1104.0002}}].

\bibitem{Senatore:2012nq}
L.~Senatore and M.~Zaldarriaga, {\it {On Loops in Inflation II: IR Effects in
  Single Clock Inflation}},  {\em JHEP} {\bf 01} (2013) 109,
  [\href{http://arxiv.org/abs/1203.6354}{{\tt arXiv:1203.6354}}].

\bibitem{Tanaka:2012wi}
T.~Tanaka and Y.~Urakawa, {\it {Strong restriction on inflationary vacua from
  the local gauge invariance I: Local gauge invariance and infrared
  regularity}},  {\em PTEP} {\bf 2013} (2013) 083E01,
  [\href{http://arxiv.org/abs/1209.1914}{{\tt arXiv:1209.1914}}].

\bibitem{Geshnizjani:2003cn}
G.~Geshnizjani and R.~Brandenberger, {\it {Back reaction of perturbations in
  two scalar field inflationary models}},  {\em JCAP} {\bf 04} (2005) 006,
  [\href{http://arxiv.org/abs/hep-th/0310265}{{\tt hep-th/0310265}}].

\bibitem{Urakawa:2011fg}
Y.~Urakawa, {\it {Influence of gauge artifact on adiabatic and entropy
  perturbations during inflation}},  {\em Prog. Theor. Phys.} {\bf 126} (2011)
  961--982, [\href{http://arxiv.org/abs/1105.1078}{{\tt arXiv:1105.1078}}].

\bibitem{Tokuda:2018eqs}
J.~Tokuda and T.~Tanaka, {\it {Can all the infrared secular growth really be
  understood as increase of classical statistical variance?}},  {\em JCAP} {\bf
  11} (2018) 022, [\href{http://arxiv.org/abs/1806.03262}{{\tt
  arXiv:1806.03262}}].

\bibitem{Pajer:2013ana}
E.~Pajer, F.~Schmidt, and M.~Zaldarriaga, {\it {The Observed Squeezed Limit of
  Cosmological Three-Point Functions}},  {\em Phys. Rev. D} {\bf 88} (2013),
  no.~8 083502, [\href{http://arxiv.org/abs/1305.0824}{{\tt arXiv:1305.0824}}].

\bibitem{Tanaka:2013caa}
T.~Tanaka and Y.~Urakawa, {\it {Loops in inflationary correlation functions}},
  {\em Class. Quant. Grav.} {\bf 30} (2013) 233001,
  [\href{http://arxiv.org/abs/1306.4461}{{\tt arXiv:1306.4461}}].

\bibitem{Tanaka:2014ina}
T.~Tanaka and Y.~Urakawa, {\it {Strong restriction on inflationary vacua from
  the local $gauge$ invariance III: Infrared regularity of graviton loops}},
  {\em PTEP} {\bf 2014} (2014), no.~7 073E01,
  [\href{http://arxiv.org/abs/1402.2076}{{\tt arXiv:1402.2076}}].

\bibitem{Senatore:2012ya}
L.~Senatore and M.~Zaldarriaga, {\it {The constancy of $\zeta$ in single-clock
  Inflation at all loops}},  {\em JHEP} {\bf 09} (2013) 148,
  [\href{http://arxiv.org/abs/1210.6048}{{\tt arXiv:1210.6048}}].

\bibitem{Assassi:2012et}
V.~Assassi, D.~Baumann, and D.~Green, {\it {Symmetries and Loops in
  Inflation}},  {\em JHEP} {\bf 02} (2013) 151,
  [\href{http://arxiv.org/abs/1210.7792}{{\tt arXiv:1210.7792}}].

\bibitem{Seery:2010kh}
D.~Seery, {\it {Infrared effects in inflationary correlation functions}},  {\em
  Class. Quant. Grav.} {\bf 27} (2010) 124005,
  [\href{http://arxiv.org/abs/1005.1649}{{\tt arXiv:1005.1649}}].

\bibitem{Tanaka:2013xe}
T.~Tanaka and Y.~Urakawa, {\it {Strong restriction on inflationary vacua from
  the local gauge invariance II: Infrared regularity and absence of secular
  growth in the Euclidean vacuum}},  {\em PTEP} {\bf 2013} (2013), no.~6
  063E02, [\href{http://arxiv.org/abs/1301.3088}{{\tt arXiv:1301.3088}}].

\bibitem{Hu:2018nxy}
B.-L. Hu, {\it {Infrared Behavior of Quantum Fields in Inflationary Cosmology
  -- Issues and Approaches: an overview}},
  \href{http://arxiv.org/abs/1812.11851}{{\tt arXiv:1812.11851}}.

\bibitem{Goldberger:2013rsa}
W.~D. Goldberger, L.~Hui, and A.~Nicolis, {\it {One-particle-irreducible
  consistency relations for cosmological perturbations}},  {\em Phys. Rev. D}
  {\bf 87} (2013), no.~10 103520, [\href{http://arxiv.org/abs/1303.1193}{{\tt
  arXiv:1303.1193}}].

\bibitem{Berezhiani:2013ewa}
L.~Berezhiani and J.~Khoury, {\it {Slavnov-Taylor Identities for Primordial
  Perturbations}},  {\em JCAP} {\bf 02} (2014) 003,
  [\href{http://arxiv.org/abs/1309.4461}{{\tt arXiv:1309.4461}}].

\bibitem{Tanaka:2015aza}
T.~Tanaka and Y.~Urakawa, {\it {Conservation of $\zeta$ with radiative
  corrections from heavy field}},  {\em JCAP} {\bf 1606} (2016), no.~06 020,
  [\href{http://arxiv.org/abs/1510.05059}{{\tt arXiv:1510.05059}}].

\bibitem{Cohen:2020php}
T.~Cohen and D.~Green, {\it {Soft de Sitter Effective Theory}},
  \href{http://arxiv.org/abs/2007.03693}{{\tt arXiv:2007.03693}}.

\end{thebibliography}\endgroup

\end{document}